\documentclass{emulatemyapj}
\pdfoutput=1
\usepackage{amsmath,wasysym,pifont}
\usepackage[colorlinks=true,breaklinks=true,citecolor=blue,linkcolor=blue,urlcolor=blue]{hyperref}
\newcommand{\yes}{\ding{51}}
\newcommand{\no}{\ding{55}}

\begin{document}

\shorttitle{On the detection of Exomoons}
\shortauthors{Michael Hippke}
\title{On the detection of Exomoons: A search in \textit{Kepler} data for the orbital sampling effect and the scatter peak}

\author{Michael Hippke}
\email{hippke@ifda.eu}
\affil{Luiter Stra{\ss}e 21b, 47506 Neukirchen-Vluyn, Germany}

\begin{abstract}
Despite the discovery of thousands of exoplanets, no exomoons have been detected so far. We test a recently developed method for exomoon search, the \textit{orbital sampling effect} (OSE), using the full exoplanet photometry from the \textit{Kepler} Space Telescope. The OSE is applied to phase-folded transits, for which we present a framework to detect false positives, and discuss four candidates which pass several of our tests. Using numerical simulations, we inject exomoon signals into real \textit{Kepler} data and retrieve them, showing that under favorable conditions, exomoons can be found with \textit{Kepler} and the OSE method. In addition, we super-stack a large sample of \textit{Kepler} planets to search for the average exomoon OSE and the accompanying increase in noise, the \textit{scatter peak}. We find a significant OSE-like signal, which might indicate the presence of moons, for planets with $35d<P<80d$, having an average dip per planet of $6\pm2$ppm, corresponding to a moon radius of $2120\substack{+330\\-370}$km for the average star radius of 1.24$R_{\odot}$ in this sample.
\end{abstract}


\section{Introduction}
Our own solar system hosts 8 planets and 16 large ($>1000$km) orbiting moons. Although more than 1,500 exoplanets have been confirmed to date, no exomoon has been detected. The question of the existence of exomoons is interesting, as it will sharpen our understanding of how ``normal'' our own solar system is in the interstellar context. The frequency of exomoons can also guide planetary formation theory. Lastly, some exomoons might even be habitable \citep{Heller2013, Heller2014b, Forgan2014, Hinkel2013}.

Various search methods have been proposed (see e.g. \citet{Heller2014}, and references therein). In what follows, we focus on two photometric methods, dubbed the ``orbital sampling effect'' (OSE, \citet{Heller2014}), and the ``scatter peak'' (SP, \citet{Simon2012}). We will show that both complement each other and that \textit{Kepler}-quality photometry is sufficient to detect individual OSE-like signals.

\begin{figure}
\includegraphics[width=\linewidth]{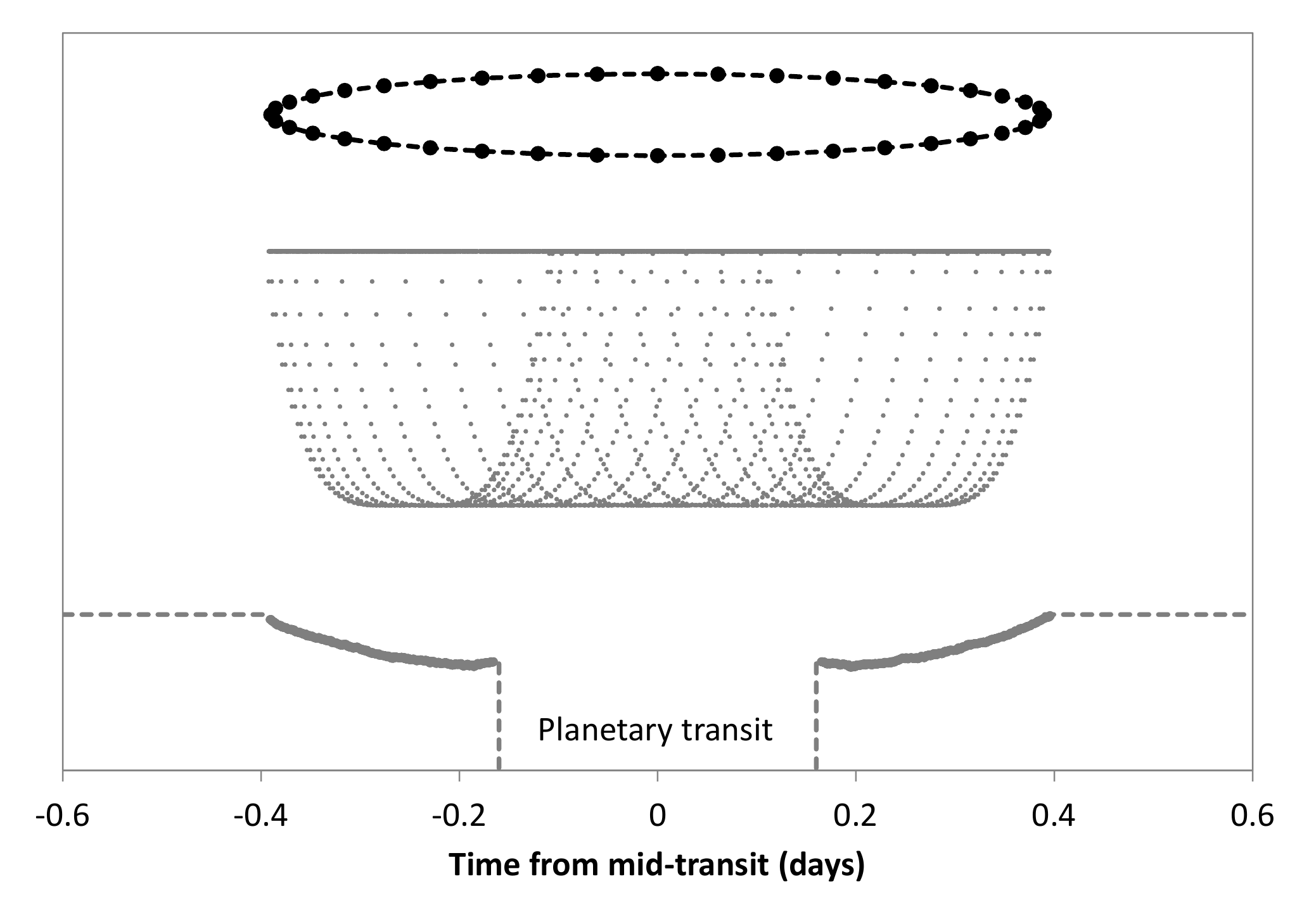}
\caption{\label{fig:OSEsketch2}Numerical simulation of the OSE in a phase-folded plot. The circular moon orbit (top) causes higher sampling at times away from planetary mid-transit. Each dot on the orbit triggered one transit in our simulation, at the given time, over plotted in the middle. When averaging the flux over time, the OSE emerges (bottom). Just before (and after) planetary transit, the OSE curve shows a slightly weaker flux decrease, which can be an indication of real exomoon presence in data, if detected. In this example, the moon orbit is circular and has zero eccentricity and inclination, resulting in a mirror-symmetrical configuration. The jitter in the OSE curve is due to the sampling of ``only'' 40 transits.}
\end{figure}

\section{Method}
\label{sec:method}
Planet transits are generally stacked to increase the signal-to-noise ratio. The OSE and SP also make use of stacking: After adding up many randomly sampled observations, a photometric flux loss appears in the phase-folded transit light curve, reflecting the moon's blocking of light. The effect depends mostly on the moon's radius and planetary distance. The OSE can be used to detect a significant flux loss \textit{before, during and after} the actual transit (if present), which might be indicative of an exomoon in transit. The basic idea is that at any given transit the moon(s) must be somewhere: They might transit before the planet, after the planet, or not at all -- depending on the orbit configuration \citep{Sato2009}. But by stacking many such transits, one gets, on average, a flux loss before, during and after the exoplanet transit (see Figure 3 in \citet{Heller2014}). The depth of this flux loss then reflects the total radii and orbits of the exomoon(s) (Figure~\ref{fig:OSEsketch2}), assuming high enough data quality. This is a simplified sketch, as it shows only a single-moon configuration. In multiple-moon configurations, an additional feature might be added through mutual moon eclipses.

A varying fraction of the flux loss occurs during planetary transit, as will be discussed in Section~\ref{shape}. Unfortunately, it is very difficult to measure these ``lost photons". The effect is very similar to a slight increase in planetary radius (transit depth). Furthermore, limb darkening is an unknown parameter and needs to be estimated. Both factors, limb darkening and the unknown planetary radius, easily mask the flux loss during planetary transit. For example, in our Sun-Earth-Moon configuration, the Earth produces an 84ppm dip, and the Moon OSE contributes $\sim$3ppm (of 6ppm transit depth) during Earth transit. Instead of a planetary radius of 6378km, one might simply mistake the combined effect for a planet of radius 6490km, a 1.8\% difference in planetary radius. While the change in transit depth itself might be borderline detectable in favorable cases, one would instead need to distinguish the transit shapes between the two scenarios. This will require further modeling (including the effect of limb darkening uncertainties), which we strongly encourage, but it is beyond the scope of this paper.

On the negative side, the OSE can only detect a sub-sample of all moon configurations: It requires the moon to be in a near sky-coplanar orbit to allow for the moon transiting the star \citep{Sato2010}. Among the large solar system moons, 4 out of 16 have inclinations over 1$^{\circ}$: Earth's moon (5.5$^{\circ}$), Triton (130$^{\circ}$), Iapetus (17$^{\circ}$) and Charon (120$^{\circ}$). If this rate is typical, then the OSE would work for the other $\sim$75\% of cases. 

\begin{figure}
\includegraphics[width=\linewidth]{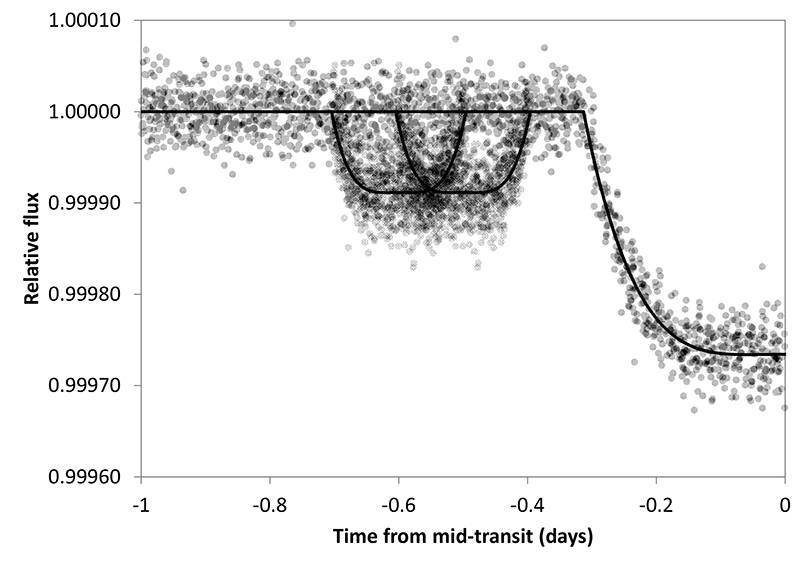}
\caption{\label{fig:ScatterSim}Simplified sketch of the SP effect in a phase-folded plot showing a stack of 3 epochs. During one epoch, no exomoon transit occurred (straight line at flux=1). During the other two epochs, exomoon transits occurred before planetary ingress, centered at -0.6d and -0.5d. Data points with noise (dots) show higher scatter during this phase-folded time. After stacking many transits, the scatter peak would be sampled as described by the photometric OSE. For clarity, the noise shown here is much lower (by a factor of 10) than ideal \textit{Kepler} noise.}
\end{figure}

While the presence of a significant flux loss might be an indication of a transiting body, it could also be caused by other sources, such as imperfect detrending of data, or time-correlated (``red'') stellar noise. It is therefore beneficial to have an (almost) independent method at hand for vetting purposes. The SP method is based on the fact that the geometrical exomoon configuration is very likely different during every exoplanet transit: On some transits, the moon might be ahead of the planet, on other transits behind it. When stacking many transits, at a given phase folded time, one gets a flux loss in some cases, and not in others. This results in increased scatter (photometric noise) when compared to out-of-transit times (Figure~\ref{fig:ScatterSim}). It is important to note that the OSE is only sensitive to \textit{flux loss} (and not the scatter), and the SP includes a median-filter to subtract the actual flux loss (if any), and is thus only sensitive to the \textit{scatter}. This makes both methods independent and complementary. Future theoretical advantages might allow simulations which take both effects into account by using Monte-Carlo runs to simulate various planet-moon configurations and the resulting OSE+SP, and providing a best-fit model for the given data, taking into account both flux loss and the corresponding increase in variance (Ren\'{e} Heller (2014), priv. comm.).

\subsection{Maximizing exomoon detection efficiency}
A literature search has brought up only one example for an application of the OSE to real data \citep{Heller2014b}; a null result for KOI189.01\footnote{After publication of the work by \citet{Heller2014b}, the system was shown to be an eclipsing stellar binary, rather than a planet, by \citet{Diaz2014}.}. For the SP, no application could be found. Thus, no significant experience on potential error sources is available. In the following, we have tried to account for all relevant potholes, but (as is always the case), there might be further tests required. We consider the methods and tests in this work as not matured enough to claim any exomoon detection. We will present our test results using several candidates as examples. The advantage of our approach is that only minimal CPU time and human labor ($<1$hr per candidate) is required, after the required tests are established and implemented in software. Then, our framework can select candidates from a large sample, and suggest these for an in-depth analysis using photodynamic modeling as established by \citet{Kipping2011}. This modeling requires $\sim$50yrs of modern CPU time for a single moon, plus considerable human effort to assess the Bayesian evidence of the result. By pre-selecting promising candidates for the modeling, the total output (per human labor time and per CPU-year) of exomoon detections can be maximized.

\subsection{Data preparation}
We employed the largest database available: High precision time-series photometry from the \textit{Kepler} spacecraft, covering 4 years of observations \citep{Caldwell2010}. Based on a list of all confirmed (821) and unconfirmed (3,359) transiting \textit{Kepler} planets \citep{Wright2011}\footnote{www.http://exoplanets.org, list retrieved on 18-Nov 2014}, we downloaded their \textit{Kepler} long-cadence (LC, 30min) datasets. For the OSE, LC data is equally well suited to SC (short cadence, 1min) data, because the effect is spread over a long ($\gg$30min) time. The stacking of many transits and  binning makes it irrelevant where exactly the ``missing photons'' are placed. For the SP however, SC data is preferable, due to the smearing effect \citep{Simon2012}. The background is that, on average, half of the flux loss during ingress and egress is shared with nominal flux, if the bin size is too large. The subsequent scatter measure is then less sensitive. 

As LC data only requires $1/30$ of the SC space and processing time, we have only processed SC data for those candidates that triggered an OSE. The bluring effect also causes additional errors on a planet (and to a lesser extent moon) fit; but the errors are small and most relevant for short transit durations. Usually they cause an error of $\sim$0.01\% for the planetary radius, much smaller than the formal errors from an adjusted fit \citep{Gilliland2010}.

\subsection{Automatic Data processing}
The detrending for our fully automatic routines was very conservative by only fitting a line to each \textit{Kepler} quarter and star (while masking times of transits), and normalizing the pre-search data conditioning (PDCSAP) \citep{Smith2012} flux. Only for interesting candidates we have manually detrended the raw SAP flux data, to check for potential PDC conversion errors. The result was visually inspected, and all those stars deviating from a flat baseline were directly rejected from further analysis. This decreased the number of candidates, but avoided complicated (and often ambiguous) detrending of e.g. (often partially) variable stars, or complicated instrumental trends.

For the remaining 2,328 exoplanets, phase-folds and scatter measures were calculated in 10 bins before and after the transits. The bin width was 0.5 transit lengths, with the exception of the two bins directly before/after the planet transit, which were shortened to avoid contamination from transit timing or transit duration variations.

We will present the results from processing all planets in section~\ref{sec:all}. Afterwards, we will discuss the framework of tests for false positives (section~\ref{sec:framework}) and apply this toolbox to example candidates in section~\ref{sec:application}.

\section{Results for processing all planets}
\label{sec:all}
We created a combined phase-fold of all useful data, to search for an average exomoon effect. For the 2,328 transiting exoplanets with good data quality, the transit duration was transformed to unity, and the flux was measured in 10 bins before and 10 bins after the phase-folded transits. Then, we added a buffer of 10\% of the transit duration both before and after the transits to be excluded from any further analysis, so that any transit duration uncertainty, or transit timing/duration variations are also excluded from mismatch with exomoon flux loss. By visually inspecting all phase-folds, we removed several candidates that showed excessive TTVs/TDVs. We find that planets which exhibit $>$10\% TTVs/TDVs (more than our buffers) are visually easy to identify; it might however be a valid approach in future processing to calculate individual TTVs/TDVs for each planet (and transit), and adjust the stack accordingly.

We tried normalizing the transit depths, but found this only added noise.

To sum up, all transits of all planets were transformed in width to one big meta phase-fold (``super-stack''). We have then examined the flux of this before and after the actual transit, in order to search for flux loss caused by exomoons, as described by the orbital sampling effect.

\subsection{Data cleansing}
The resulting data were strongly dominated by outliers. This is understandable as many factors contribute to different transit depths, and noise levels: The size and apparent brightness of the host star, the size of the planet, stellar variability on a scale lower than rejected by our visual inspection, instrumental differences, and others. We decided on a set of filters to remove outliers, and included the stellar brightness ($<13$mag in \textit{J} as measured by 2MASS), the scatter per star (we kept the better half), and the data completeness (we required 80\% of the \textit{Kepler} spacecraft duty cycle). Accepting all stellar brightnesses, and all data completeness levels does only slightly deteriorate the results. This is a trade-off hard to quantify, as one adds both signal (more stars) and noise (lower quality stars). We found, however, that the scatter of the individual lightcurve is of great influence. Our results are also consistent for the case of removing only the few very strongest outliers, so that we consider the data to be robust.

It is important to mention a selection effect from this filter choice: When rejecting dimmer and/or noisier stars, the average stellar radius changes. Smaller stars (e.g. M-dwarfs) exhibit more stellar noise \citep{Basri2013} and are usually less luminous. Consequently, our sample is shifted towards larger radii. While our sample has an average stellar radius of 1.24$R_{\odot}$, the sample of planetary periods 35--80d, without further filtering, has an average of 1.11$R_{\odot}$. In the future, more refined exomoon studies might take this into account.

\begin{figure*}
\includegraphics[width=0.5\linewidth]{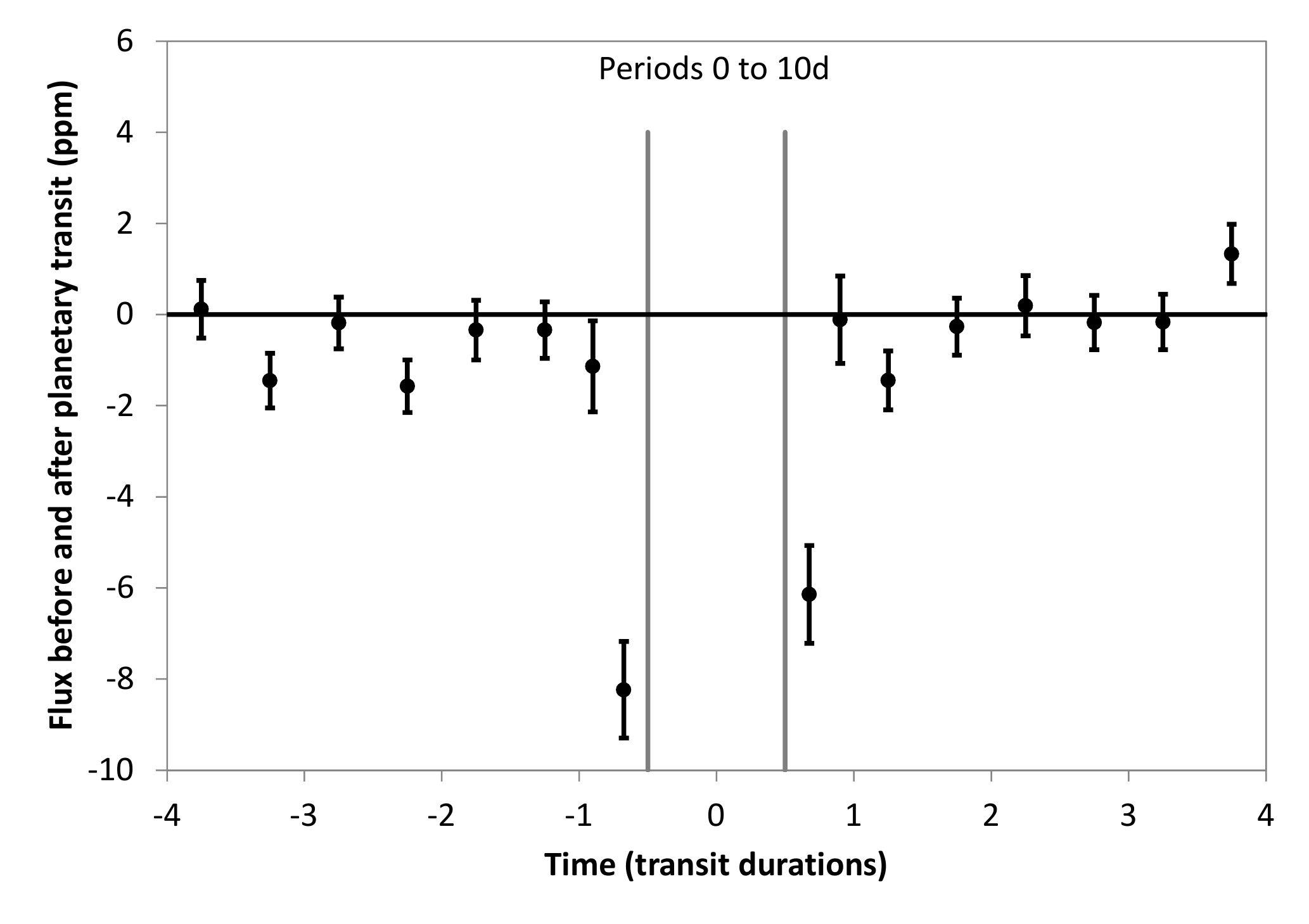}
\includegraphics[width=0.5\linewidth]{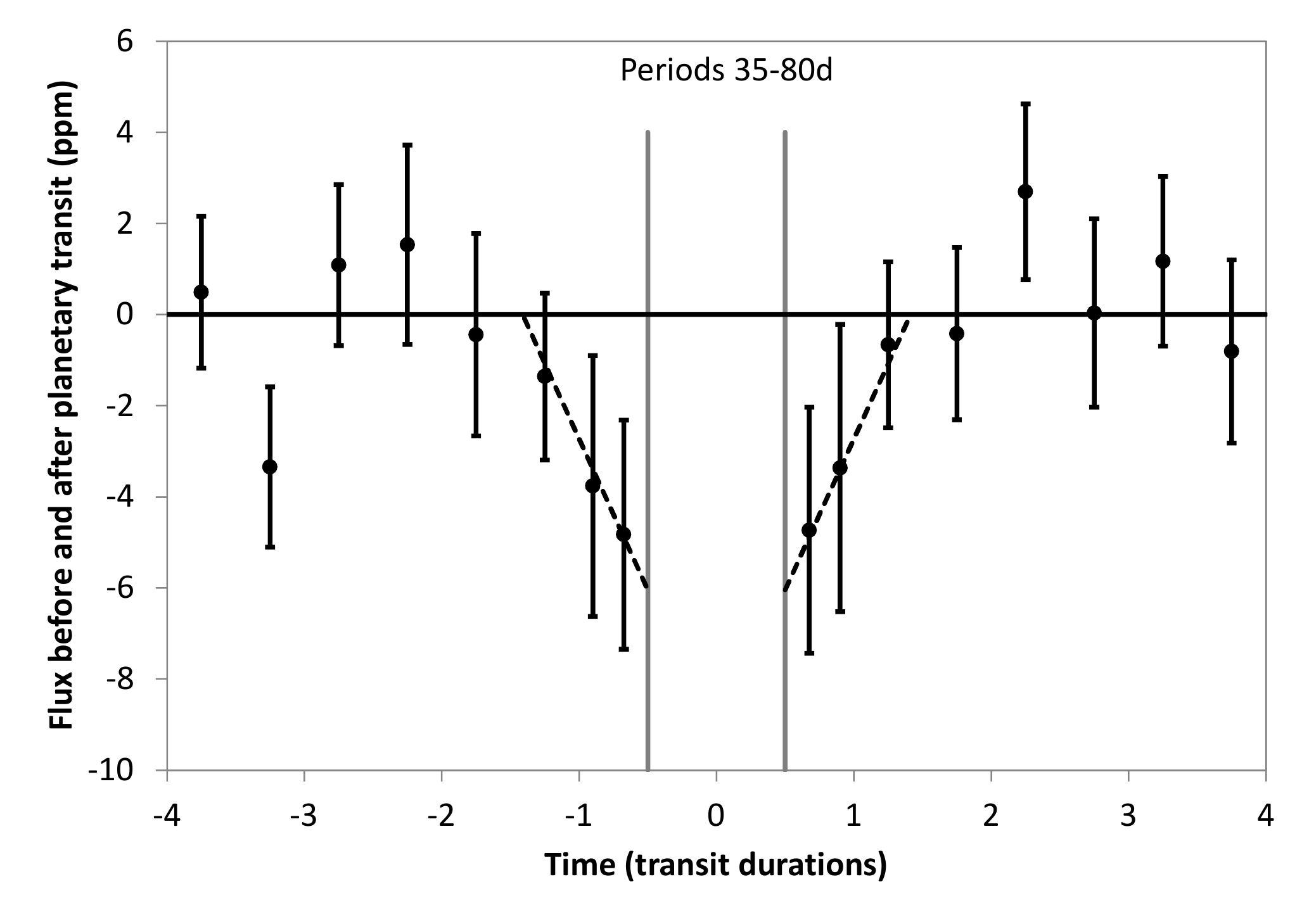}

\includegraphics[width=0.5\linewidth]{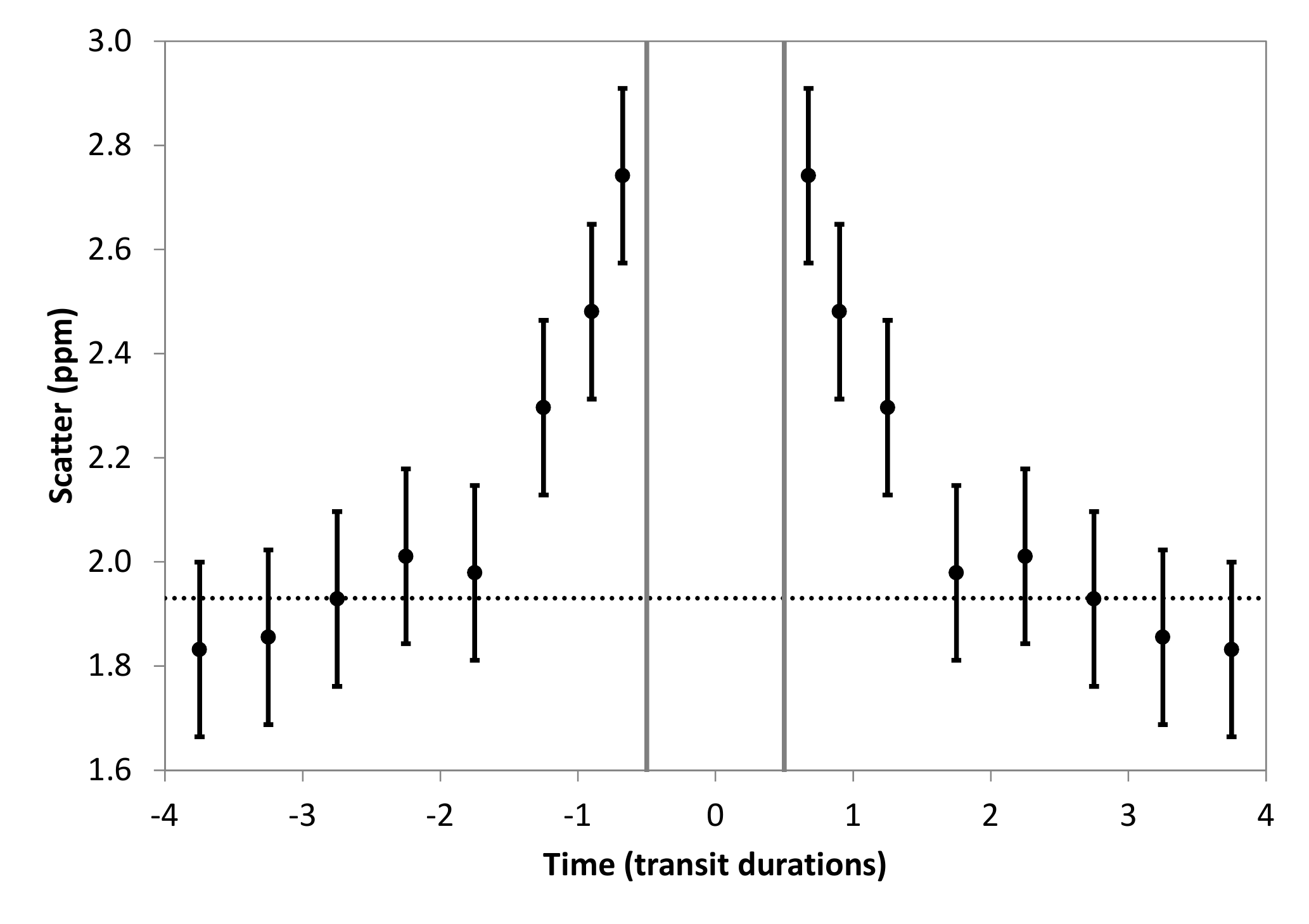}
\includegraphics[width=0.5\linewidth]{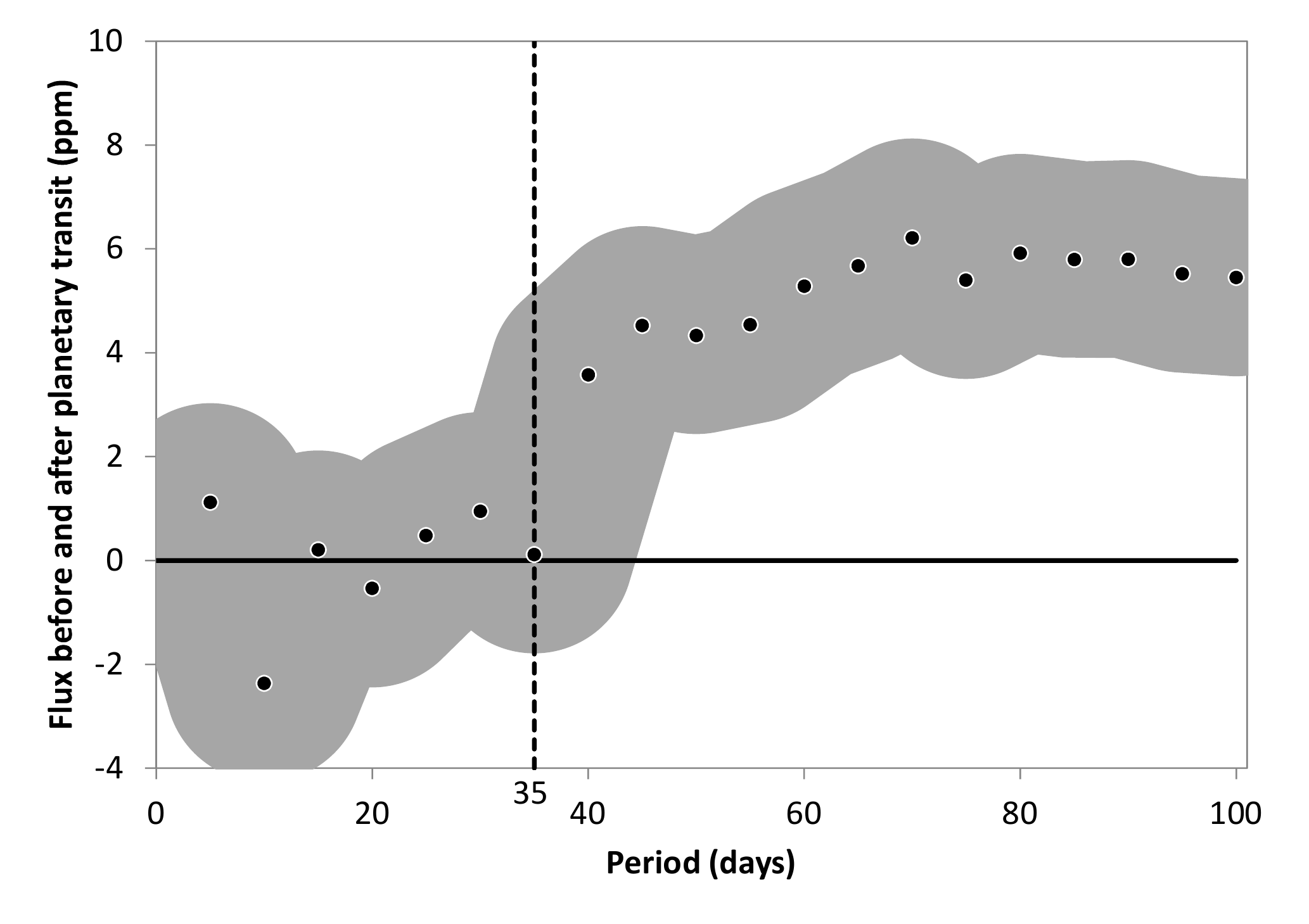}

\includegraphics[width=0.5\linewidth]{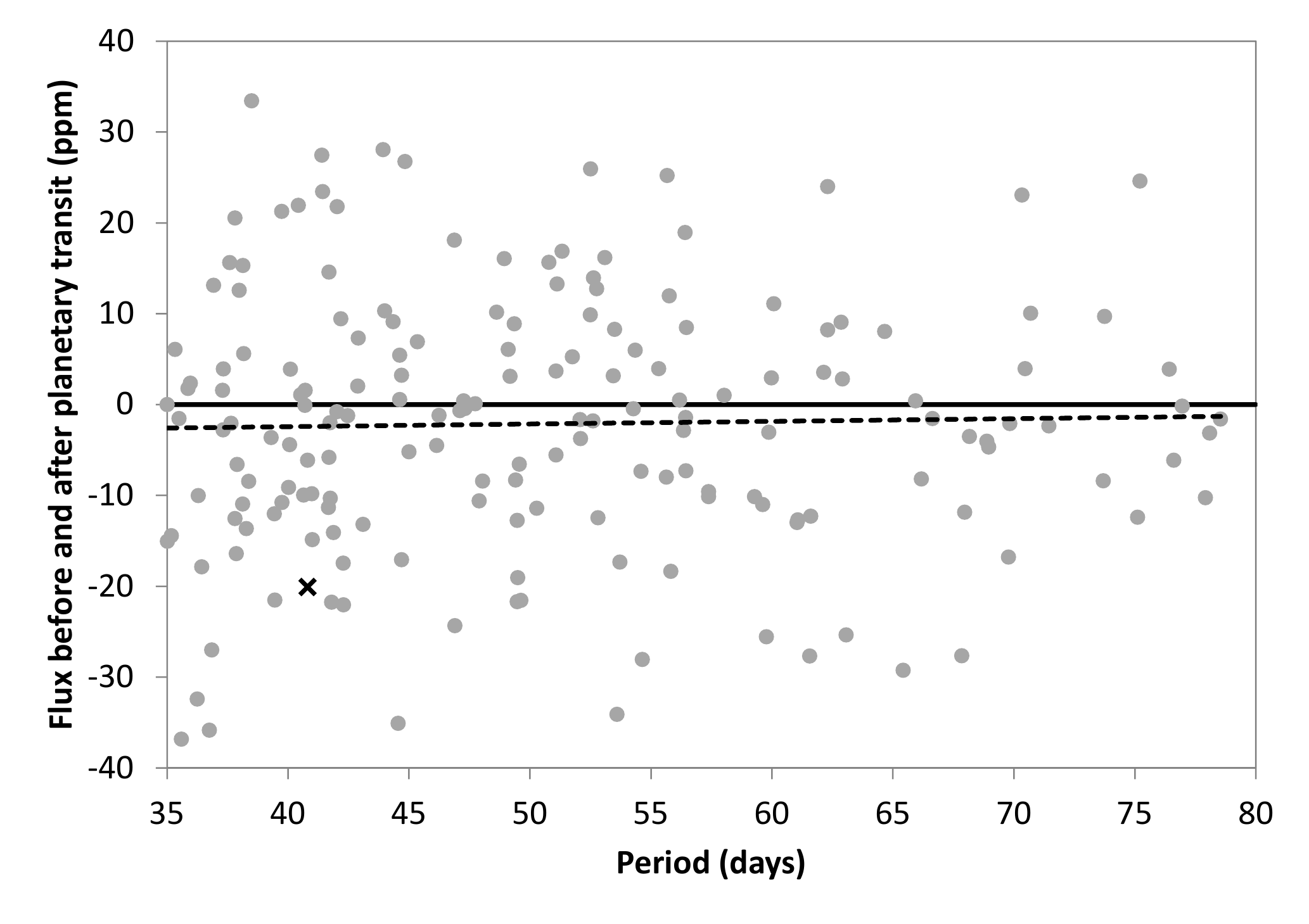}
\includegraphics[width=0.5\linewidth]{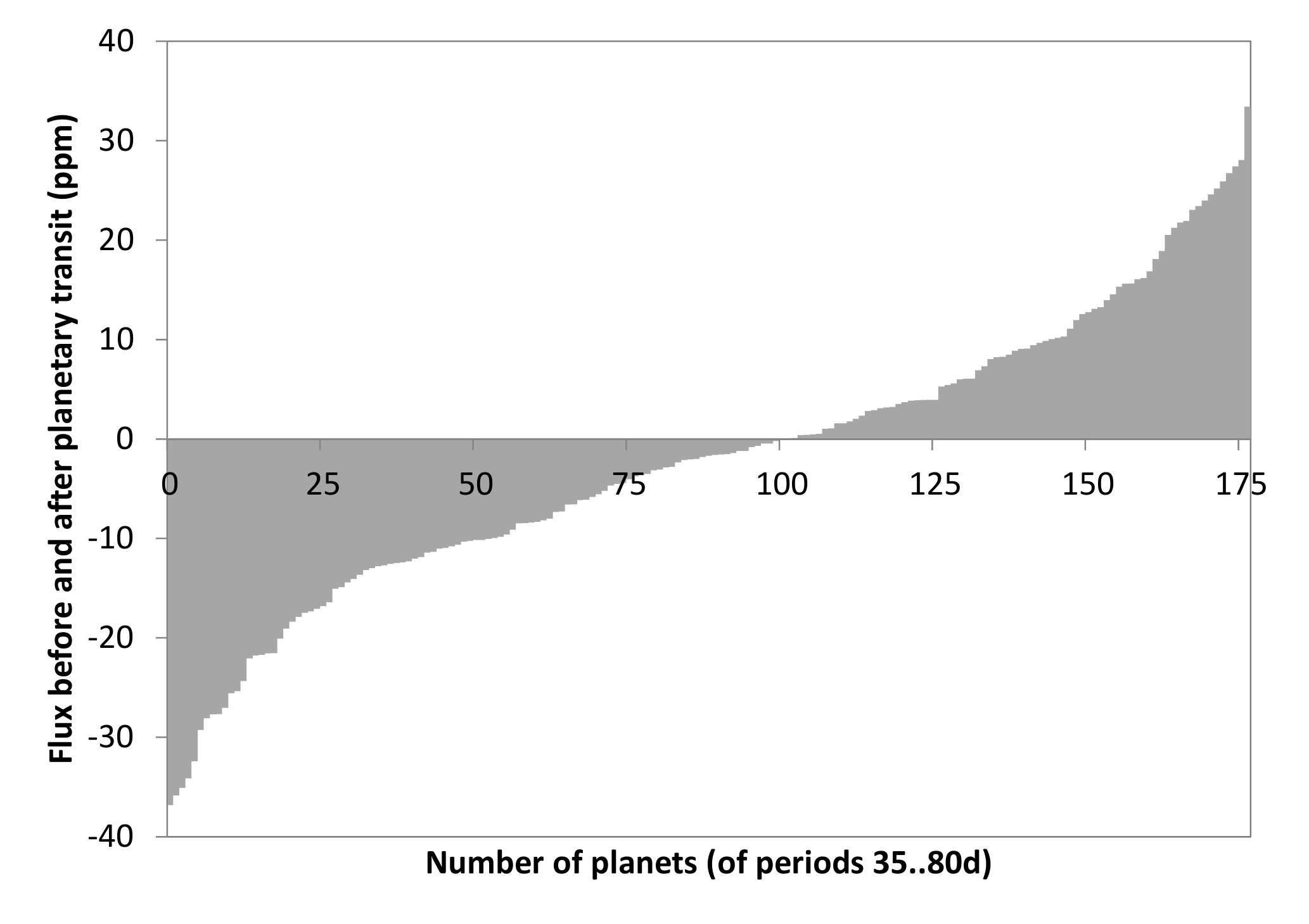}
\caption{\label{fig:stack-data}Top left: Stacked phase fold for periods $<10d$. Flux loss is only apparent very close to the planetary transit. Top right: Periods $35d..80d$ with flux loss in neighbouring bins. Dashed line shows a symmetrical linear regression. Middle left: Scatter peak for periods $35d..80d$ is clearly detected, here shown in symmetrical form. Middle right: Cumulative flux loss (positive values) depends strongly on the planetary period. The exomoon effect from OSE is visible for P$>35d$ (vertical line), exceeding the uncertainty (grey shading) for P$>40d$. Bottom left: Planetary period versus flux loss/gain. While any flux gain is likely caused by noise, it shows the high overall scatter of this work. The average (linear trend as dotted line) shows significant ($p=0.048$) flux loss. Kepler-264b is marked with a cross. Bottom right: Histogram of planets between 35 and 80d periods. Of the 176 planets, 101 show a flux loss (57\%). While many of these might be false positives (also on the plus side), the distribution is already significantly skewed in a binomial test ($p=0.029$).}
\end{figure*}

\subsection{Results for the ``super-stack''}
We find a strong correlation of flux loss to the planetary orbital period. Figure~\ref{fig:stack-data} shows our stacked flux for 652 planets with periods $<10d$ (left) and for 177 planets with periods $35d<P<80d$ (right). Short-period planets are not expected to possess moons \citep{Cassidy2009}, so that this group can be used as a reference point. However, a significant flux-loss just before and after the planetary transit is also apparent in the $<10d$ group (and most prominent in the very shortest period planets), perhaps due to a slight leak of transit flux loss into the neighboring bins, despite our buffers. Indeed, when neglecting the bins closest to the planetary transit, the dips are gone. 

When plotting only the 177 longer-period ($35d<P<80d$) planets, an additional flux loss appears, starting and ending at $\sim 2$ planetary transit durations. As will be explained in section~\ref{sub:c1}, at least 15-20 transits are required before the OSE can build up in a single case. When sampling many longer-period ($>80d$) planets, a similar flux loss should appear, which we do not detect. We attribute this to our sample size: While we have a total of 4,631 transits (from 176 planets) with periods $35d<P<80d$, we have only 727 transits from 77 planets in the group for longer-period planets. The smaller number is caused by two reasons: First, \textit{Kepler} is less sensitive in detecting longer period planets (at a given planetary radius), as less transits can be stacked. This reduces the sample sizes. From this reduced sample, only 224 planets survived our visual rejection of excessively noisy candidates, as discussed in the previous section. Finally, when applying the formal criteria (brightness, jitter, data completeness), only 77 remain. This lower number of only linearly detrended light curves might not be sufficient for the detection of the OSE in this group. We encourage future studies to try different selection criteria, and detrending methods, in order to explore the parameter space of flux loss for all planetary periods.

For the planets with periods of $35d<P<80d$, the effect is measured as an average flux loss of $6\pm2\times10^{-6}$ (6ppm), corresponding to a moon radius of $2120\substack{+330\\-370}$km for the average (1.24$R_{\odot}$) star in this sample. Also, a significant scatter peak is detected, with the bins affected from the OSE having a scatter of $2.7\pm0.2$ ppm, compared to the baseline of $1.9$ppm. The detection of both effects, the OSE-like signal and the SP, may indicate exomoons as the underlying cause.

\section{Framework for individual candidates} 
\label{sec:framework}
A simple dip in the data is necessary, but not necessarily sufficient for the detection of an exomoon. As will be explained in section~\ref{sub:sensitivity}, we have found 4 out of 56 planets to exhibit such a dip at the 95\% significance level, likely not caused by TTVs/TDVs. One method would be to refer these candidates to photodynamical modeling, but this is very expensive in terms of labor and CPU-time. Therefore, we have developed a framework of tests for false-positives. This can either reduce the number of candidates, or at least rank and prioritize them for further analysis.

\subsection{Required results from the OSE}
\label{sec:req-ose}
For the presence of a moon, we propose the following criteria to be checked:

\begin{itemize}
\item[{\textbf{OSE1}}] A significant ($p=95\%$) flux loss before and after the planetary transit.
\item[{\textbf{OSE2}}] A significant ($p=95\%$) trend towards more flux loss at closer to the transit than further away.
\item[{\textbf{OSE3}}] A significant ($p=95\%$) flux loss shall not appear when only a small number of transits ($n<10$) is used.
\item[{\textbf{OSE4}}] The dips before ingress and after egress shall be the largest dips in the dataset.
\item[{\textbf{OSE5}}] The dips should not be caused by star-spots.
\item[{\textbf{OSE6}}] The dips should not be caused by rings.
\item[{\textbf{OSE7}}] The amount of flux loss needs to be physically plausible.
\item[{\textbf{C1}}] The signal shall not be located in a small part of the data only, but be contained in $>50\%$ of it.
\end{itemize}

The choice of significance level, $p=95\%$ ($\sim$2$\sigma$) can of course be varied, depending on the data quality and the capacity for follow-up analysis. To calculate this significance, our null hypothesis is a normalized nominal flux. We then perform a two-sample mean-comparison test with the potential flux loss data as one group, and the reference data (long-out-of-transit) in the other group:

\begin{equation}
	t = \frac{(\bar{x}_{1}-\bar{x}_{2})-\Delta} 
	         {\sqrt{ \frac{ {s_1}^2} {n_1} +  \frac{ {s_2}^2} {n_2} }}    
\end{equation}

where $\bar{x}_{1}$ and $\bar{x}_{2}$ are the means of the two samples, $\Delta$ is the hypothesized difference between the population means, $s_1$ and $s_2$ are the standard deviations and $n_1$ and $n_2$ are the sizes of the two samples.

This mean-comparison test gives the two-sample $t$ statistic, for which we calculate the $p$-values using the $t(k)$ distribution. The test slightly underestimates the OSE significance, as it does not account for the non-linearity in slope of the OSE.

We have considered a similar measure, the signal-to-noise ratio for transits \citep{Jenkins2002,Rowe2014}, which compares the depth of the transit mode compared to the out-of-transit noise:

\begin{equation}
\label{eq:sn}
{\rm S/N} = \sqrt{N_{T}}\frac{T_{dep}}{\sigma_{OT}}
\end{equation}

with $N_{T}$ as the number of transit observations, $T_{dep}$ the transit depth and $\sigma_{OT}$ the standard deviation of out-of-transit observations. We find that both measures give similar results, and choose to continue giving significance levels ($p$-values) in the following. For future studies, we encourage the creation of a significance measure which fully accounts for the non-linear curve shape of the OSE. 

When not noted otherwise, the bin length for the out-of-transit group was taken to be one half transit duration before, and after planetary ingress (egress). 

Planets on very short orbits are increasingly unlikely to host moons, due to mass loss \citep{Cassidy2009}, very small Hill radii, and orbital decay due to tides \citep{Barnes2002}. As the range of stellar and planetary temperatures, masses and radii is large, a single limit cannot be given; estimates are in the range of 5--15 days for the minimum orbital period of the host planet. For long-period planets, there are not enough transits in the data for the build-up of the OSE. For a minimum of 20 transits within the 4 years of \textit{Kepler} observations, the maximum period for possible detection is then 73 days. Taking only planets with periods of 15--73 days, we have 311 confirmed planets plus 958 ``Kepler Objects of Interest'' (KOIs). Using the traditional 95\% level significance level, and detecting a dip both before and after planetary transit, results in a total false positive probability (per candidate) of $2\times(1-0.95)=0.0025$. Thus, one can assume a total of $0.0025\times1269=3$ false positives, a number which can further be reduced with the other tests in our framework. Therefore, we suggest to be relaxed towards the requirements and also accept partially negative test results, especially as the tests sometimes yield \textit{false negative} results.

Regarding OSE1, it is useful to treat ingress and egress separately, and also binned together (section~\ref{sub:ose1}). As shown in Figure~\ref{fig:OSEsketch2}, the dip has a defined shape, most strongly dependent on the number of moons, their radii and semi-major axes. As these parameters can only be estimated, we suggest to perform a test (OSE2, section~\ref{sub:ose2}) for the slope: The flux loss should be greater towards planetary transit, than further away from it. We use a linear regression to test for a slope, knowing that it only roughly approximates the true OSE curve. To quantify the result of the linear regression, we again give $p$-values to judge the rejection of our null hypothesis of a constant flux (zero slope parameter). 

Furthermore, such a dip emerges only after sampling a sufficient number of transits. This can be tested (OSE3, section~\ref{sub:ose3}), together with numerical simulations for the number of transits required. In this section, we will select (many) randomly chosen subsection of the data, and check if they are consistent (within the errors) with the total data, so that most, and not just a few data segments contain the potential exomoon signal.

Then, when the dips have been established as genuine, we need to make sure that such is not common within the dataset: The two dips should be, ideally, the deepest in the whole dataset (OSE4, section~\ref{sub:ose4}). 

After this step, in case the dips could not be explained as stellar or instrumental noise, we shall consider a physical cause. Dips can be caused by star spots (OSE5, section~\ref{sub:ose5}) or, more interestingly, by rings (OSE6, section~\ref{sub:ose6}). We will explain our methods in detail in these sections.

If these can be ruled out (or appear unlikely), we shall seriously estimate the actual star-planet-moon configuration and its stability, to check whether the system is physically possible, plausible and stable (OSE6, section~\ref{sub:ose7}). These tests include the discussion of long-time stability, and an orbit within the Hill-radius and outside the Roche-lobe. 

Inspired by \citet{Kipping2014}, we also employ a useful data consistency check, demanding that the signal should be spread broadly in the data, and not be located in only a part of it (C1, section~\ref{sub:c1}).

\subsection{Required results from the SP}
Equally, we define the following criteria for the detection of a scatter peak:
\begin{itemize}
\item[{\textbf{SP1}}] A significant ($p=95\%$) scatter peak before, at and after the transit.
\item[{\textbf{SP2}}] Shifting the period shall not significantly decrease the scatter peak.
\item[{\textbf{SP3}}] Accounting for TTVs and TDVs.
\item[{\textbf{SP4}}] The strength of the scatter peak shall be astrophysically plausible, and scale with the amount of data used.
\item[{\textbf{C2}}] The signal shall not be located in a small part of the data only, but be contained in $>50\%$ of it.
\end{itemize}

While SP1 is self-explanatory (section~\ref{sub:sp1}), SP2 and SP3 (section~\ref{sub:sp2}) are more complicated as the transit folding might not be perfect due to uncertainties, TTVs and TDVs. Lastly, as for the OSE, the amount of scatter should correspond to the estimated star-planet-moon configuration, and it should be be spread over all of the data (section~\ref{sub:c2}).

\begin{table*}
\center
\caption{Detrending parameters for the candidates\label{tab:detrendingmethods}}
\begin{tabular}{lcccc}
\tableline
Parameter                              & 241b      & 241c      & 264b      & 367.01\\
\tableline
Median boxcar [days]                   & 1.0 - 2.0 & 1.0 - 2.0 & 2.0 - 3.0 & 0.2 - 1.0 \\
Masking [\% of transit durations]      & 150 - 300 & 150 - 300 & 150-200   & 150 - 300 \\
Visual fit quality                     & Very good & Very good & Bad       & Excellent \\
Same result from polynomial detrending & Yes       & Yes       & Yes       & No        \\
\tableline
\multicolumn{5}{l}{Parameter ranges give very similar results.}

\end{tabular}
\end{table*}

\begin{figure*}
\includegraphics[width=0.5\linewidth]{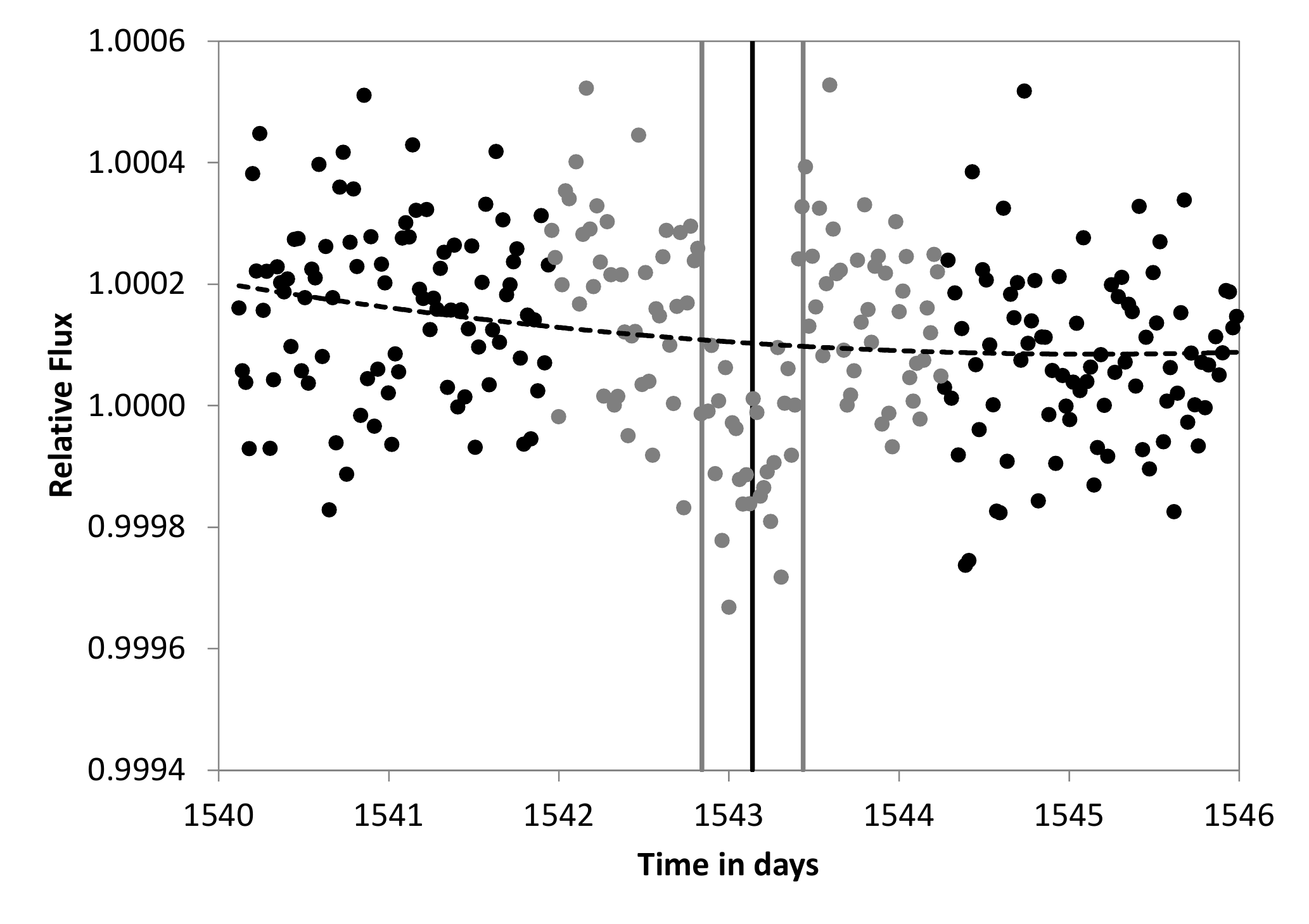}
\includegraphics[width=0.5\linewidth]{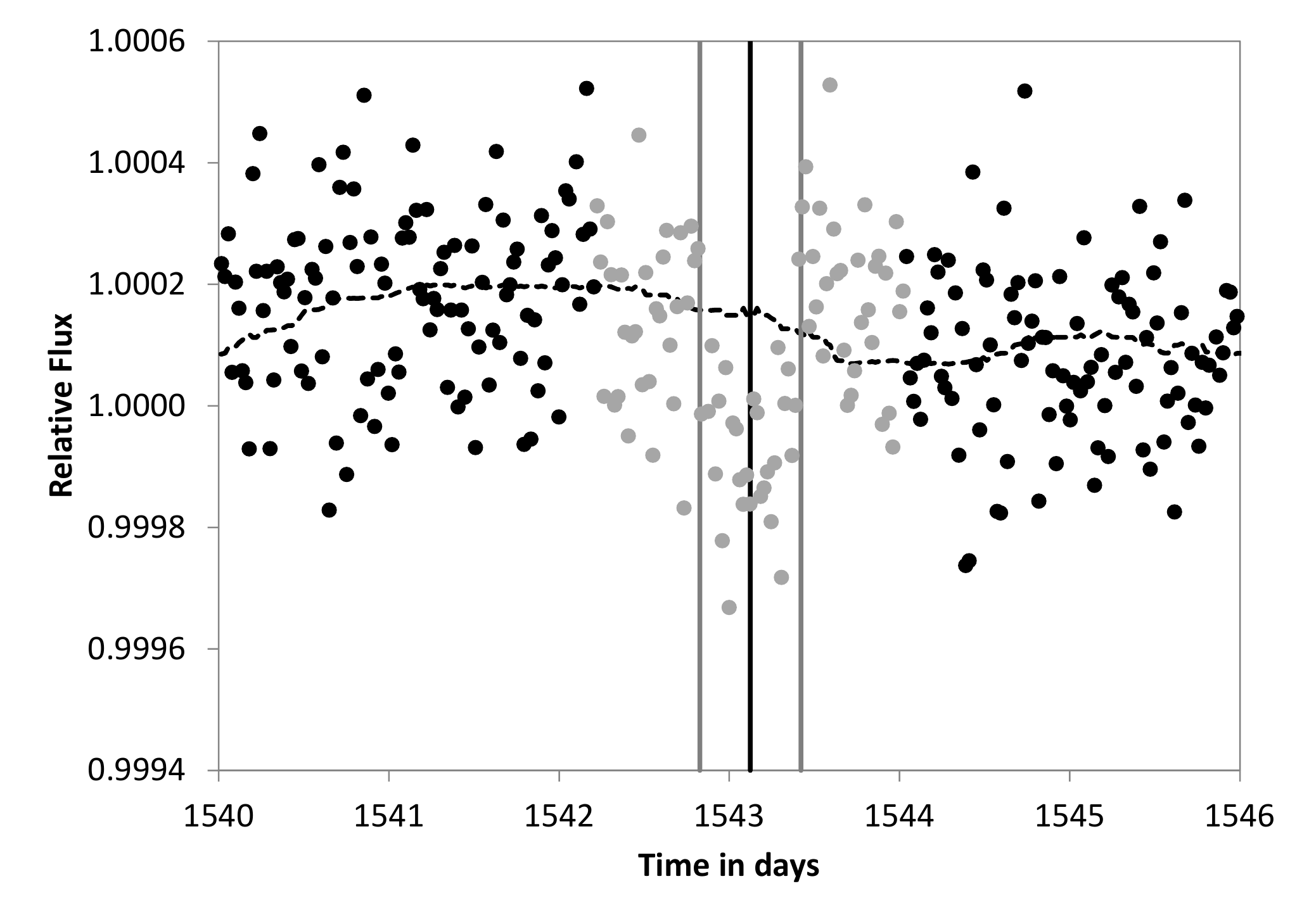}

\includegraphics[width=0.5\linewidth]{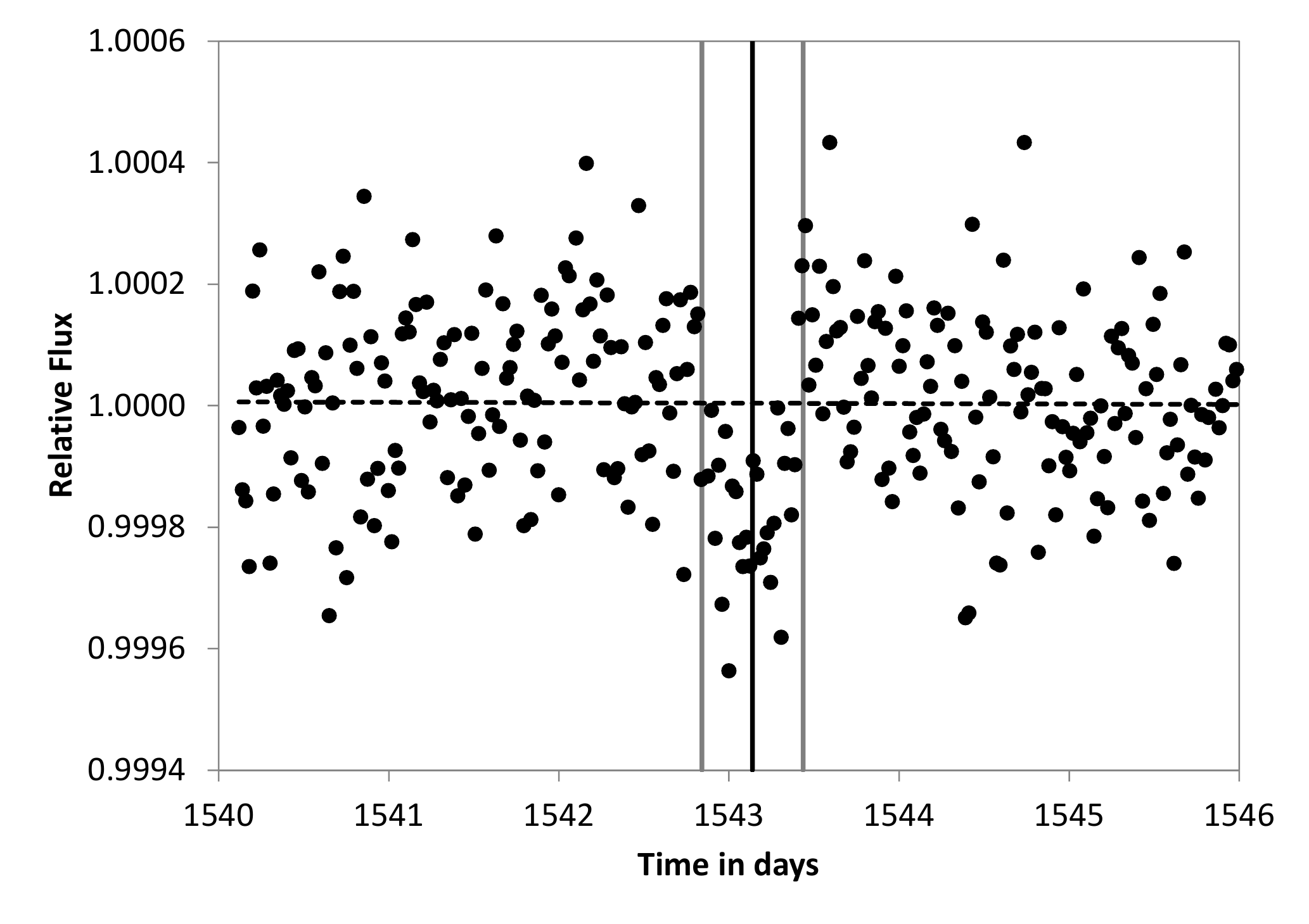}
\includegraphics[width=0.5\linewidth]{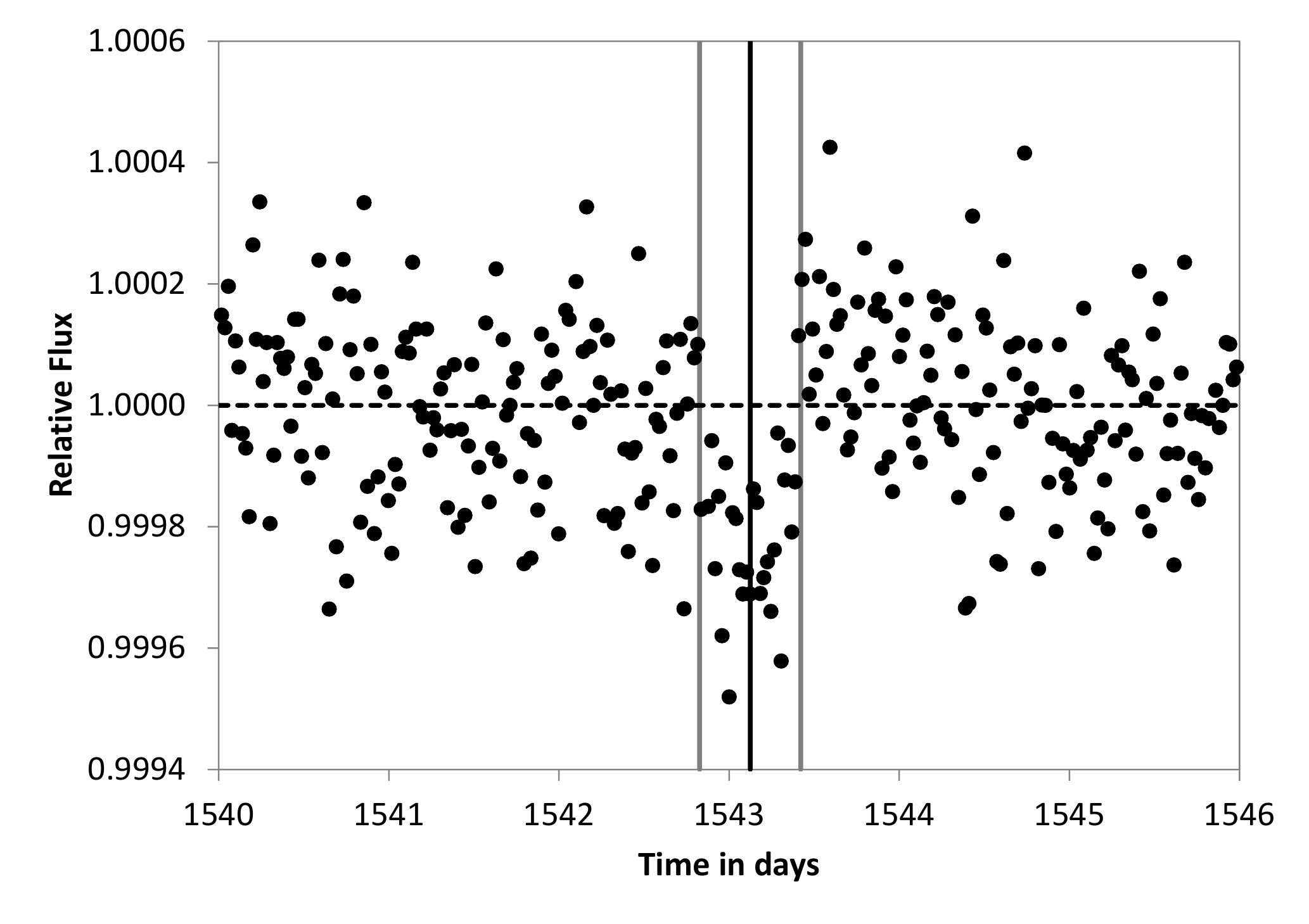}
\caption{\label{fig:detrending}Detrending with a parabola (left) and sliding median (right) of the last recorded transit of Kepler-264b at Barycentric Julian Date $\sim1542$ (MJD-2454833). Grey dots are excluded from the fit to protect the potential flux loss from the OSE (top). Bottom: The parabola result has an average normalized flux (excluding transit and potential exomoon time) at slightly above 1, which has to be normalized for every single transit individually, so that the flux for the time long before and after the transit is set to 1. Note that the results from both methods are very similar, which is a useful consistency check.}
\end{figure*}

\subsection{Manual detrending and baselining}
\label{sub:detrending}
The initial automatic search process had only employed a linear detrending and normalization, which is sufficient to reject the vast majority of negative candidates. For the remaining interesting candidates, however, great care must be taken regarding detrending and baselining. Usually, the baseline used for transits is the time immediately before and after the transit. This method, by definition, renders the orbital sampling effect invisible. Instead, we need to derive the baselines from times long ($>2$ transit durations) before and after the actual transits. The detrending must also protect not only the times of transit, but also at least two transit duration before and after the transit, in order not to smooth out any flux loss. Wrongly detrending right though the (moon and/or planet) transit dips destroys the OSE: The fitted curve goes deeper (as influenced by the dip), and thus the resulting detrended data is higher in flux, so that any flux loss from (averaged, stacked) OSE is lost.

We examined both the raw \textit{Kepler} data (Simple Aperture Photometry, SAP) and the Presearch Data Conditioning (PDC) result of the \textit{Kepler} data analysis pipeline. The latter tries to remove discontinuities, outliers, systematic trends and other instrumental signatures, while preserving planet transits and other astrophysical signals. It uses a Bayesian Maximum A Posteriori (MAP) approach ``where a subset of highly correlated and quiet stars is used to generate a cotrending basis vector set which is in turn used to establish a range of reasonable robust fit parameters.'' \citep{Smith2012}. 

For the SAP data, we removed outliers with a sliding median and a 3$\sigma$ filter. Afterwards, we tried two common detrending methods with both data sources. First, we subtracted a least-squares parabola fit to an appropriate time before and after mid-transit (e.g. \citet{Santerne2014,Gautier2012}). For Kepler-264b, we used 3 days before and after. Alternatively, we applied a sliding median to the whole dataset (e.g. \citet{Carter2012,TalOr2013}). For both methods, we removed data points affected by other transiting planets in the system, if present. Also, to protect the very slight putative moon flux loss, we blinded the detrending algorithm for the time span of 1.5 transit durations from mid-transit, so that it could not remove such a dip. For Kepler-264b and the median detrending, a boxcar length of 2-3 days was optimal, as it steps over the blind period and still adapts to trends on the order of a few days. For best results, these times need individual adjustments for each candidate. It is important to ensure that an adequate range of detrending parameters gives similar results, in order to qualify the chosen detrending methods and parameters as robust. A useful cross-check is to slowly shrink the protected flux area, and/or shorten the boxcar length of a sliding median detrending. As a consequence, the OSE must vanish, as the detrending adapts more and more to putative slight moon dips, and removes them. The choice of the size and duration of the boxcar is also a crucial aspect for the detection of a SP
signal. For a detailed discussion, we refer to \citet{Simon2012} (their section 4.2).

We find that the results are consistent with SAP and PDC-SAP data, but not always for parabola (polynomial) and median detrending, as listed in Table~\ref{tab:detrendingmethods}. The PDC-SAP data gives less noise, as is expected due to the additional co-trending vectors which effectively remove instrumental trends. Most of the noise in PDC-SAP is likely due to astroseismic jitter -- section~\ref{sub:ose5} will discuss the (likely) rotational period of e.g. 11.1d for Kepler-264, accounting for an amplitude of 81ppm with strong sidelobes.

For reference, we show the detrending results for PDC-SAP data and both detrending methods for Kepler-264b in Figure~\ref{fig:detrending}. The results  differ only slightly. Unfortunately, these detrending methods cannot be used for the scatter peak, as will be discussed in section~\ref{sub:sp1}. We list the range of adequate detrending parameters in Table~\ref{tab:detrendingmethods}. We also release an interactive spreadsheet\footnote{\url{http://jaekle.info/ose/detrending.html}} for the interested reader. In this sheet, masking times and boxcar length can be modified, and the resulting flux loss is displayed instantly. The accompanying website provides an introduction to the parameters, and explains possible conclusions.

\begin{figure*}
\includegraphics[width=0.5\linewidth]{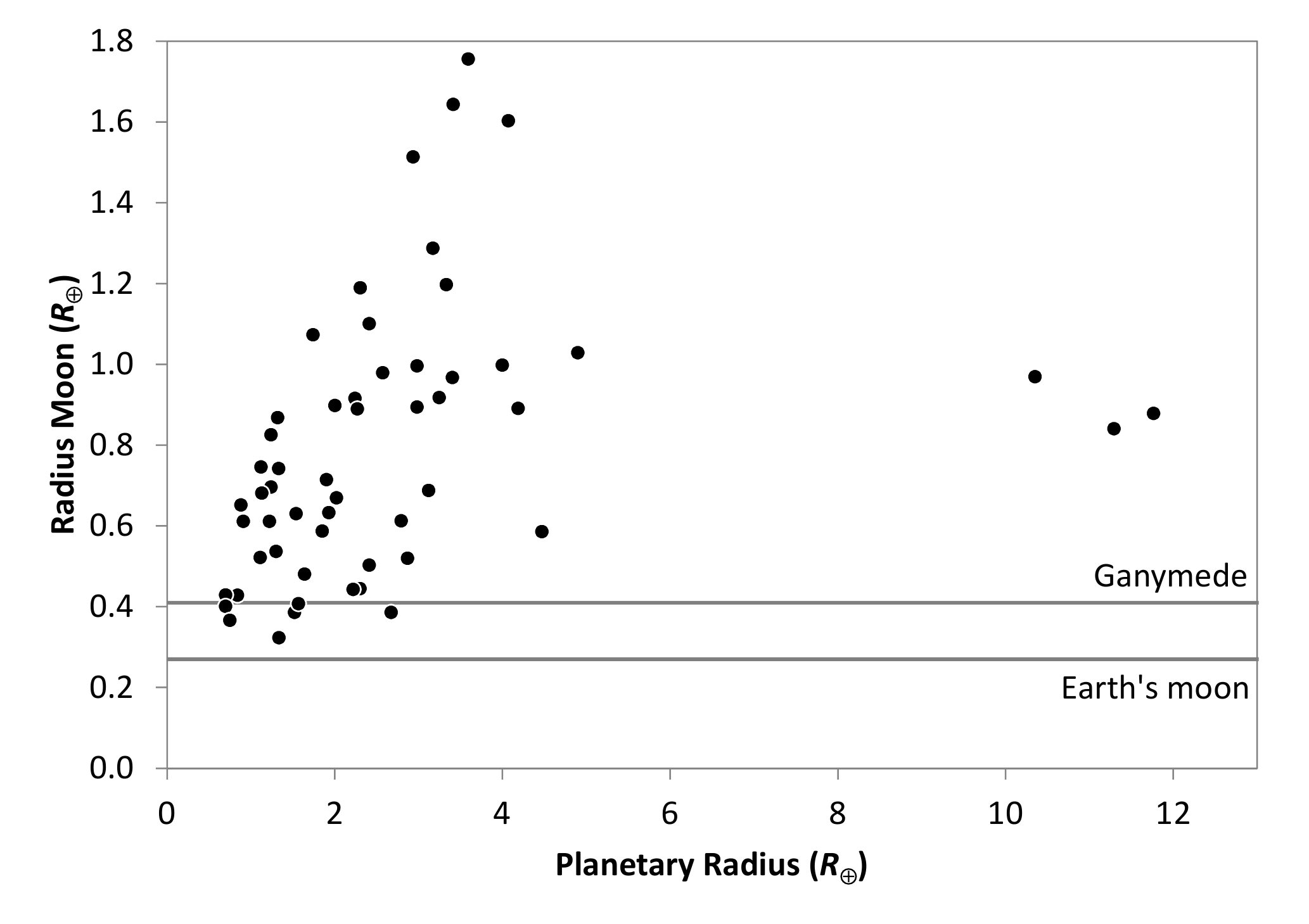}
\includegraphics[width=0.5\linewidth]{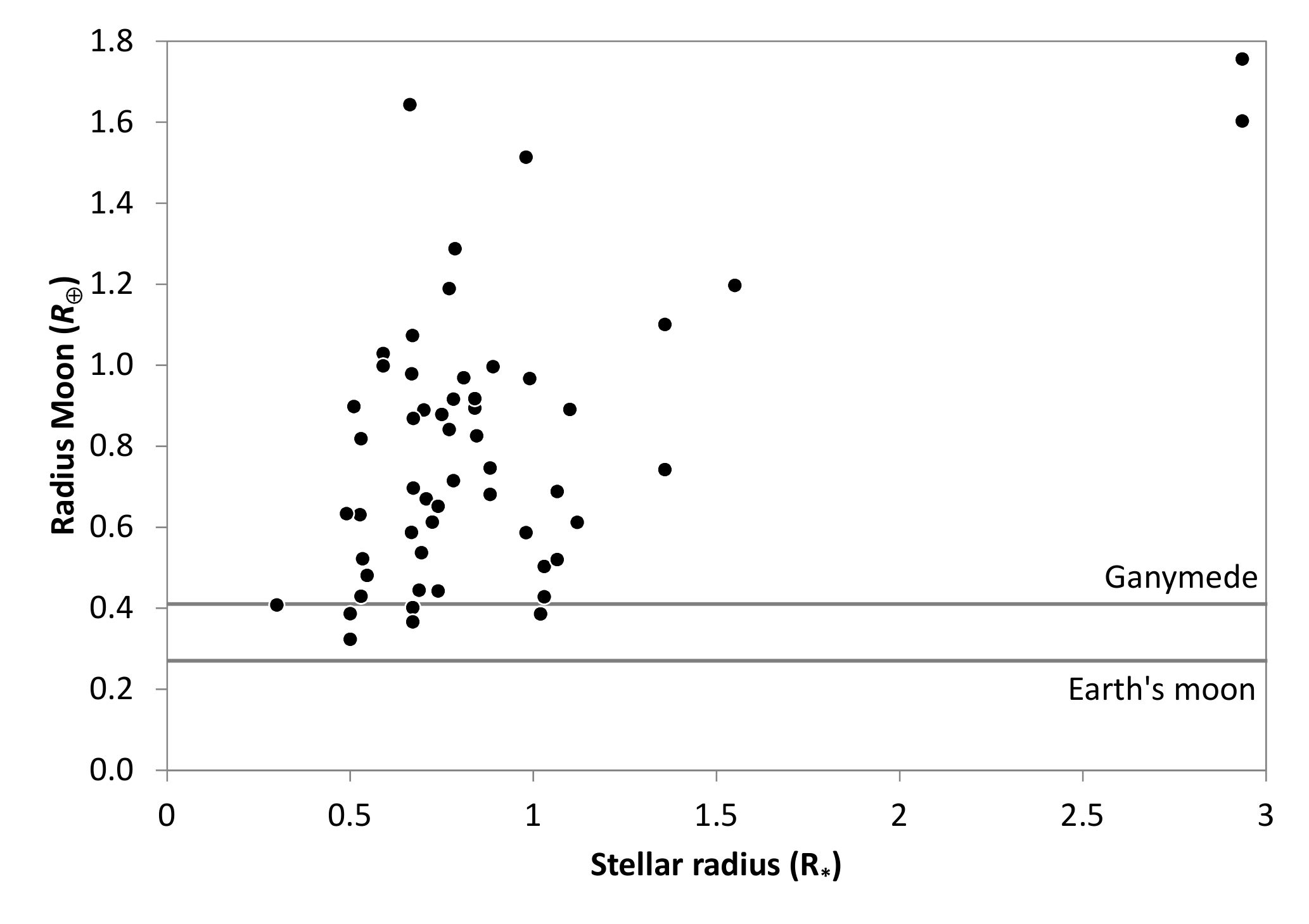}

\includegraphics[width=0.5\linewidth]{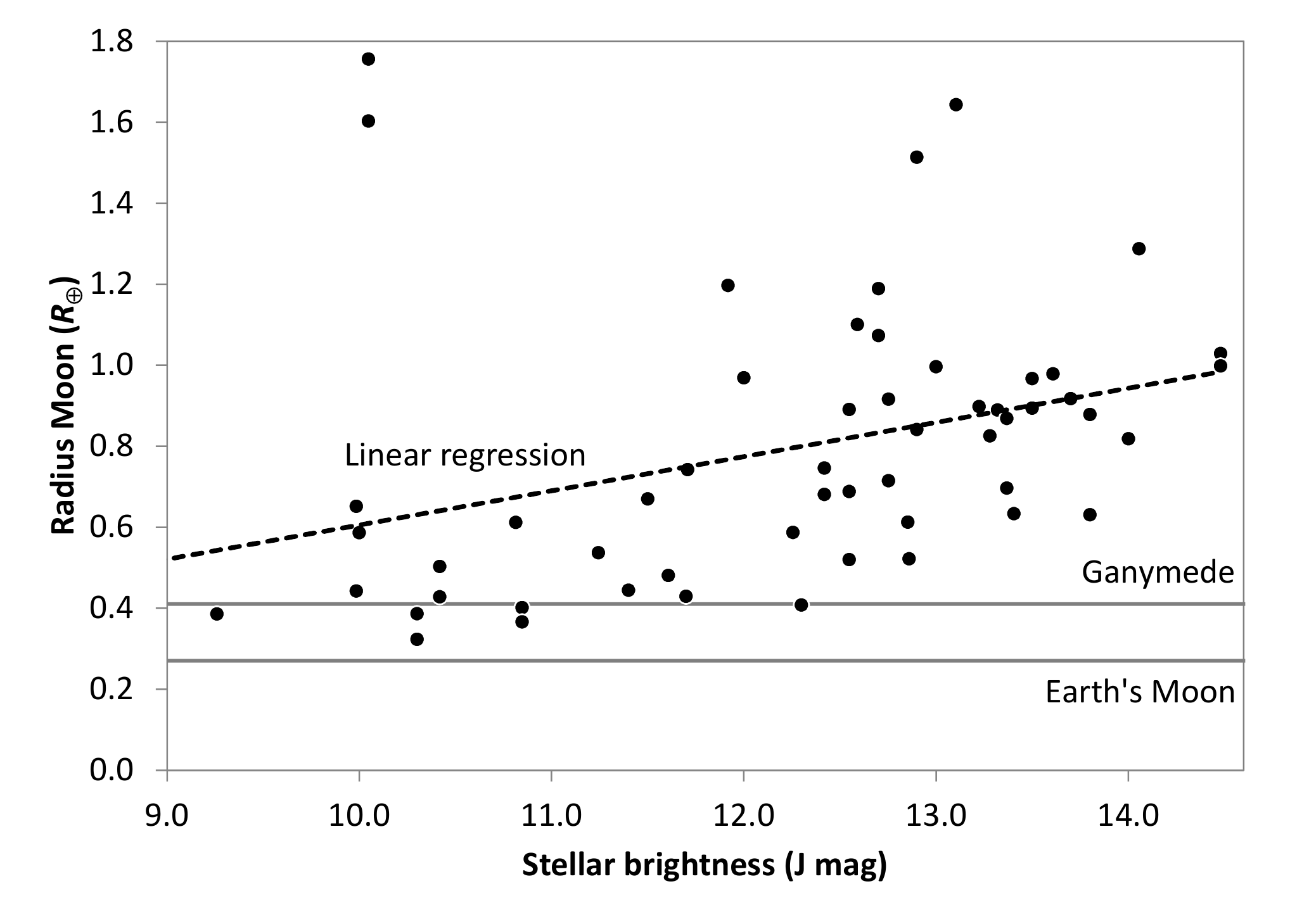}
\includegraphics[width=0.5\linewidth]{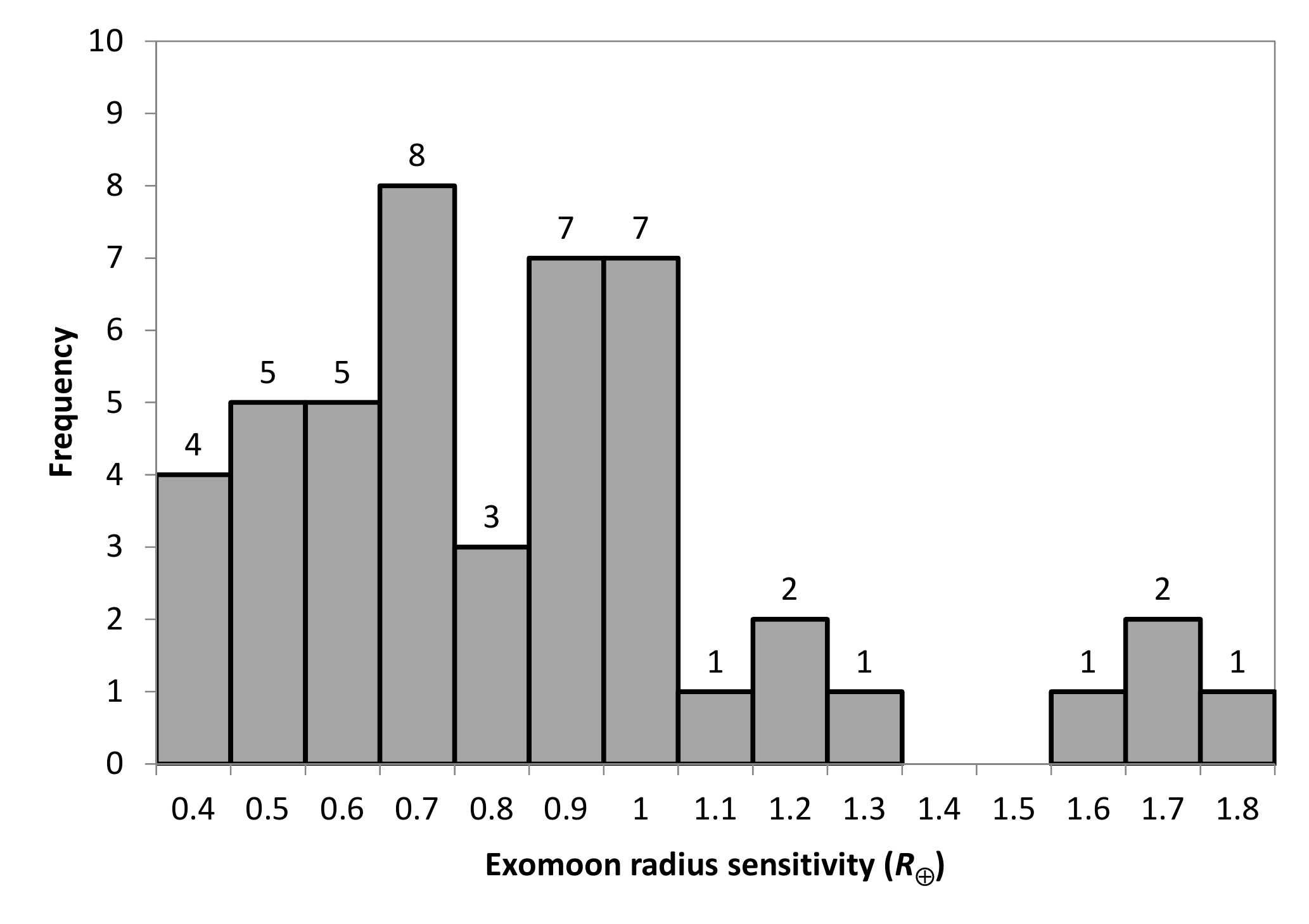}
\caption{\label{fig:sample}Upper limits (2$\sigma$) of moon radii for a sample of 56 planets analyzed in-depth; and histogram of the distribution.}
\end{figure*}

\begin{figure}
\includegraphics[width=\linewidth]{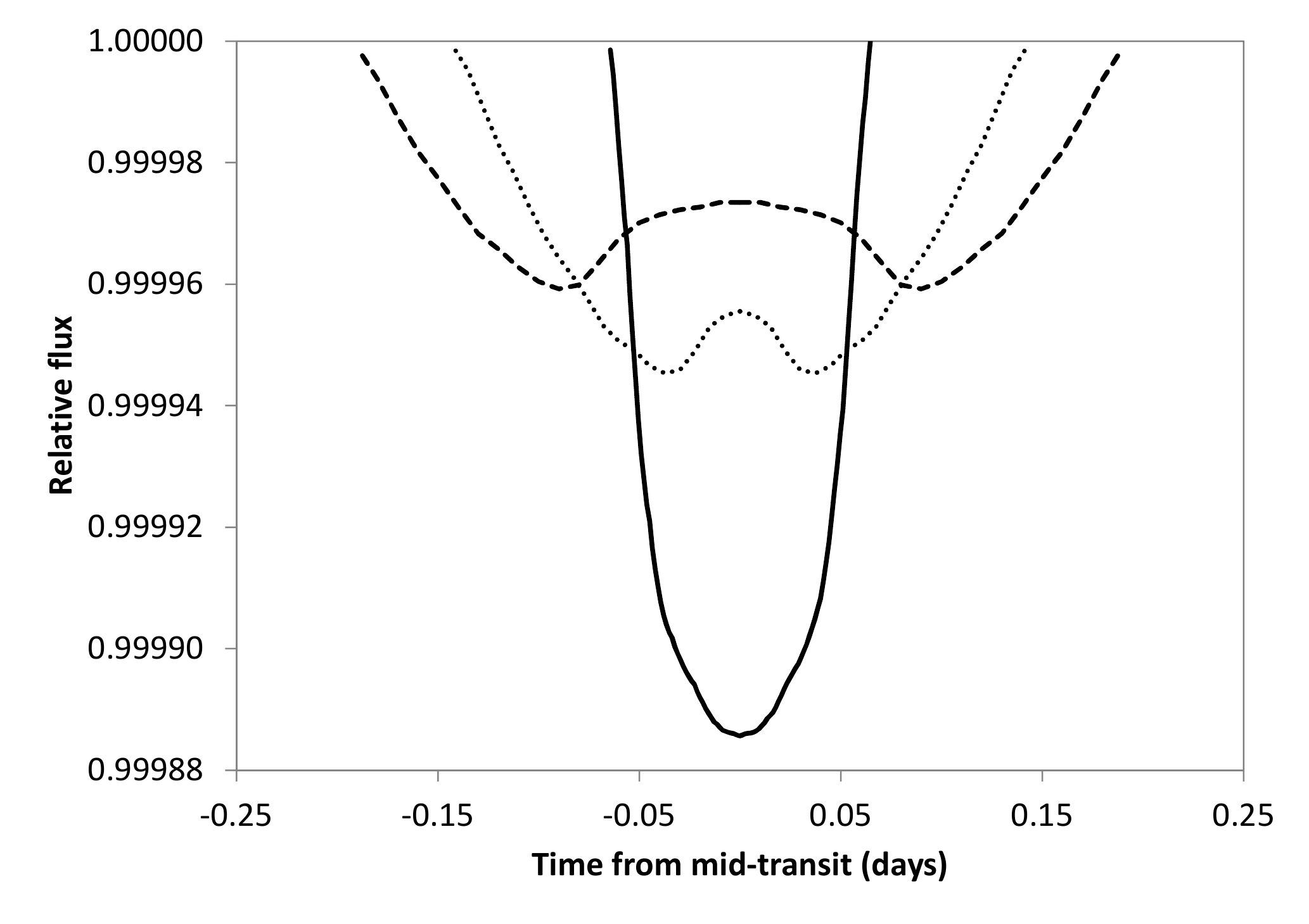}

\includegraphics[width=\linewidth]{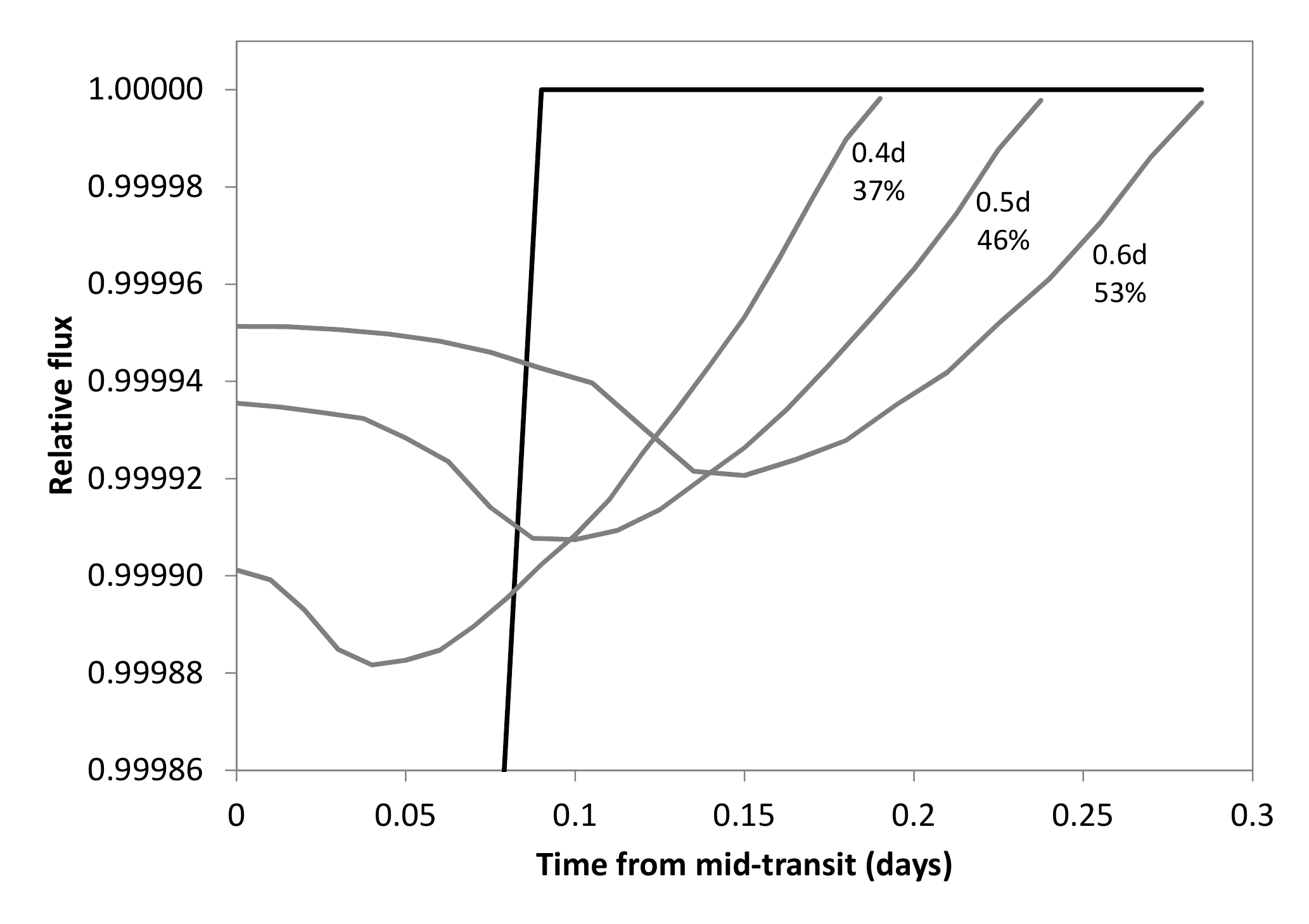}
\caption{\label{fig:OSEwidth}Top: Comparison of a single moon transit (solid line) and the corresponding OSE, for a wide (dashed) and a medium-wide (dotted) orbit. The radius is kept constant. The integral for all three curves must be identical. Bottom: OSE dips caused by different semi-major axes of a putative exomoon $R_{\leftmoon}=1.1R_{\oplus}$ (grey lines) orbiting Kepler-241c (black line). Plot shows egress with moons of transit durations 0.4 -- 0.6 days. The area under each moon integral is identical, but the percentages of flux loss occurring outside of planetary transit is different: Wider orbit moons are more easily detected.}
\end{figure}

\begin{figure}
\includegraphics[width=\linewidth]{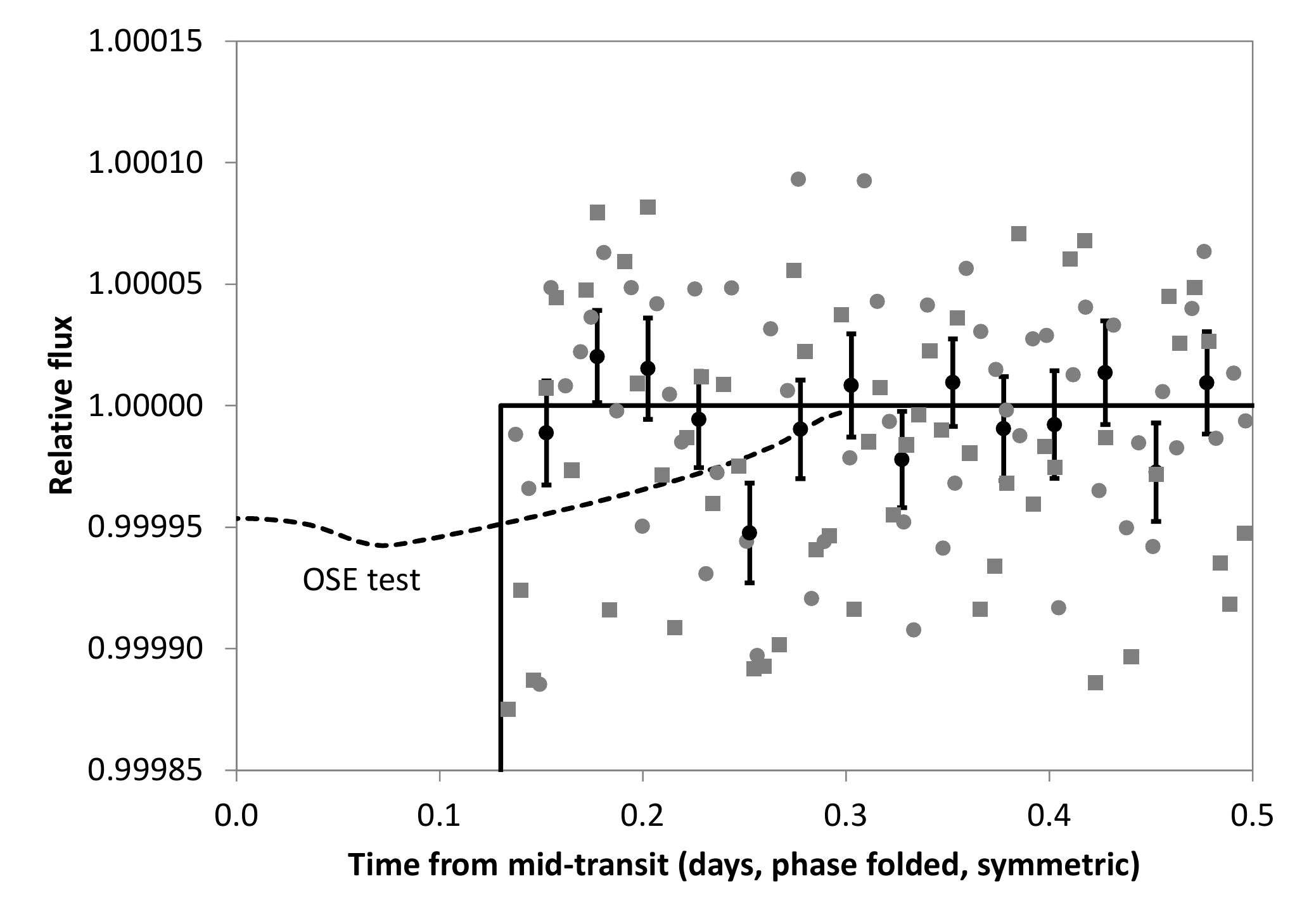}
\caption{\label{fig:OSE-KOI189-01}Null result (straight line) for KOI-189.01. A hypothetical OSE (dashed line) with $R_{\leftmoon}=0.84R_{\oplus}$ can be rejected with 2$\sigma$ confidence. This reflects the level of significance chosen for our sensitivity results. Dotted and square symbols show data from ingress and egress, respectively; data points with uncertainties are combined bins of 30min length.}
\end{figure}

\subsection{Sensitivity}
\label{sub:sensitivity}
We have performed a deep analysis for a sample of 56 planets, in order to derive limits for the sensitivity to an exomoon with respect to the stellar radius, stellar brightness and host planet size. We included interesting suitable examples from previous searches \citep{Kipping2013a, Kipping2013b, Heller2014b, Kipping2015a} with periods between 15d--80d. These were Kepler-231c (KOI-784.01), KOI-314.01, KOI 314.02, KOI-463.01 and KOI-189.01; all with null results as expected. In addition, we selected mostly planets around M-dwarfs, as they exhibit deeper transit depths due to their smaller radii, plus some randomly chosen planets in order to explore the parameter space of sensitivity. This doubles the sample of currently examined planets \citep{Kipping2015b}. 

For the limits, we assumed a circular moon with zero inclination and eccentricity, on a medium-wide orbit of two planetary transit durations, which corresponds, on average, to a semi-major moon axis as 30\% of the Hill radius. For every candidate, we have calculated the synthetic OSE light curve for such a moon, and varied only one parameter: Its radius, which affects the depth of the OSE curve. Then, we have compared the real flux per candidate, using the null hypothesis as described in criteria OSE1: A normalized, nominal flux with ``no dip'', representing no moon flux loss, and tested with a two-sample mean-comparison test. Again, this underestimates sensitivity during ingress/egress, but we prefer to have the error on this side.

We started our iterative search with a large (2R$_{\oplus}$) moon, which produces a deep OSE curve. For this initial configuration (and all following) we then performed the two-sample mean-comparison tests, asking whether the real data and the simulated OSE curve are significantly different (at the 2$\sigma$ level). This was the case for all 56 planets, therefore rejecting the presence of a (2R$_{\oplus}$) moon for all planets in the sample. We have then gradually decreased the tested moon radius (in 0.01R$_{\oplus}$ steps) and repeated the tests, until the real data and the resulting OSE curve crossed our 2$\sigma$ threshold. This crossing is the result shown in Fig~\ref{fig:sample}, giving the sensitivity at which a moon is \textit{just} rejected with 2$\sigma$ confidence in a two-sample mean-comparison test of the two datasets.

We have cross-checked this method with our full signal-injection and retrieval process as detailed in section 5.8.2. In this section, we will perform a full numerical modeling of the curves, inject these data and retrieve the OSE curve. The sensitivities with both methods are identical, as is expected. Therefore, we judge our sensitivity limits to be robust.

To clarify definitions, in this paper we define orbits with semi-major moon axes $>$50\% of the Hill radius as ``wide'' orbits, and those $>$25\% as ``medium-wide''. Compared to the Galilean moon system around Jupiter, where the innermost moon Io and the outermost moon Callisto orbit Jupiter between 0.8\% and 3.5\% the Jovian Hill radius, these limits are actually very large. As can be seen in Fig~\ref{fig:OSEwidth}, detection sensitivity drops for small orbit moons, so that these cases will be harder to detect. Therefore, one can expect a strong detection bias in the present, and future work towards wider-orbit moons.

We have used a custom median detrending for each candidate. Figure~\ref{fig:sample} shows the results, presenting 2$\sigma$ (95\% confidence interval) upper limits. In the best case, the OSE method is sensitive down to $R_{\leftmoon}\sim0.32R_{\oplus}$, which is almost the size of Earth's moon. In 6 of 56 cases, the method is sensitive to a Ganymede-sized moon, in agreement with \citet{Heller2014}. We find a correlation of sensitivity and stellar brightness: A linear regression returns a slope $>$0 with significance $p=0.011$, touching the $1\%$ (3$\sigma$) significance level. Furthermore, the test by \citet{Breusch1979} for heteroskedasticity in this regression is also significant, with $p= 0.001$. As is visually evident from Figure~\ref{fig:sample}, the scatter in sensitivity increases for stars with lower apparent magnitude. We speculate that the true slope is not linear, but quadratic or logarithmic; however due to the small number of data points we will not perform an in-depth analysis in this paper. It is for future studies to determine the underlying parameter space.

To illustrate the data quality and approach, Figure~\ref{fig:OSE-KOI189-01} shows the null result for KOI-189.01, confirming the results from \citet{Heller2014b}, and now giving a sensitivity limit. The 2$\sigma$ upper limit in this case is a high $R_{\leftmoon}<0.84R_{\oplus}$, suffering from the relatively dim host star ($K_{P}$=14.4). As mentioned, this KOI has been shown to be not a planet (but an eclipsing binary star), but for this illustration the cause for the transit dip does not matter.

Furthermore, we have created numerical simulations for a range of moon semi-major axes, in order to explore the parameter space with regards to detectability. We kept all parameters constant and only varied the semi-major axis. As can be seen in Figure~\ref{fig:OSEwidth}, wider moon orbits result in longer transit durations (in the stacked OSE sense), but with lower transit depth. This is because the total flux loss (in photons of all photometry) is always the same, so that the integral, i.e. the area under each curve is identical. To illustrate the effect, let us assume a moon with a very small semi-major axis just outside the Roche lobe. The corresponding OSE curve will then mostly overlap with the stacked planetary transit. As we have argued in section~\ref{sec:method}, it is very hard to detect flux deltas during stacked planetary transit, due to limb darkening uncertainty, and the unknown planetary transit depth. Outside of planetary transit, however, the putative flux loss of a moon can be compared to the nominal, normalized flux. Thus, we argue that moons with wider orbits are more easily detected. We show the fraction of flux loss that occurs outside of planetary transit, for one specific example, in Figure~\ref{fig:OSEwidth} (lower panel). It is also worth mentioning that all but the smallest orbit moons exhibit a feature dubbed ``right and left wings'' by \citet{Heller2014}. This gives a certain shape to the OSE that makes it, when detected, easier to distinguish from a simple linear flux drop, e.g. due to stellar red noise.

Depending on the orbit configuration, an unknown fraction of the signal coming from moon transits occurs during planetary transits, which is harder to detect. As the distribution of exomoons is unknown, the average of this fraction is also unknown; it can be high (e.g. 70\%) if moons tend to have small orbits, or lower (e.g. 30\%) for wider orbits. If no modeling for the OSE during planetary transit is made, as in this present work, this certain fraction of the OSE signal is lost. For an ideal, co-planar case, and assuming such modeling can be done with future theoretical advances, up to 100\% of the OSE signal can be retrieved. This puts the OSE method at an advantage, when compared to photodynamical modeling by \citet{Kipping2009}: The OSE can leverage stacking to reduce red stellar noise, which has been shown to be particularly useful for short-time ($<$10hrs) time-correlated noise by \citet{Hippke2015}. Photodynamical modeling, however, has to rely on individual transits, of which each is prone to time-correlated red noise. Of course, this applies only to moons where the OSE is sensitive to, i.e. the coplanar case, as discussed in section~\ref{sec:method}.

We found six candidates with a significant flux loss, as defined in section~\ref{sec:framework} by using the two-sample mean-comparison test, and discuss these in-depth in section~\ref{sec:application}. Out of these, two or three are identified as likely false positives. For reference, we have also found 3 planets to exhibit a significant (95\% confidence interval) flux \textit{gain}, consistent with statistical expectations.

\begin{figure}
\includegraphics[width=\linewidth]{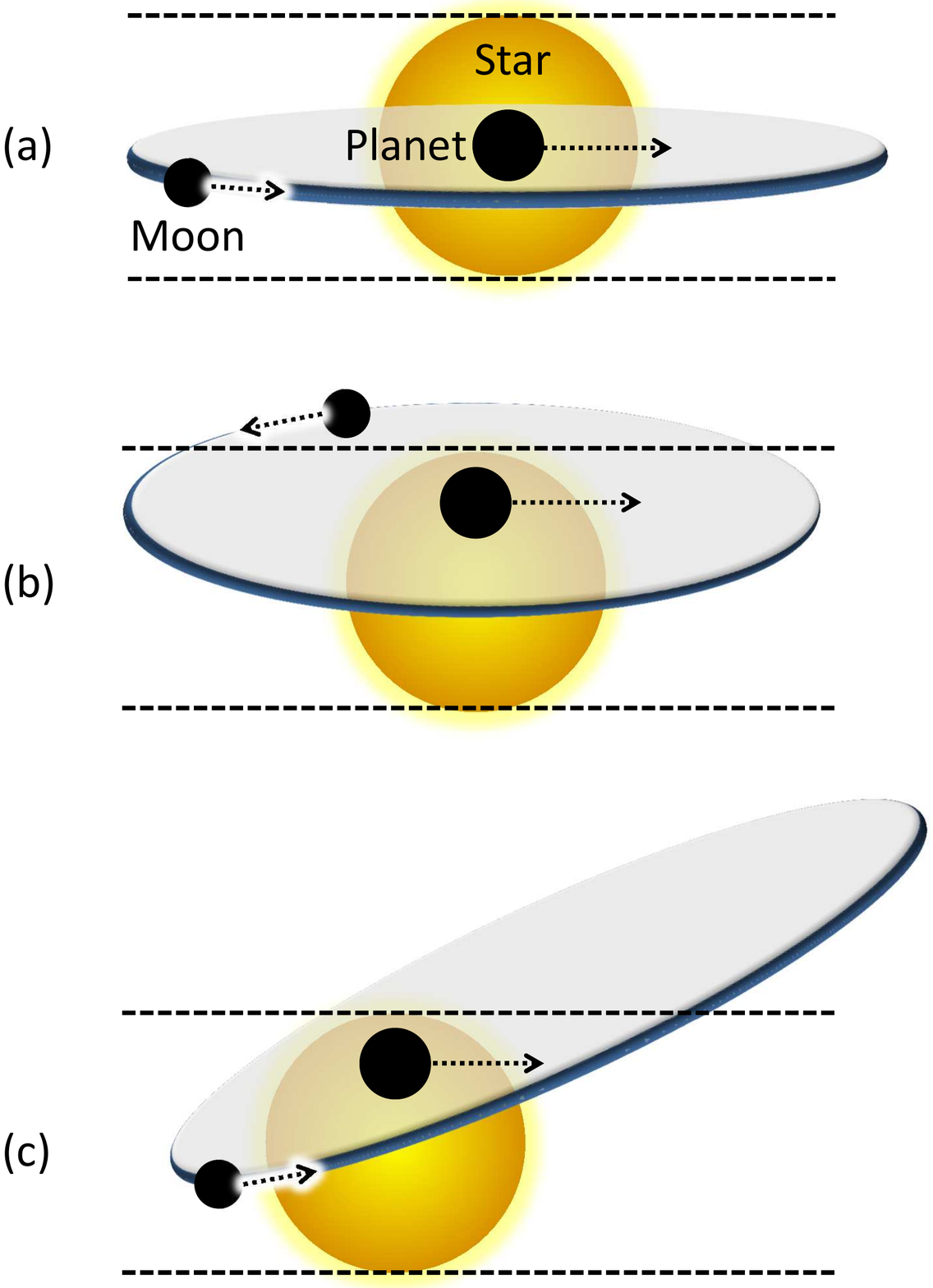}

\includegraphics[width=\linewidth]{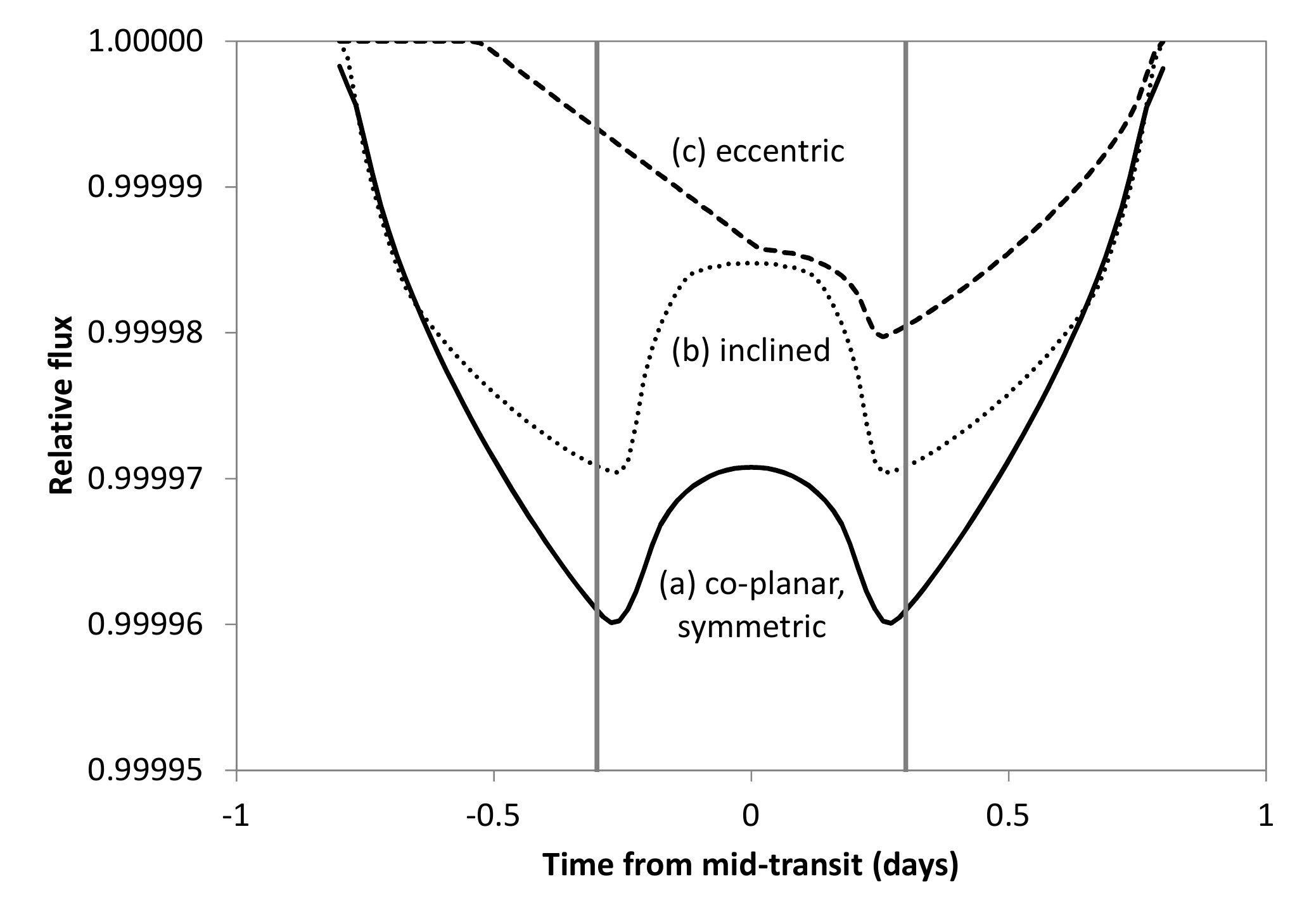}
\caption{\label{fig:OSEconfigs}Top: Orbital configurations showing star-planet-moon: Co-planar, near-zero inclination (a); inclined (b) and eccentric (c) so that part of the moon transit is not observed (outside dashed horizontal lines). All three situations show the planetary mid-transit. In plots (a) and (c), the moon transits after the planet. In plot (b), the moon does not transit in this specific epoch. Individual transits are almost always snapshots of a given orbital configuration: The moon orbits the planet, but on a much longer timescale (days) than the transit duration (hours). Bottom: Inclined orbits (dotted) are still symmetrical, but less deep. Eccentric orbits (almost) lack flux loss on one side, depending on the eccentricity. Grey vertical lines show the planetary transit duration.}
\end{figure}

\subsection{Shape of the OSE}
\label{shape}
The shape of the phase-folded orbital sampling effect varies mainly with the moon's radius, its semi-major axis and the orbital configuration of the system. For a single-moon case, Figure~\ref{fig:OSEconfigs} shows the three main cases and their OSE curves: A near co-planar configuration is the simple case, and can serve as the standard hypothesis. This is what we find among 75\% of large solar system moons, as discussed in the introduction. It has a symmetrical shape varying in depth and width depending on the semi-major axis (compare also Figure~\ref{fig:OSEwidth}). 

The second case is an inclined moon, tilted towards the observer. This configuration still results in a symmetric OSE, but reduced depth and thus detectability. The greatest difference in shape is during planetary transit, making this configuration hard to distinguish from the co-planar case (this is because the OSE is harder to observe during planetary transit).

A more severe case is eccentricity (or an inclination shifted laterally from the observer), as shown in the bottom panel. The resulting OSE is highly asymmetric, showing that cases with flux loss on mainly one side are not necessarily false positives. In practice, with noise added, one would find \textit{no dip} on one side (ingress or egress), and a \textit{regular} dip on the other side with the best possible \textit{Kepler}-quality data.

These results can be reproduced with the interactive spreadsheet (section~\ref{fitter}), by removing the data points (i.e. setting them to flux=1) from the numerical sampling which are are not observed due to inclination or eccentricity. This introduces a slight error for the eccentric case, as it does not account for the change in velocity of the moon. We encourage modelers to advance the theory in order to account for this effect.

With \textit{Kepler}-class photometry, we do not expect to be able to make significant distinctions between these cases. This might be possible with future instruments such as \textit{PLATO 2.0}. However, in case of a moon found with other methods, for instance using photodynamical modeling, the postulated configuration can be cross-checked with its expected OSE curve. Both should agree, within the errors, to make it a consistent exomoon detection.

\section{Results for the application of the framework to individual candidates}
\label{sec:application}
Our in-depth analysis of 56 stars (as explained in section~\ref{sub:sensitivity}) gave six candidates worth further examination. The following sections will give a brief overview to these systems, and then proceed to the individual tests.

\begin{table*}
\center
\caption{Best-fit moon results\label{tab:moons}}
\begin{tabular}{lccccccl}
\tableline
System      & $K_{P}$ & $R_{*} [R_{\odot}]$ & $P_{P}$ [days]         & $R_{P}$ [$R\oplus$] & $R_{\leftmoon}$ [$R\oplus$] & $a_{\leftmoon P}$ [$10^4$km] & Reference \\
\tableline
Kepler-241b & 15.4    & $0.668\pm0.289$     & $12.718092\pm0.000035$ & $2.30\pm1.03$       & $1.1\pm0.1$                & $450\pm90$ & \citet{Rowe2014}\\
Kepler-241c & 15.4    & $0.668\pm0.289$     & $36.065978\pm0.000133$ & $2.57\pm1.11$       & $1.4\pm0.1$                & $320\pm60$ & \citet{Rowe2014}\\
Kepler-264b & 13.0    & $1.550\pm0.310$     & $40.806230\pm0.000540$ & $3.33\pm0.74$       & $1.6\pm0.2$                & $660\pm120$ & \citet{Rowe2014}\\
KOI-367.01  & 11.1    & $0.980\pm0.185$     & $31.578659\pm0.000007$ & $4.47\pm0.55$       & $0.8\pm0.1$                & $150\pm50$ & \citet{Rowe2015}, DV\footnote{\textit{Kepler} Data Validation Pipeline, http://exoplanetarchive.ipac.caltech.edu} \\
Kepler-102e & 11.5    & $0.740\pm0.020$     & $16.145700\pm0.000100$ & $2.22\pm0.07$       & $0.5\pm0.1$                & $300\pm40$ & \citet{Marcy2014}\\
Kepler-202c & 13.9    & $0.667\pm0.034$     & $16.282493\pm0.000038$ & $1.85\pm0.10$       & $0.7\pm0.1$                & $480\pm90$ & \citet{Rowe2014}\\
\tableline
\multicolumn{8}{l}{For reference: Ganymede has $R_{\leftmoon}=0.41 R\oplus$} \\
\tableline
\end{tabular}
\end{table*}

\subsection{Candidate summary}
We summarize the six exemplary candidates in Table~\ref{tab:moons} and describe them in the following sections. Kepler-102e and Kepler-202c are similar to KOI-367.01, as they show a flux loss following criteria OSE1 only when using a median detrending, but not when using polynomial (parabola) detrending. This case will therefore be discussed using KOI-367.01 as an example.

\subsubsection{Kepler-264b}
Validation of the planets \textit{b} and \textit{c} in the Kepler-264 system was done by \citet{Rowe2014} using data for Q1--Q15. We have checked these results by using the PDCSAP data with subsequent median-filtering, as described in section 4.3. Then, we have fitted the standard \citet{Mandel2002} transit model, as implemented by \textsc{PyAstronomy}\footnote{\url{https://github.com/sczesla/PyAstronomy}}. The results for the orbital and transit parameters are the same, within the errors, as in \citet{Rowe2014}. We have not searched for additional planets, and have not tried to reproduce the false-positive framework of the original authors -- our intention was to double-check the transit model result. 

We found an OSE-like signal for planet \textit{b}, but also mention planet \textit{c} in the following do give some context on this particular system. 

\begin{figure}
\includegraphics[width=\linewidth]{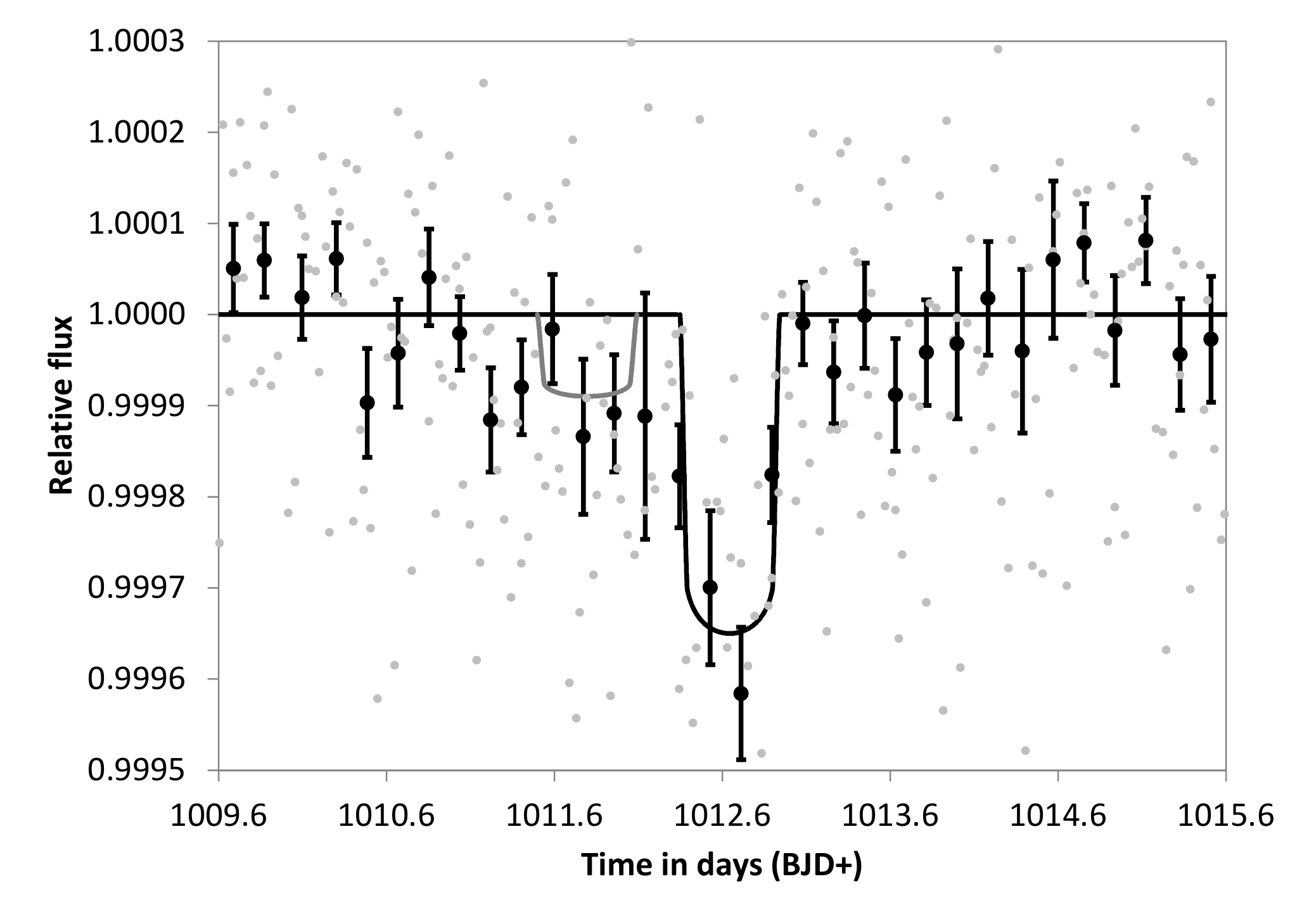}
\caption{\label{fig:singletransit}Single transit of Kepler-264b at BJD=1012.6. A potential transit curve of a hypothetical ultra-large moon (1.6$R_{\oplus}$) is shown before planetary ingress, with a flux of -90ppm. Such a small signal cannot be reliably detected in a single transit observation. Binned data (dots with error bars) are only for visual clarity.\\}
\end{figure}

Both planets \textit{b} and \textit{c} are best described as warm/hot Neptunes. As can be seen in Figure~\ref{fig:singletransit}, a single (non-stacked) transit of a hypothetical exomoon around Kepler-264b with 1.6$R_{\oplus}$ (90ppm) does not provide sufficiently low noise even for a detection of a large body. Following equation~(\ref{eq:sn}), the S/N during a single moon transit is $\sim$2.4, which is too low to claim a detection at the traditional threshold for S/N of 7 \citep{Jenkins2002,Rowe2014} or even 10 \citep{Fressin2013}. Due to this low S/N, we do not try to fit individual moon transits, as is done in photodynamical modeling. Consequently, Figure~\ref{fig:singletransit} should be understood only as an illustration.

During the operation of the \textit{Kepler} spacecraft, a total of 36 transits occurred. We could use a total of 27 transits, while 9 were lost to data missings.

As in \citet{Rowe2014}, we only find insignificant TTVs and TDVs on the order of a few minutes, for both planets \textit{b} and \textit{c}, as is expected in such relatively long-period configurations \citep{Awiphan2013}. We have tried the further processing described in this paper with and without adjusting to these insignificant variations, and get identical results, both for the OSE and the SP method. Following results are thus derived from the linear ephemeris, neglecting TTVs and TDVs. 

As an additional test for errors, we have added a buffer time of 2$\sigma$ of the uncertainty of the transit duration before and after the transit, so that an additional time of 0.028d before and after transit was excluded from further analysis. This did not change the results significantly, so that in the following we only exclude the nominal uncertainty (0.014d) from the analysis. These TTV/TDV treatments are the same for all following candidates.

\subsubsection{Kepler-241b and -c}
Two planets are known to orbit Kepler-241, a K-star with small radius ($0.668\pm0.289R_{\odot}$). They have periods of 12.7 (\textit{b}) and 36.1 days (\textit{c}), and radii of $2.3\pm1.03$ and $2.57\pm1.11R_{\oplus}$, making them both Super-Earths. Interestingly, we have detected a flux loss for both planets, with \textit{b} having a short period which might cause stability issues for moons. We have tested a wide variety of detrending parameters to check if both planets were affected by detrending issues, but found this star particularly simple to detrend: The PDCSAP flux shows only a weak ($\sim$0.1\%) variation, likely due to rotation and spots, on a timescale of $>10$ days, which is easily and cleanly removed with a sliding median. Also, parabola detrending gives virtually identical results.

To derive limits for the Hill radii of these planets, we have taken the values for the stellar radius ($0.668\pm0.289$ $R_{\odot}$) and density ($\rho_{*}=3.257\pm1.171$g~cm$^{-3}$) from \citet{Rowe2014}. This results in a best-fit stellar mass of 0.69$M_{\odot}$. For rocky planets with a density of 5g~cm$^{-3}$, the Hill radii would be 409,000km (b) and 901,000km (c). For a lower density of 2g~cm$^{-3}$, the Hill radii would be smaller, 301,000km (b) and 665,000km (c). Consequently, one might argue that a potential exomoon orbiting 241b, in a distance of 450,000km as listed in Table~\ref{tab:moons}, might be in tension with Hill stability, when neglecting stellar uncertainties. For planet c, no such conclusion can be made. In these cases, spectroscopy and radial-velocity data would help to reduce the stellar uncertainties.

\subsubsection{KOI-367.01}
KOI-367.01 is an unconfirmed planet candidate, with the risk of being a false positive. We have included this candidate due to its significant OSE-like signal, and the high photometric quality caused by the bright ($K_{P}=11.1$) host star, likely a G-star ($0.98\pm0.185R_{\odot}$) like our Sun. No other planets are known to orbit this star. Candidate \textit{.01} is Neptune-sized ($4.47\pm0.55R_{\oplus}$) on a 31.6 day orbit.

\subsection{Results for criteria OSE1: Significance}
\label{sub:ose1}
The significance of the dips before ingress and after egress can be measured easily, with the unknown factors being the duration of the dip, which depends on the orbital configuration of the moon(s); and the transit depth. After preliminary estimation of its duration, we propose two tests: One for the full length, and one for the half closer to planetary transit. This is useful because the OSE is stronger closer to planetary transit, allowing for a more secure detection. We suggest that the shorter test must be passed for ingress and egress side, and the longer test be a bonus.

For simplicity, it can also be useful to test for a bin width of one (and: one-half) planetary transit duration, as this covers the majority of moon configurations. Indeed, for our example case of Kepler-264b, we get an estimated flux loss starting at $\sim\vert0.8\vert$d from mid-transit (transit begins/ends at $\vert0.297\vert$d), so that the out-of-transit dip duration is $\sim\vert0.5\vert$d, roughly equivalent to the planetary transit duration (0.594d).

\begin{figure}
\includegraphics[width=\linewidth]{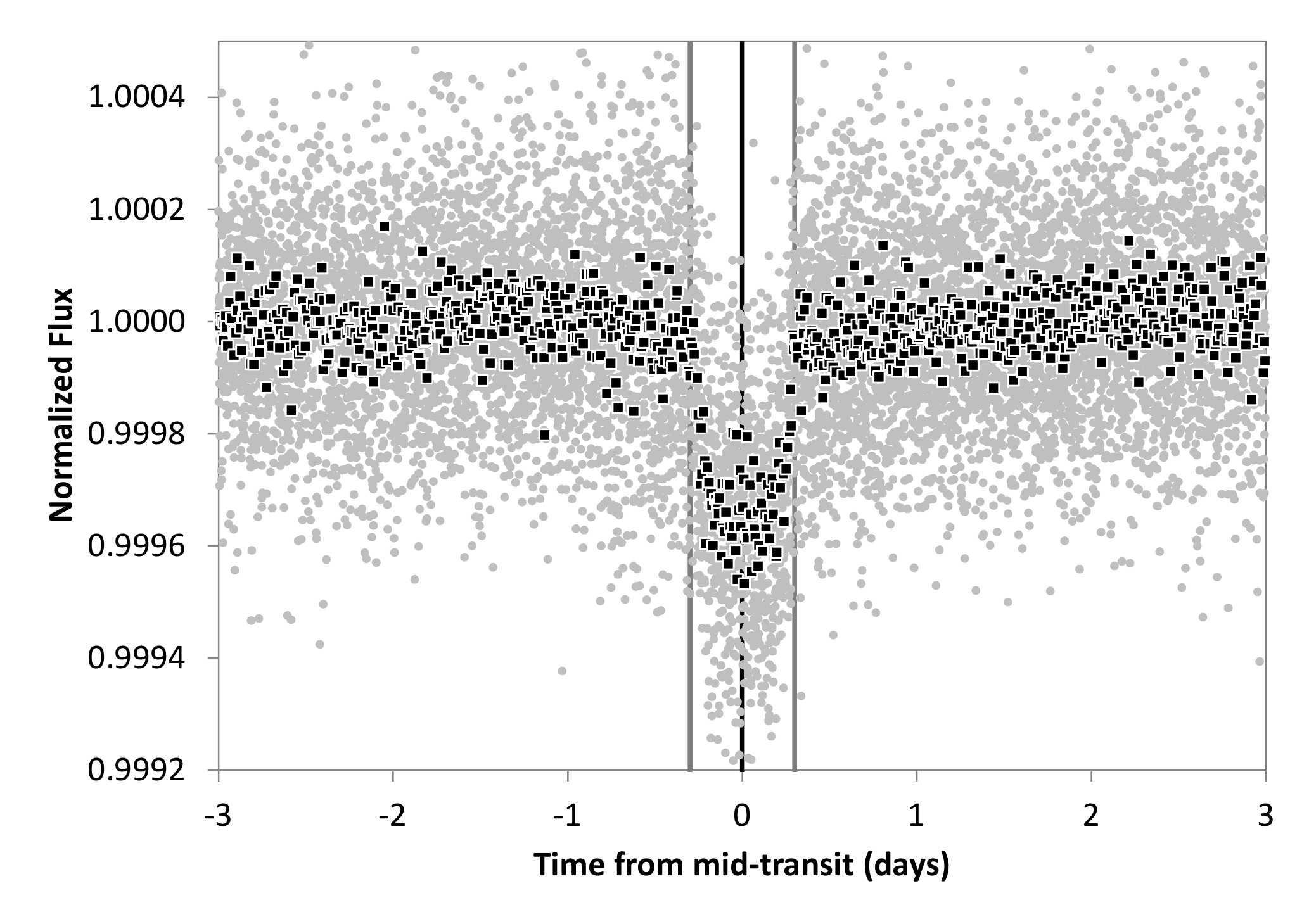}

\includegraphics[width=\linewidth]{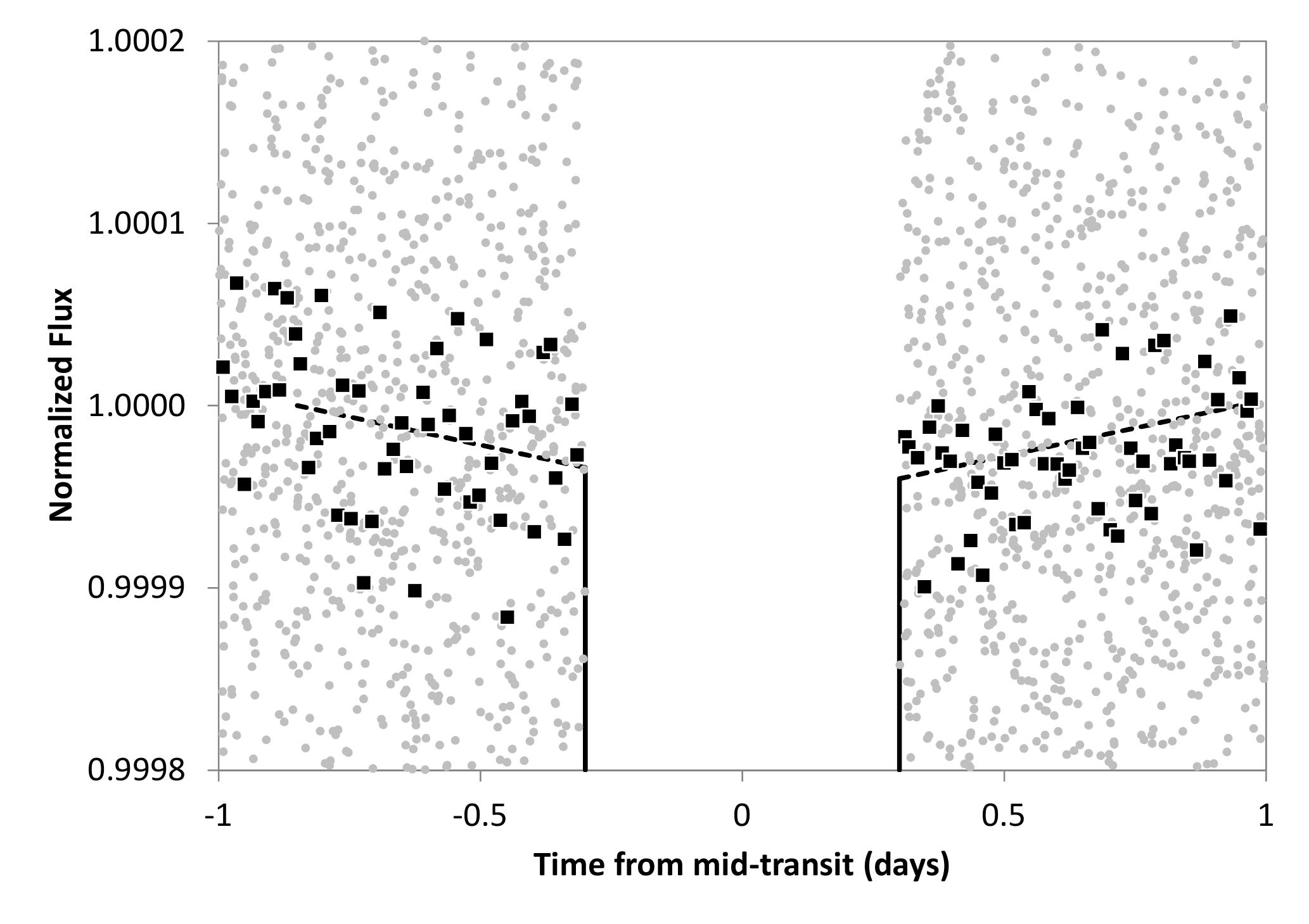}

\includegraphics[width=\linewidth]{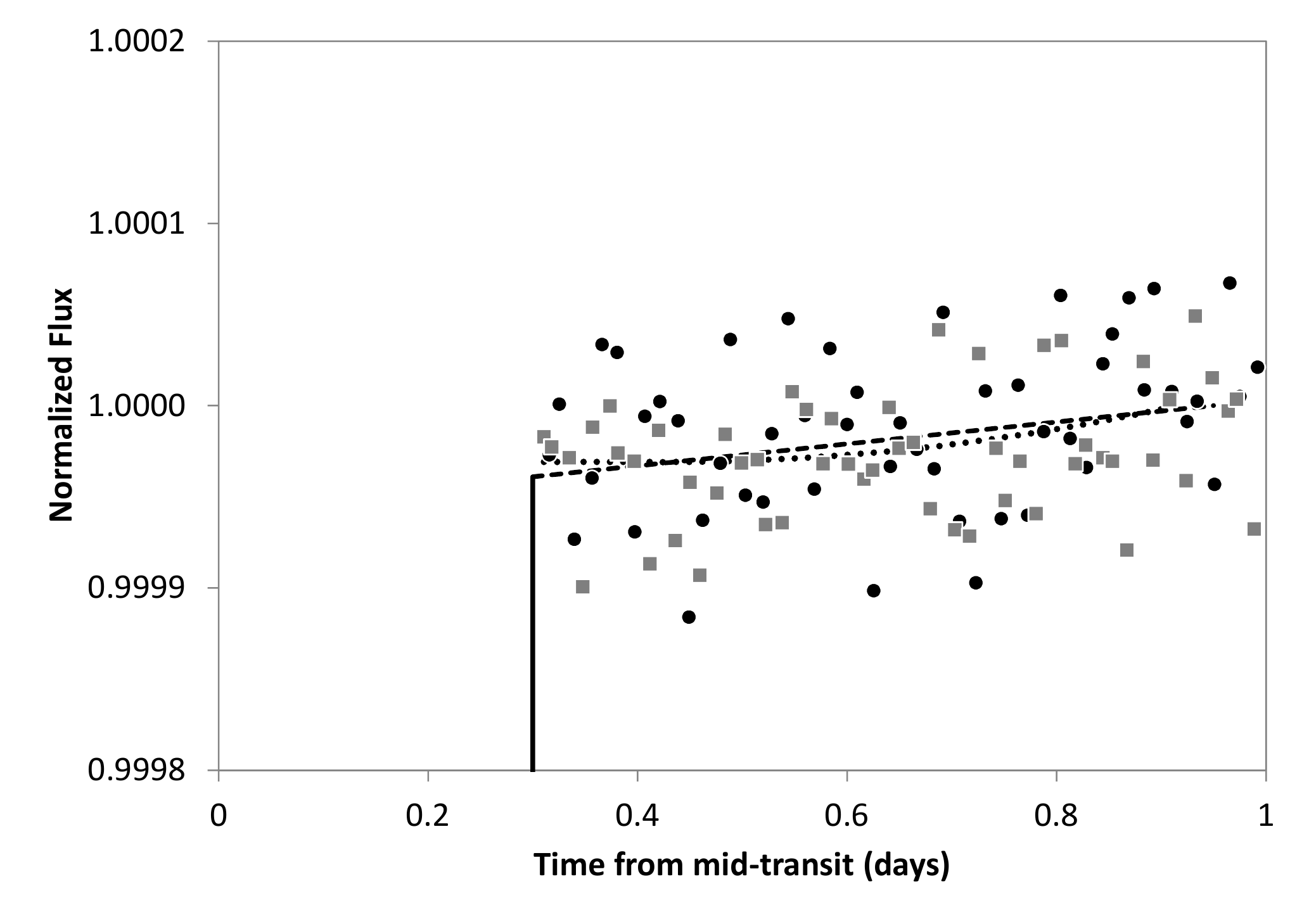}
\caption{\label{fig:flux}Kepler-264b phase-folded flux for a time of 3 days before and after planetary transit (top), and zoom into the region of one day before and after (middle). Grey dots are LC data, squares represent 20 LC bins. The bottom plot shows a symmetrical fold, with circles (squares) representing ingress (egress) 20 LC bins. Dashed line shows a least-squares linear fit, dottled line a numerical OSE fit (see also Figure~\ref{fig:flux-ose-fits}), and the vertical line the planetary transit.}
\end{figure}

For the data as presented in Figure~\ref{fig:flux}, results are virtually identical for both detrending methods. We get a significant flux loss during ingress (2.0$\sigma$), egress (4.8$\sigma$) and combined (4.8$\sigma$), measured as the mean of the first half transit duration before (after) planetary transit. When taking the bin width as one full transit duration, we get an insignificant flux loss during ingress (1.2$\sigma$) and still significant flux loss at egress (4.9$\sigma$). This means that the required test passed, and the optional test failed (marginally). 

The results for the other candidates are summarized in Table~\ref{tab:testresults}. Every one of the others fails one or more tests, but all are significant for the data at ingress and egress binned together.

\begin{table*}
\center
\caption{Test result summary for candidates, at 2$\sigma$ (95\%) significance\label{tab:testresults}}
\begin{tabular}{lccccc}
\tableline
Test                       & 241b            & 241c            & 264b           & 367.01\\
\tableline
OSE1: Significance (l,r,b) & \yes (\no) \yes & (\no) \yes \yes & \yes \yes \yes & \yes \yes \yes \\
OSE2: Slope (l,r)          & \yes \no        & (\yes) \yes     & \yes \yes      & \yes \no \\
OSE3: Stacking             & \yes            &  \yes           & \yes           & \yes \\
OSE4: Uniqueness (a,b)     & \no \no         &  \no \no        & (\yes) \yes    & \no \no \\
OSE5: Star-spots           & \no             &  \no            & \yes           & \yes \\
OSE6: Rings                & \yes            &  \yes           & \yes           & \yes \\
OSE7: Plausibility         & \yes            &  \yes           & \textbf{?}     & \yes \\
C1: Temporal spread (OSE)  & \yes            &  \yes           & \yes           & \yes \\
\tableline
\multicolumn{5}{l}{Legend: \yes Pass; \no Fail; (\no),(\yes) Marginal or partial fail/pass}
\end{tabular}
\end{table*}

\subsection{Results for criteria OSE2: Slope}
\label{sub:ose2}
From the sampling of the orbital sampling effect, the flux loss must increase towards planetary transit. This can be tested with a linear regression and t-tests (or F-tests) for different folded time durations, as explained in the previous section. For Kepler-264b, the slope parameters before ingress (3.1$\sigma$) and after egress (2.1$\sigma$) are both inclined towards planetary transit, and significant. The test fails for the egress side of Kepler-241b and KOI-367.01. For the ingress side of Kepler-241c, we get a marginal pass. The other sides and candidates pass the tests. The power of the individual test result should not be overestimated, as the photometry is already close to its limit.

\subsection{Results for criteria OSE3: Stacking}
\label{sub:ose3}
\label{sec:robust-ose}
During the examination of other candidate exomoons, we discovered some examples were only one or a few transits caused effects mimicking OSE and SP. One such case is Kepler-420b, where the $12^{th}$ transit at BJD$\sim$2455960 was affected by thermal changes on the telescope \citep{Santerne2014}. This can be tested by splitting the data in segments, and analyzing these as described above. If only part of the data seems to cause OSE and SP, it must be regarded as a spurious result.

Following the idea of \citet{Kipping2014} to use only part of the data and check the results versus the other chunk, we have created a similar test. We ask whether a randomly chosen subsection of the data is consistent (within the errors) with the total data, so that most, and not just a few data segments contain the potential exomoon signal. This test was repeated 100 times on randomly chosen parts of the data. For our candidates, we find that any 50\% of the data give the same result (a flux loss both during ingress and egress), albeit with lowered significance due to the lower number of data points. Inversely, we judge that $\sim50\%$ of the data are required for the results described here, and it is irrelevant which half is used.

In principal, OSE3 and C1 ask the same question. In this paper, test OSE3 focuses mainly on the detection of outliers, while C1 will check the actual build-up of the OSE in-depth, including numerical simulations.

\subsection{Results for criteria OSE4: Uniqueness}
\label{sub:ose4}
Ideally, there should be no other dips in the dataset that are more prominent than those before ingress and after egress. This criteria is closely connected with the actual measure of \textit{significance} of such dips, but it is not formally the same: At any given significance level, it can occur solely from noise that additional dips show up, dips even deeper than the ones caused by the OSE. In a white-noise world, one could simply set a high enough significance limit to restrict this -- and would, in return, loose a few moon detections to \textit{false negative} results. In a red-noise world, however, with data coming from instruments with trends, such is unwise, as it can eliminate valid information because of trends that might be \textit{far away} temporally (and thus irrelevant, because unconnected to the signal in question). To be precise, a signal can be perfectly valid and should not be discarded because of an instrument malfunction that occurred, for example, only once and only in a distant phase-folded time.

\begin{figure}
\includegraphics[width=\linewidth]{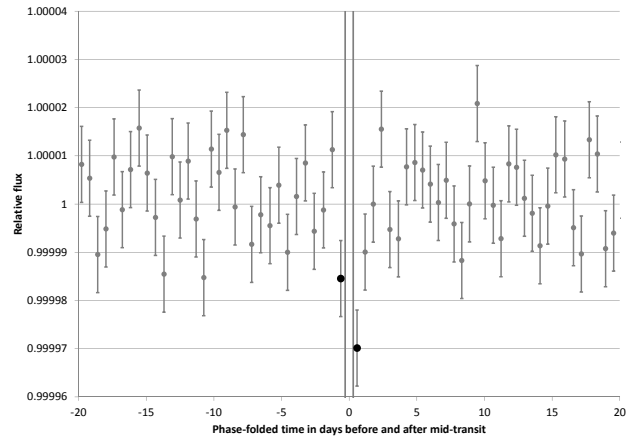}
\caption{\label{fig:dips}Phase-folded flux for Kepler-264b in bins of one transit duration, for the complete dataset. Planetary transit is between the grey lines. The two neighboring bins before ingress and after egress (black dots) have the highest flux loss of all bins, although the bin before ingress only marginally. Care must be taken that the bin width is adjusted so that no planetary flux loss leaks into the other bins in question. The bin during planetary transit is not shown for clarity (at $\sim$300ppm, it would be four times the diagram height downwards).}
\end{figure}

We suggest two tests for the uniqueness of the dips in a given dataset, one hard test and one soft test. The hard test demands that one cannot shift the (virtual) folded transit time, and then still have a significant flux loss before ingress \textit{and} after egress. To test this, one plots a phase-folded flux graph that bins the flux to a useful duration, e.g. one transit duration (of course this choice depends on the putative orbit axis). In the default configuration, one gets a dip in the first bin before and after transit. The test demands that no significant dip occurs at bin $n+1$ and bin $n-1$, with a virtual transit in a sliding bin $n$. As can be checked in Figure~\ref{fig:dips}, Kepler-264b passes this test.

The soft test is that the two binned dips before and after the (real) transit are the two largest dips that occur in the whole dataset. For Kepler-264b, this test is also successful, albeit only marginally. If this test fails, we recommend to plot a temporal map (like the river-plot, Figure~\ref{fig:river11d}) of the relevant dips. If single outliers cause other, deeper dips, we would judge it fair to remove these (few) outliers as the likely cause.

Unfortunately, the test is less useful for short-duration transits on long periods: For a hypothetical moon around Kepler-102e, the bin width is $\sim$0.05d during a period of 16.1d, giving 322 bins to compare. Also, it is tedious to check and remove numerous outliers that naturally occur from noise (16 expected at the 95\% significance level). Consequently, the other candidates fail this criteria without such clean-up. We suggest to skip the test for this category of planets.

\begin{figure*}
\includegraphics[width=0.5\linewidth]{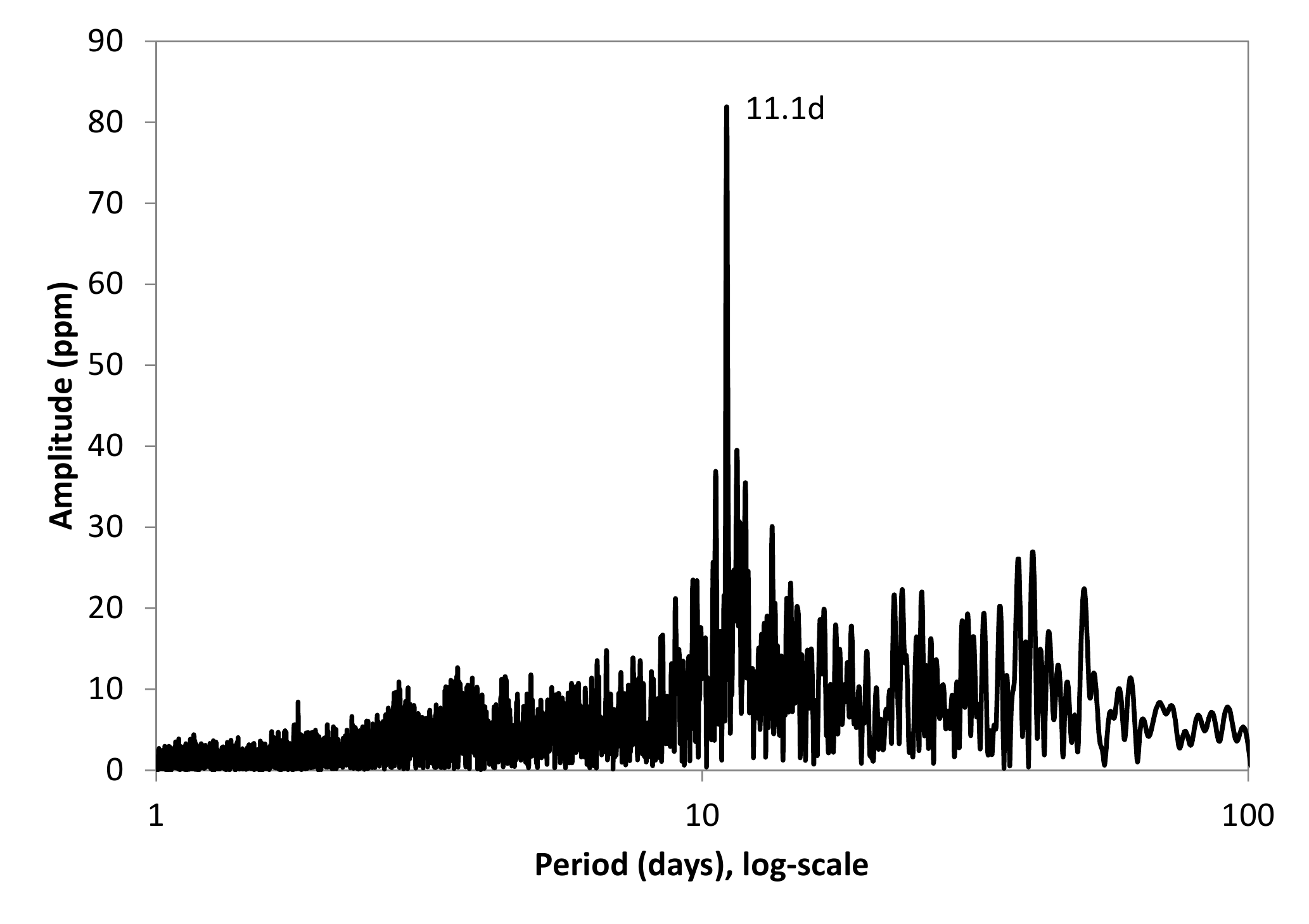}
\includegraphics[width=0.5\linewidth]{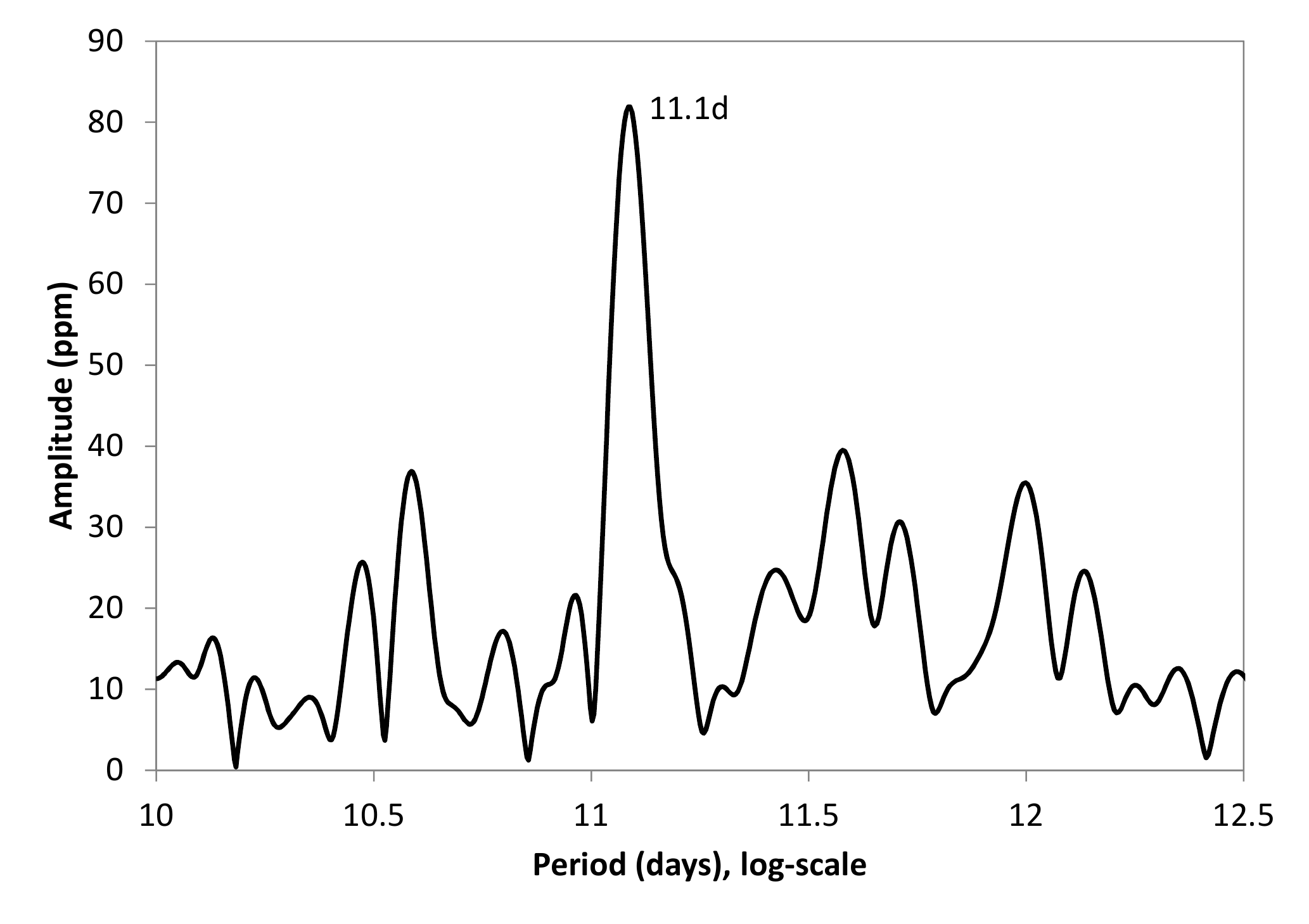}
\caption{\label{fig:rotation}Periodogram for 264b, showing the strongest peak at 11.1d (left), likely the rotational period with an amplitude of 81ppm. Zoom (right) shows sidelobes from differential rotation, a strong indicator for stellar rotation.}
\end{figure*}

\subsection{Results for criteria OSE5: Star-spots}
\label{sub:ose5}
We recommend the use of the normalized PDC-SAP data to search for the stellar rotation, as it effectively removes non-astrophysical systematics in the photometry while leaving the stellar variability intact \citep{Smith2012}. Of course, the transit events need to be removed. We computed Lomb-Scargle periodograms and Fourier transforms for our candidates. For Kepler-264b, we found a consistent highest power peak at a period of 11.1$\pm$0.07 days (with the uncertainty being the full width of the peak at half maximum). This peak is accompanied by several significant sidelobes at periods between 10.5 and 12 days (Figure~\ref{fig:rotation}), as is expected for an activity behavior reminiscent to that of our sun, where differential rotation causes a range of periods from 25 days at the equator to 34 days at the poles.

The OSE might be imitated by stellar spot crossings, in case the ``circumstellar orbital plane of the planet-satellite system were substantially inclined against the stellar equator (...) Such a geometry would strongly hamper exomoon detections via their photometric OSE.'' \citep{Heller2014}. In general, a stellar rotation period with an integer multiple of the planetary period would most strongly affect the OSE. In such a configuration, star-spots would re-occur at the same frequency as the planetary transit, creating dips in a phase-folded plot that might lie just before or after transit, resembling the OSE.

In the case of Kepler-264b, however, the closest integer multiple, $4\times11.1$d=$44.4$d, is too far off to the period of planet \textit{b} (40.8d). We find it therefore unlikely that the OSE is mimicked by star spots. The same is true for Kepler-102e. For Kepler-241c, we find several peaks, the strongest at 12.9d and 19.1d. Smaller sidelobes are around one half, and one third of the planetary period (36.1d). In this case, it is irrelevant what the stellar rotation period in fact is: There is the possibility that star-spots cause the OSE for planets orbiting this star, and thus the test is considered to be failed for both Kepler-241b and \textit{c}.

\subsection{Results for criteria OSE6: Rings}
\label{sub:ose6}
Rings can form when a body enters inside a planet's Roche radius, where the body is ripped apart by tidal forces and its particles become a planetary ring. The Roche radius can be calculated as $2.4R_{P}(D_{P}/D_{S})^{1/3}$, with $D$ denoting the relative densities of planet and satellite. For plausible configurations using the values from section~\ref{sub:stability} for Kepler-264b, the Roche lobe for a planetary density of 2.9g cm$^{-3}$ is $\sim$50,000km, an order of magnitude smaller than the putative moon orbit. For lower densities, and for the other candidates which have smaller radii, the Roche lobe is even smaller.

A simplified search for Saturnian rings has been performed by \citet{Tusnski2013}, using synthetic \textit{Kepler} data and a dark Saturn-model. Rings, however, also produce diffractive forward-scattering and consequently a flux \textit{gain}, which helps to distinguish them from moons. In a more refined simulation, \citet{Hippke2015} examined \textit{our} Saturn transiting \textit{our} quiet Sun through a simulated \textit{PLATO 2.0} instrument (see their Figure 9). The study used real solar data and the ring model from \citet{Barnes2004}, together with calibration data for Saturn's rings from the 1989 occultation of 28 Sgr by Saturn \citep{French2000}. By signal injection and retrieval, it was shown that both the flux gain and loss from ring occultations can be detected. Full modeling of such transits is a topic of recent interest, because rings cause an increase in transit depth that ``may lead to misclassification of ringed planetary candidates as false-positives and/or the underestimation of planetary density'' \citep{Zuluaga2015}.

Depending on the ring composition (ice and/or rock) and density, forward-scattering might be strong or (very) weak. We encourage further study of this timely and interesting topic.

\subsection{Results for criteria OSE7: Plausibility and orbital parameters}
\label{sub:ose7}
It is useful to check whether a putative star-planet-moon configuration is astrophysically possible, plausible and stable over the long time. While the possibility of any configuration is essential, the situation is less clear for the plausibility: Before the discovery of Hot Jupiters, such planets would have seemed implausible, and their discovery \citep{Mayor1995} was challenged as they were found to be incompatible with theories of planetary formation \citep{Rasio1996}. 

The same is true for the stability. While we should assume that most moons are stable over long (Gyr) times, there might be configurations for which this is not the case. One related example are Saturn's rings, which would be observable with \textit{Kepler}-class photometry \citep{Barnes2004}, and are known to be unstable on timescales of $<$100 Myr \citep{Dougherty2009}. Thus, instability should raise strong doubts about the presence of a moon (and less so about the presence of a ring), but does not proof their non-existence.

We suggest to derive estimates for putative planet-moon configurations, as will be explained in the following section, and then to assess the system.

\begin{figure*}
\includegraphics[width=0.5\linewidth]{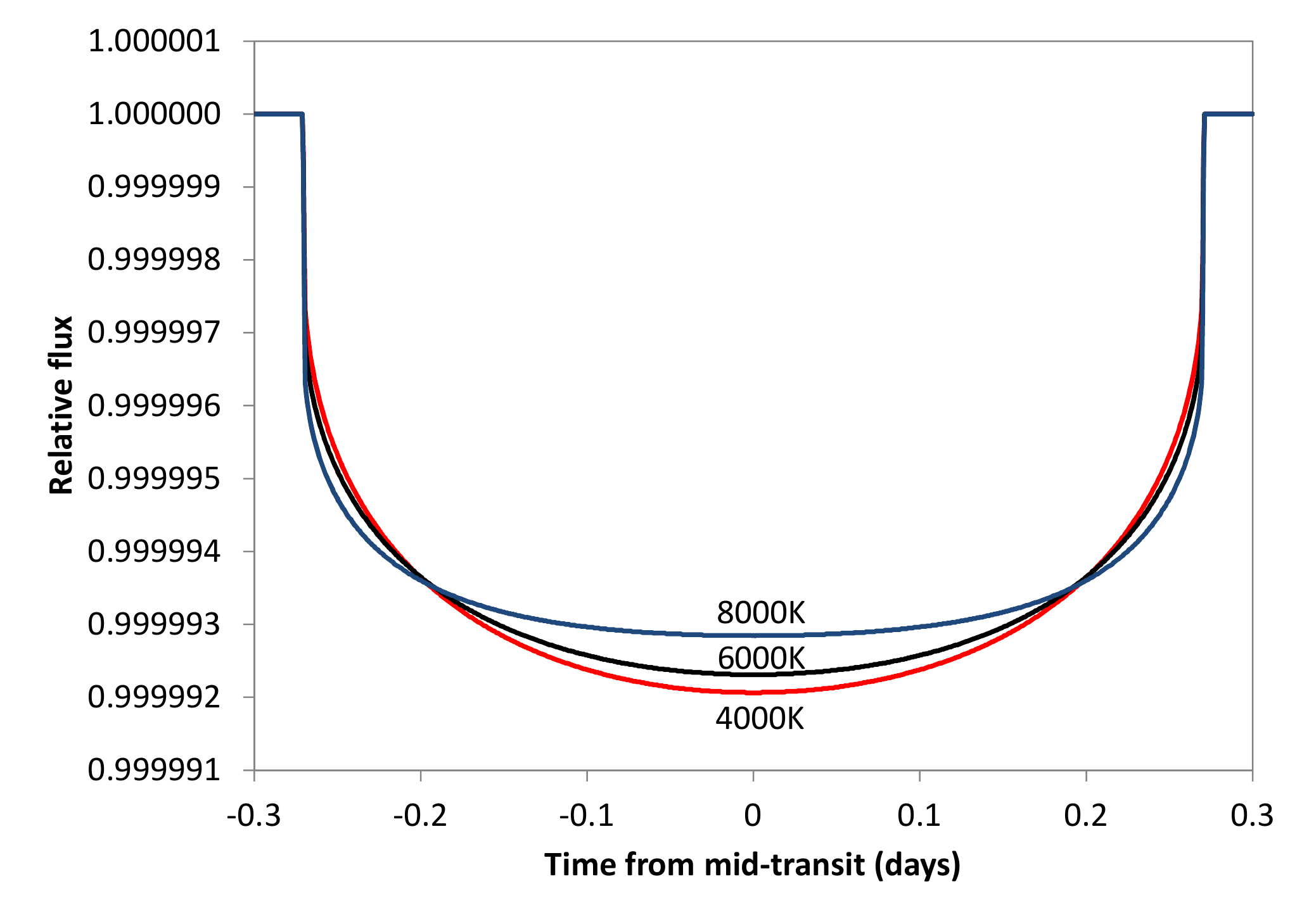}
\includegraphics[width=0.5\linewidth]{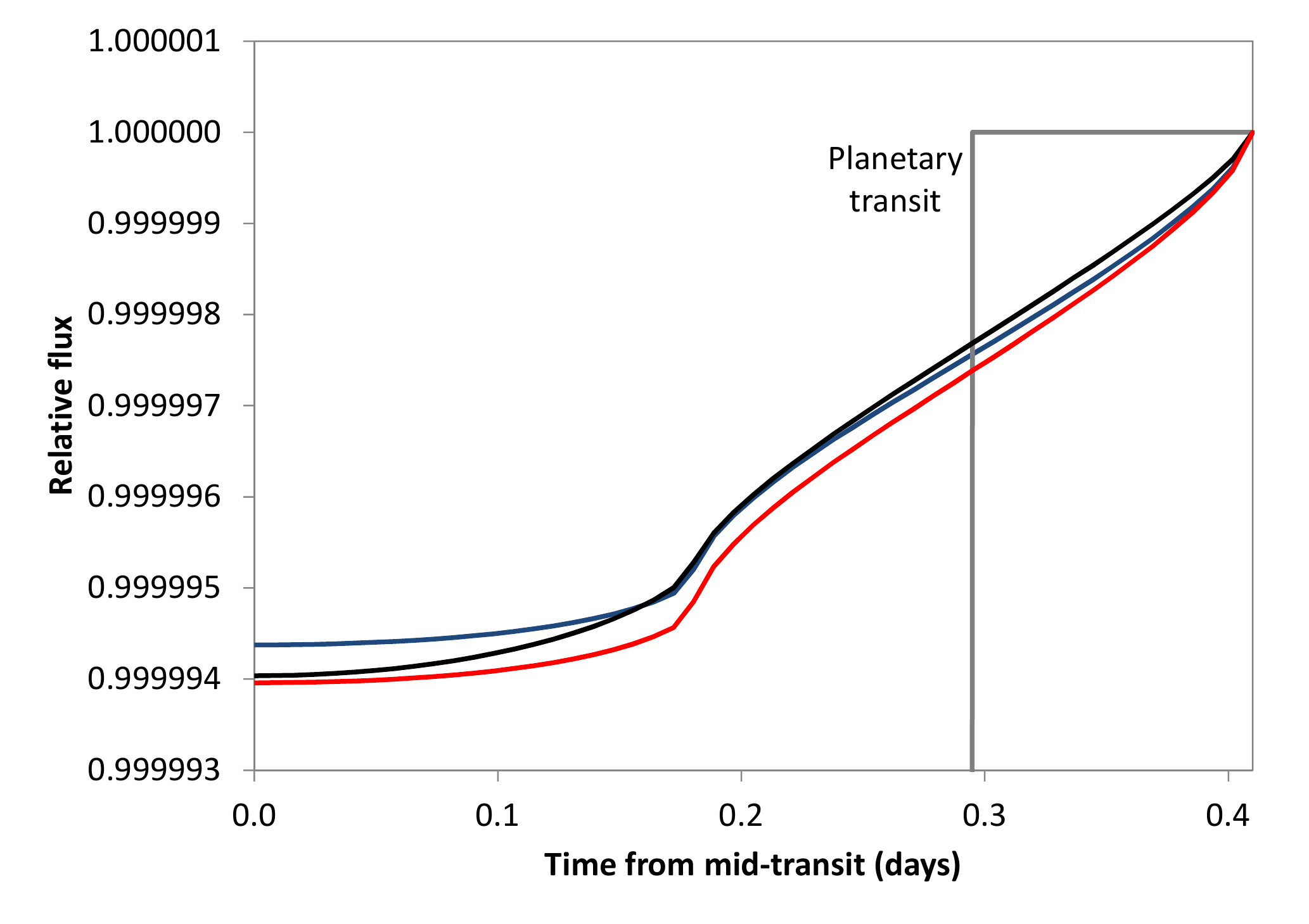}
\caption{\label{fig:limb}Effects of limb-darkening on the OSE. Left: Synthetic transit shape of Earth's moon transiting our Sun (black line, 6000K). While keeping all else identical, we have changed only the quadratic limb darkening parameters to match temperatures of 4000K (red) and 8000K (blue). Right: The resulting OSE for these three transit curves, shown as a zoom into egress. The differences are clearly visible, but very small compared to expected \textit{Kepler} noise levels.}
\end{figure*}

\begin{figure*}
\includegraphics[width=0.5\linewidth]{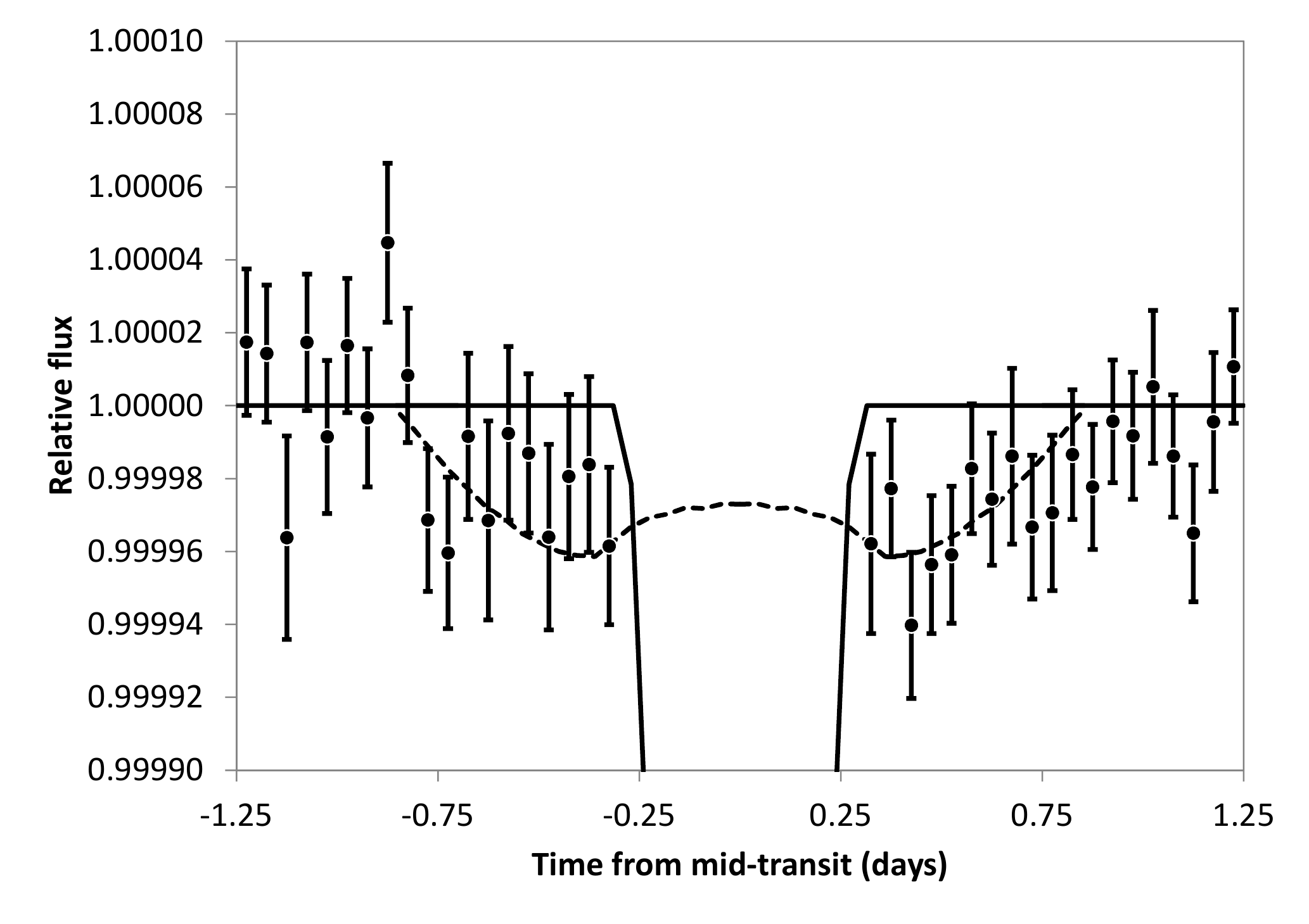}
\includegraphics[width=0.5\linewidth]{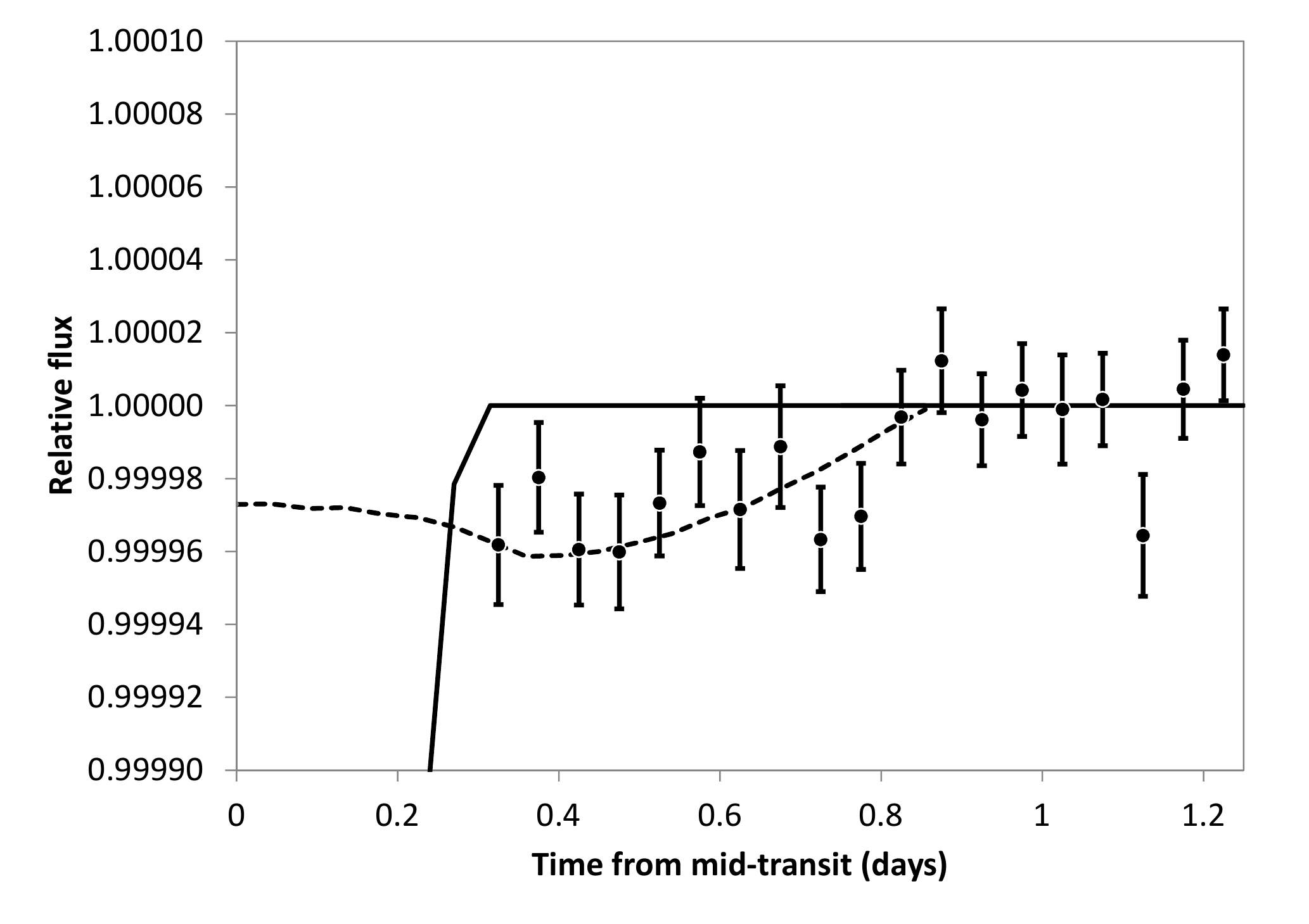}
\caption{\label{fig:flux-ose-fits}OSE fit for Kepler-264b with $1\sigma$ uncertainties. Dashed line is a numerical OSE fit. Left: Separate data for ingress and egress. Right: Binned together assuming symmetry. Best fit gives $R_{\leftmoon}=1.6\pm0.2 R_{\oplus}$, semi-major axis 660,000km.}
\end{figure*}

\begin{figure*}
\includegraphics[width=0.5\linewidth]{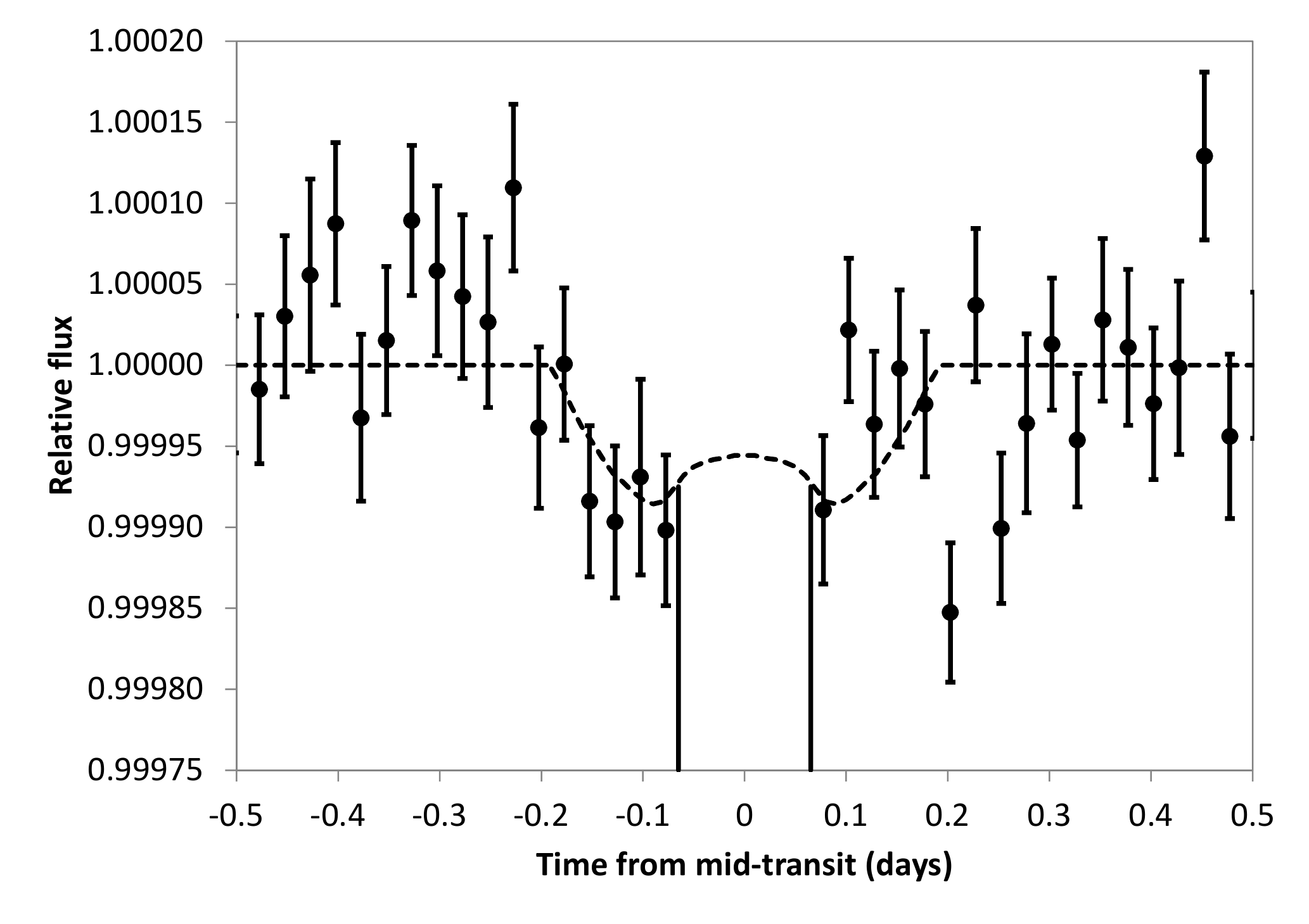}
\includegraphics[width=0.5\linewidth]{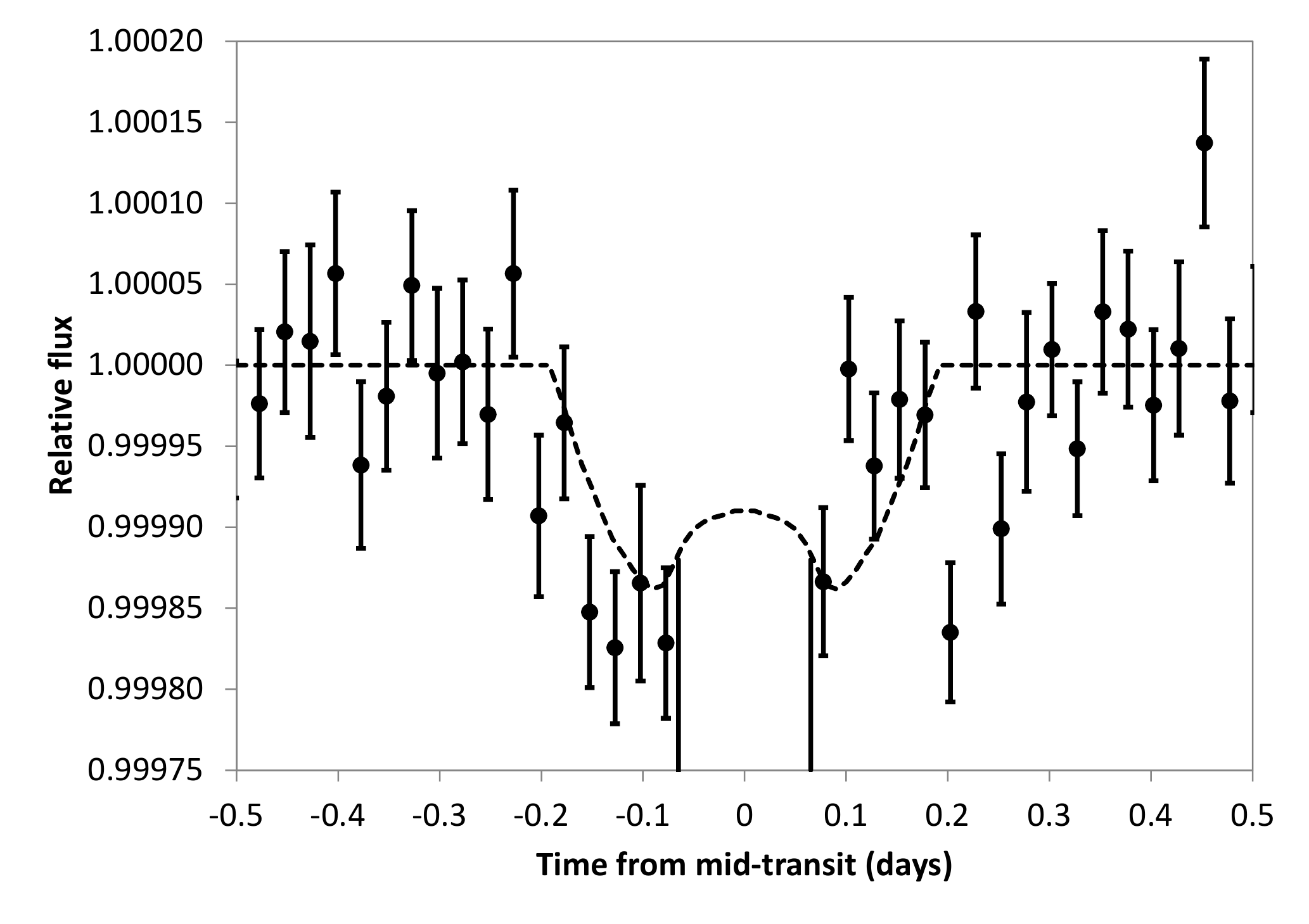}
\caption{\label{fig:241b}OSE fits for Kepler-241b, comparing two detrending options. Same scales for comparison. Left: Median filtering, giving best fit $R_{\leftmoon}=0.8\pm0.1 R_{\oplus}$, semi-major axis 450,000km. Right: Parabola detrending, giving best fit $R_{\leftmoon}=1.1\pm0.1 R_{\oplus}$, same semi-major axis. This shows the uncertainty which is caused by the choice of detrending.}
\end{figure*}

\subsubsection{Fitting the data to the OSE}
\label{fitter}
A fit of the data to various exomoon parameters can be done using the analytical equations from \citet{Heller2014} with an iterative $\chi^2$ minimization. The parameter space of such a Monte-Carlo run can also include the test for multiple moons versus a single moon. With \textit{Kepler}-data, however, we judge the differences (a few ppm) as too small for a try.

Alternatively, one can create numerical simulations, either on picture/pixel level, or on photometric level, as we have done. The principle of stacking is shown in Figure~\ref{fig:OSEsketch2}, and we make our software template available for the interested reader\footnote{\url{http://jaekle.info/ose/}}. In this paper, we have only fitted single moons on circular, symmetrical cases. Using the method as demonstrated in the spreadsheet, we have numerically sampled OSE curves using 10,000 samples, and a time resolution of 1min. For every system, we have created these curves for varying moon radii (in 0.01 R$_{\oplus}$ steps) and semi-major axes (in 10,000km steps). This results in a large ($10^5$) number of possible solutions. To find the best fit, we have used the unbinned data, and performed a least-squares fit to the collection of synthetic curves, choosing the one with the smallest squared residuals as the best fit. To determine uncertainties, we have adopted the full width at half mean.

This method introduces two slight errors. It adopts the transit curve (and thus transit duration) of the planet for the sampling of the moon OSE. This neglects the difference in transit duration due to the smaller body, and thus introduced a slight error ($\sim$1\%). It has, however, the advantage to naturally account for the best-fit limb darkening, as this is included in the best-fit planet transit. The second error is that in the perfect sky co-planar configuration, a small number ($<$6.4\%, see section 5.9.1.) of mutual planet-moon eclipses occurs, which the sampler ignores. This could be accounted for by removing an estimated fraction (e.g. 3\%) of samples. To keep this work simple, we have neglected the slight error. Note that there are near-sky-coplanar configurations (towards inclination), where no mutual eclipses occur. So this choice is in fact one possible configuration.

We have estimated the effect of an error in limb darkening values, by calculating different transit curves, and resulting OSEs, for various limb darkening parameters (and keeping all else equal). For this, we calculated synthetic light curves for a Sun-Earth-Moon configuration using \textit{PyAstronomy}, and varied only the temperature of the star. While the uncertainty in stellar temperature is $\sim$100K for the candidates in this paper, we have used values of 4000K, 6000K (approximately our Sun) and 8000K. Obviously, for a star of the size of our Sun, the temperature range cannot be that large. In order to explore the extreme ends of limb darkening, however, we have chosen these values. The corresponding coefficients are taken from \citet{Claret2011}. Afterwards, we calculated the corresponding OSE. As can be seen in Figure~\ref{fig:limb}, the effect is small, but clearly visible. For a smaller error in temperature, e.g. the typical uncertainty of 100K, the effect on the OSE is very small. Therefore, we can safely ignore limb darkening uncertainties, and continue with the best-fit values from the planetary transit models.

\begin{figure}
\includegraphics[width=\linewidth]{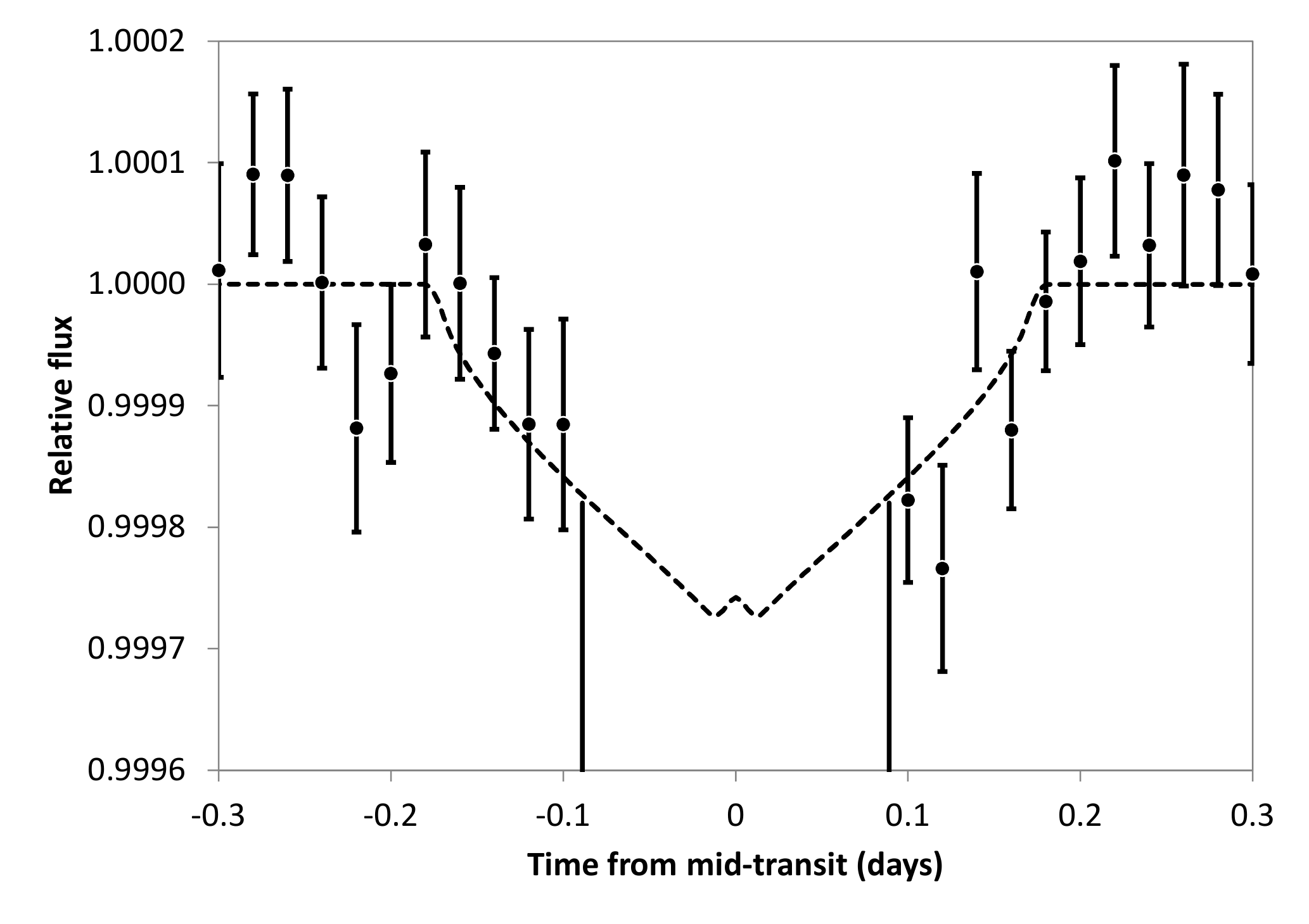}
\caption{\label{fig:241c}OSE fit for Kepler-241c using median filtering. In this case, the OSE has a distinctively different shape at planetary mid-transit, due to the smaller semi-major axis.}
\end{figure}

\begin{figure*}
\includegraphics[width=0.5\linewidth]{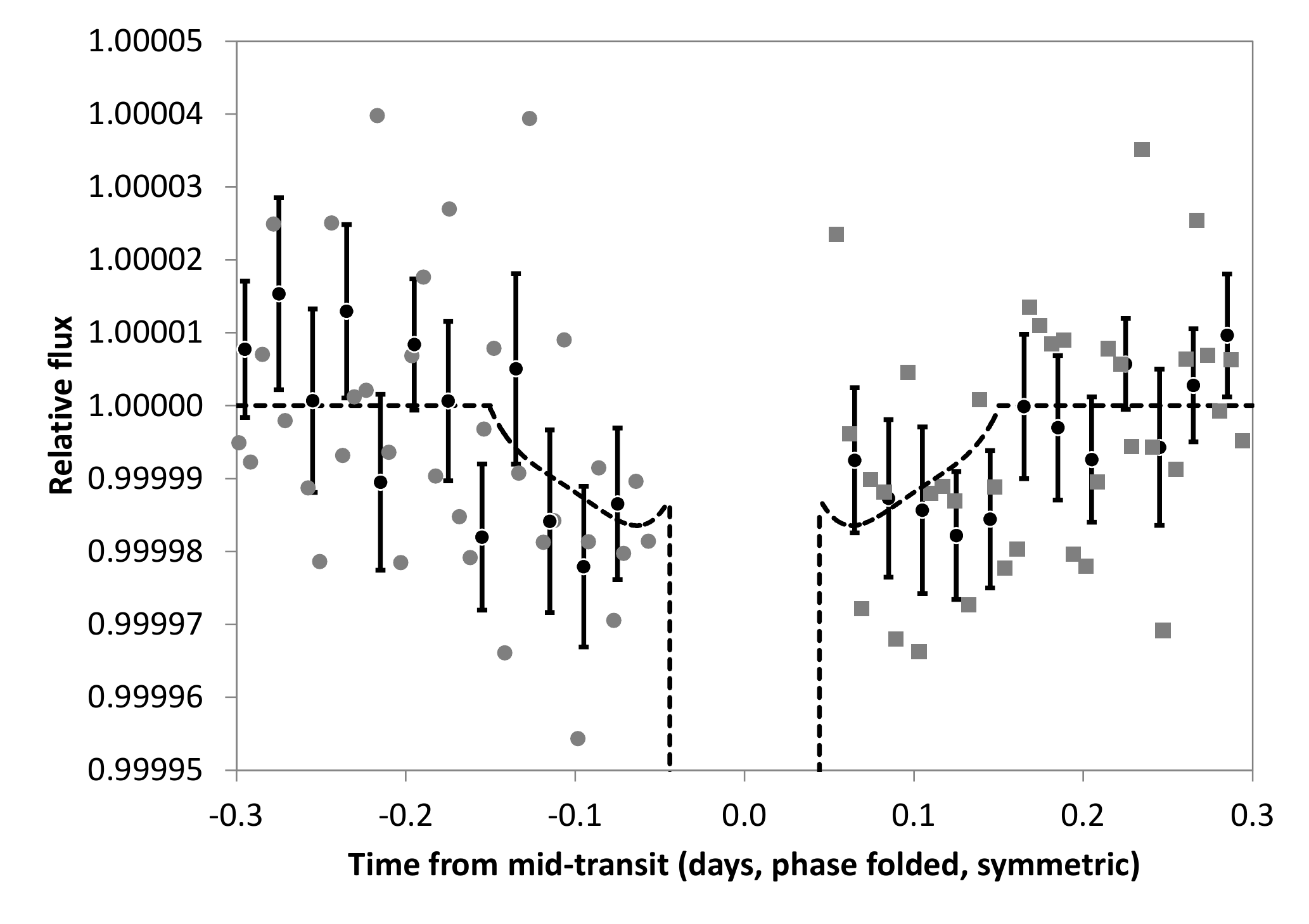} 
\includegraphics[width=0.5\linewidth]{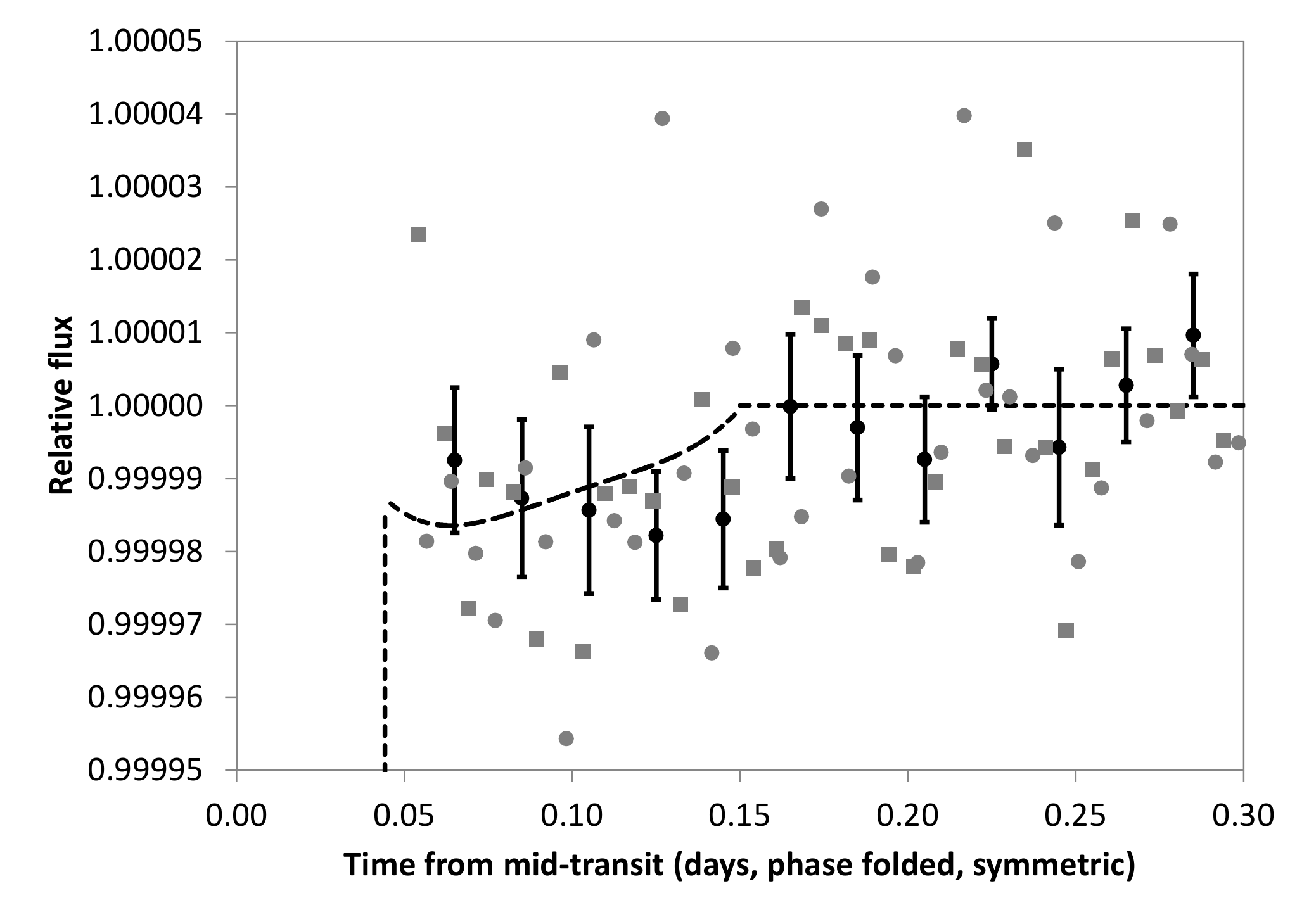}
\caption{\label{fig:flux-ose-fits367}KOI367-01. Left: Ingress and egress shown separately for illustration. On the egress side, the slope test (OSE2) has failed, as the flux should decrease towards planetary transit. Right: Combined data giving a best fit for $R_{\leftmoon}=0.8\pm0.1 R_{\oplus}$, semi-major axis 150,000km.}
\end{figure*}

We have performed these simulations for all four systems in question. The best-fit results are shown for Kepler-264b (Figure~\ref{fig:flux-ose-fits}), Kepler-241b (Figure~\ref{fig:241b}) and \textit{c} (Figure~\ref{fig:241c}), and for KOI-367-01 in Figure~\ref{fig:flux-ose-fits367}.

In addition, we have performed a signal-injection and retrieval, as will be explained in the following section.

\subsubsection{Signal injection and retrieval}
\label{sub:injection}
A useful test is to use real data (including its red noise), without any transit events, and inject simulated moon (and planet) transits into it. Afterwards, both results can be compared; they should be consistent within the errors.

We have performed this for Kepler-264b, using the same number of transits (26) and randomly chosen plain time segments of Kepler-264b data. Into these data, we have injected the moon dips described in the previous section. Afterwards, we have retrieved and folded them to the same OSE. The result is virtually identical to the real data, and visually indistinguishable from Figure~\ref{fig:flux-ose-fits}. We have repeated this test 10 times using different segments of data, and all look very similar -- some a bit above in flux compared to the real result, some a bit below. This is of course no proof that a detection is real, but it is a necessary test to check that a detection \textit{can} be possible with the given data quality. Clearly, the real Kepler-264b data is sufficient for a 1.6 $R_{\oplus}$ exomoon detection.

It would be beneficial to repeat this test for all candidates in this work. However, the signal-injection and retrieval has proven very time consuming. One needs to generate the best-fit transit shapes for planet and moon, select appropriate time segments, scale all values and inject the data. Then, all transits need to be checked manually: Are there any gaps in the dataset that reduce the number of transits required? As the number of transits should be the same in the real data and in the injected data, one needs to manually shift the plain data until this is resolved. Afterwards, the full OSE find-and-retrieve is required. Finally, we recommend to repeat this procedure several times, as in Monte-Carlo sampling, to avoid choosing ``better'' or ``worse'' plain data (in terms of noise) than the original. We encourage the development of a software tool for this task, which will be valuable for future OSE validations.

\subsubsection{Plausibility of the OSE fits}
From the results shown in Table~\ref{tab:moons}, it is apparent that the putative total moon radii are all larger than single solar-system moons. It is useful to put these results in context. We will take Kepler-264b as an example, as this candidate has the largest putative moon.

Assuming a solar radius of $R_{*}=1.55R_{\astrosun}$ for Kepler-264b and a planetary radius $R_{P}=3.3\pm0.74R_{\oplus}$, the total exomoon radii would then be $\sim$10,000km (1.6 $R_{\oplus}$) on a semi-major axis of $a_{SP}/R_{P}=31\pm11$ planetary radii, i.e. $\sim$660,000km. For comparison, the Earth/Moon orbit is at 384,400km; Jupiter/Ganymede at 1,070,000km. The uncertainty of the semi-major moon axis is dominated by the planetary impact parameter $b=0.93\substack{+0.01 \\ -0.64}$ and the host star's radius, $R_{*}=1.55\pm0.31R_{\astrosun}$, which is reflected in the planetary radius uncertainty, 3.3$\pm$0.74R$_{\oplus}$. It is important to note that these parameters quantify the total OSE, and it is unclear whether this comes from a single moon or multiple moons.

\begin{figure}
\includegraphics[width=\linewidth]{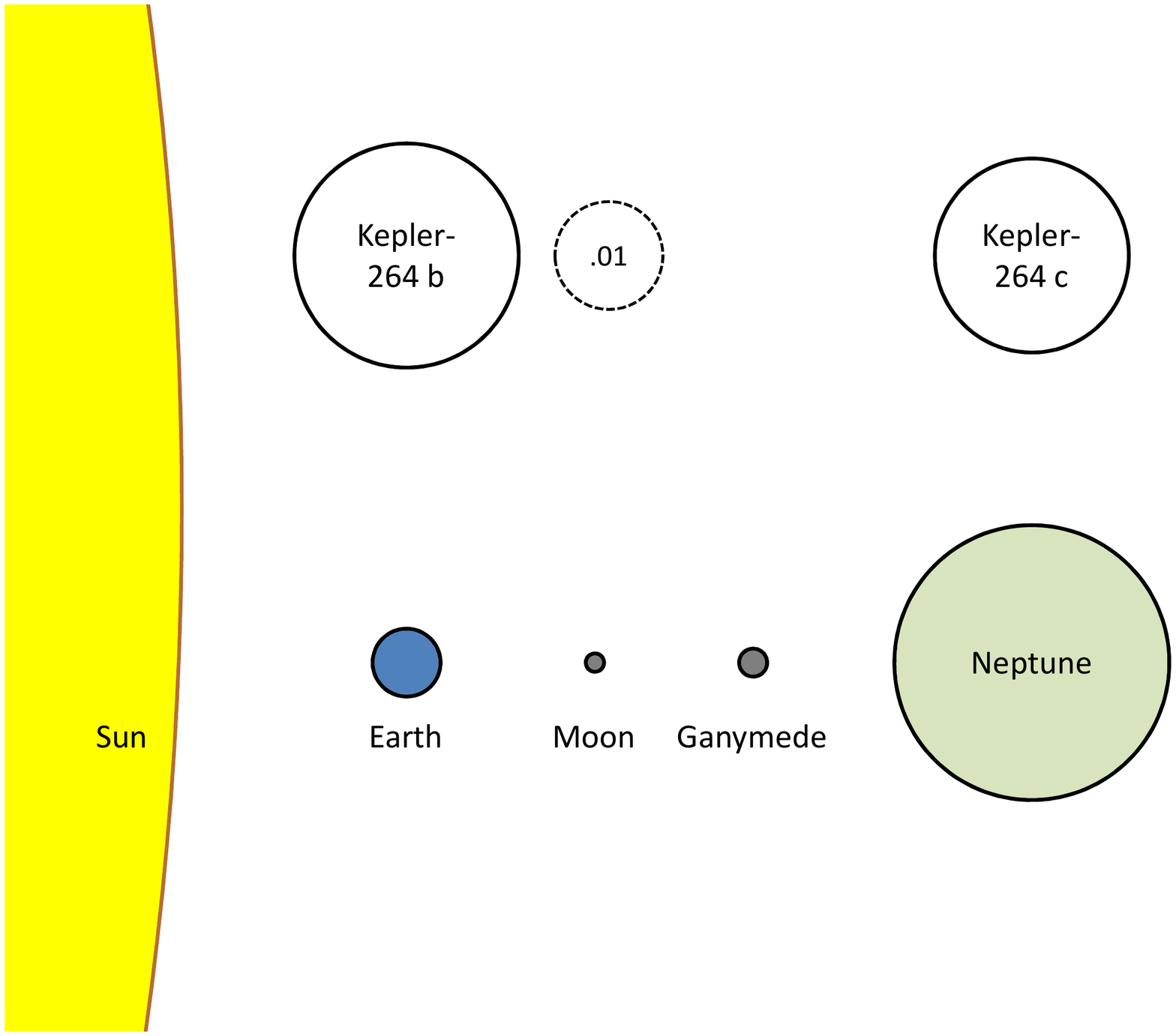}
\caption{\label{fig:size}Size comparison for Kepler-264 planets a and b, together with potential exomoon .01. The bottom row shows some solar system objects for comparison. Distances are not to scale.}
\end{figure}

For comparison, Jupiter's largest moons Ganymede (2,632 km, 5.6ppm), Callisto (2,410 km, 4.7ppm), Io (3,660 km, 2.7ppm) and Europa (1,560km, 2ppm) would account for a flux loss of 15ppm, that is $\sim$50\% of the required 31ppm, in case they were orbiting Kepler-264b on short orbits. Kepler-264b is almost (86\%) the size of Neptune, thus making the argument for several moons more plausible. Figure~\ref{fig:size} shows a sketch of the potential exomoon size for the case of a single moon.

Following formation theories, such large moons would likely not have formed in situ \citep{Canup2006}, but be captured \citep{Williams2013}. It has also be shown that Neptune-sized planets can capture Earth-sized moons \citep{Porter2011}.

\subsubsection{Shape of the OSE}
The data quality is not sufficient for a detection of the ``wings'', where moon egress/ingress takes place just before/after planetary transit. This occurs for single moons on wide orbits. A detection of this feature might be a powerful tool used with better photometry in the future: It requires the data to not only exhibit some dip, but follow a distinct dip shape. Judging from Figure~\ref{fig:flux-ose-fits}, $\sim $2-3 $\times$ lower noise is required for a 2$\sigma$ detection, indicating that such a detection is possible with \textit{PLATO 2.0}-class photometry. Indeed, in a recent study by \citet{Hippke2015}, this was demonstrated by signal-injection and retrieval for a 2R$_\oplus$ planet with a 0.4R$_\oplus$ (Ganymede-sized) moon orbiting a 0.5R$_\odot$ M-dwarf.

\subsubsection{Orbital stability}
\label{sub:stability}
Kepler-264b is the most critical case in terms of stability, as its fit gives the largest radius and the widest orbit. Therefore, we discuss orbital stability for this example. 

Neglecting a potential eccentricity, the Hill sphere can be calculated as $r=a\sqrt[3]{M_{P}/3M_{\astrosun}}$, where $r$ is the Hill radius and $a$ the distance between star and planet. As the mass of the planet cannot be measured from the photometry directly, we can only derive limits for the planetary density with regards to the Hill sphere. For a Hill radius of 660,000km, the planetary density needs to exceed 1.75g cm$^{-3}$, which is slightly above the density of Neptune (1.64g cm$^{-3}$). For a heavier composition of e.g. 2.9g cm$^{-3}$, as in the comparably sized Kepler-20c \citep{Gautier2012}, the Hill sphere would extend to 1,700,000km, so that the potential exomoon would lie at 39\% of the Hill radius. 

Following Kepler's 3rd law, one can derive limits on the planetary mass due to the presence of a moon. Previous studies proposed transit timing variation \citep{Kipping2010} and the OSE \citep{Heller2014} to measure the mass of a planet, using its moon. There is a third test: Hill stability. Depending on the mass (density) of the planet, 264b's orbital period for a 660,000km orbit would be in between 12.9 days (2.9g cm$^{-3}$) and 17.1 days (1.64g cm$^{-3}$). This shows that exomoons can reveal a planet's mass from Hill stability, using only photometry.

Retrograde moons are argued to be stable up to nearly (93\%) of the Hill radius \citep{Domingos2006}. On the other hand, prograde moons (which provide the vast majority among the principal solar system moons) are only stable out to about 0.5 Hill radii. A prograde moon at 39\% of the Hill radius would be at the edge of orbital stability. And, second, it would be much farther away from its planet than any major moon in the solar system (in terms of Hill radii, not absolute distances). So the moon interpretation in this case is doubtful.

To sum up, an exomoon around Kepler-264b could be stable for the (less common) retrograde case, and if the planet's density exceeds that of Neptune, which is plausible for a planetary radius of 3.3 $R_{\oplus}$.

\subsection{Results for criteria C1: Temporal spread}

\subsubsection{River-plot}
\label{sub:c1}
\begin{figure}
\includegraphics[width=\linewidth]{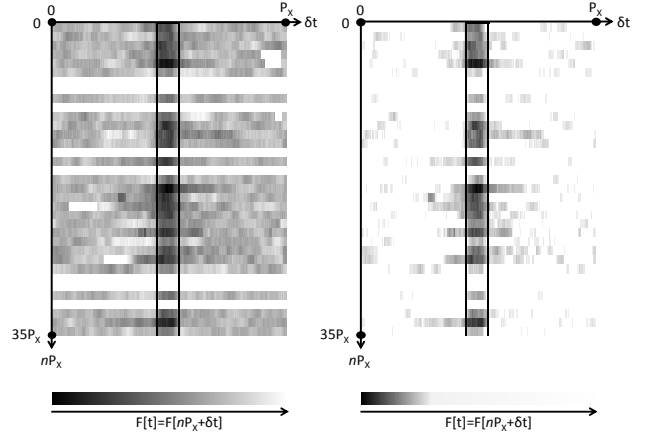}
\caption{\label{fig:river}Left: River-plot for Kepler-264b. Each row is one transit (in between vertical lines); time runs from top to bottom. Darker colors are lower flux, whites in the left plot are missings. Both top plots show the same data, but with different shadings for clarity. On the right plot, it can be readily seen that the flux loss clusters before ingress and after egress. It is also evident that this flux loss is not located in only part of the data (e.g. in only one line, in only the first half), but is present all over the dataset.}
\end{figure}

In the case of \textit{one} moon, it should only block light on some transits (not all), and only on one side (ingress or egress) per transit. Unfortunately, the photometry is at the very limit for exomoon dips on the ppm level in a single transit. We suggest to make a ``river plot'' \citep{Carter2012, Nesvorny2013}, and show this for Kepler-264b in Figure~\ref{fig:river}. In the stretched version (right panel), one tends to believe that the dips are indeed one-sided. For instance, the first row (period) shows a flux loss before ingress, while the second has only a flux loss after egress. Also, a few transits seem to lack any moon dip, as expected for those cases when the moon is in front or behind the planet, or not transiting at all (e.g., due to a non-zero inclination). However, there are also rows that show features both before ingress and after egress (e.g. row 5). What is more, some rows show features longer than possible for a moon transit (e.g. row 13). 

\begin{figure}
\includegraphics[width=\linewidth]{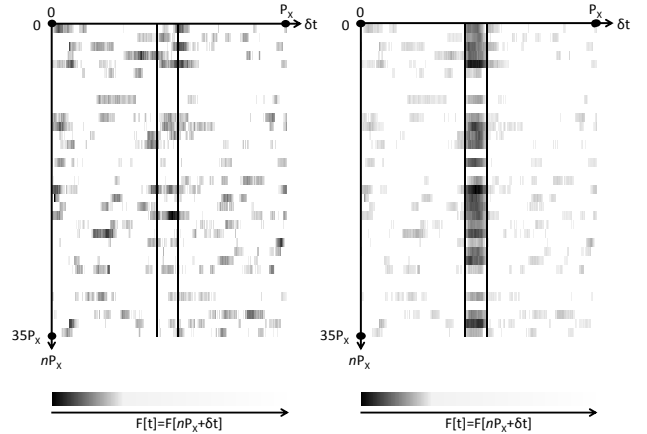}
\caption{\label{fig:river11d}Left: River plot null test for the phase folded time -11 days before planetary transit, where the third largest dip (after ingress and egress) in the dataset occurs, caused by noise. Flux loss seems to concentrate on the first rows. Right: Same time and data, after injection of the transit dip. Flux loss is not apparent anymore, indicating a noise-only dip, as is expected at this time of no transit.}
\end{figure}

Such noise features are also present in a null test (Figure~\ref{fig:river11d}) for the phase folded time -11 days before planetary transit, where the third largest dip (after ingress and egress) in the dataset occurs, caused by noise. Without photodynamical modeling, which is beyond the scope of this paper, we find it hard to decide which of these features have an individual astrophysical cause, and which are instrumental trends or simple noise. 

As can be seen in Figure~\ref{fig:river}, 8 out of 27 transits (30\%) do not show a potential moon transit outside the planetary transit. This can be compared to the theoretical upper limit of $\sim$6.4\% ($\sim$2 transits), as derived by \citet{Heller2012} using the Roche stability criterion. Thus, the missing 6 transit events indicate that the cause for the dips is not a moon. To have a 30\% probability of a moon eclipsing its planet in the given star-planet-configuration, the moon would need to be well within the planetary Roche radius, breaking the moon apart. Moreover, if the moon's structure would be stable due to a ridiculously high moon density, then the moon would actually show multiple eclipses during one stellar transit \citep{Heller2014c}. There are two ways around this problem, and both are problematic. The first is to assign the missing transits to red stellar noise, as a single moon transit S/N is $<$3. This would however mean that the \textit{positive} events (flux loss events) could equally be caused by red noise, with the opposite sign. The second way around it is to assume a non-coplanar configuration, e.g. inclined (compare Figure~\ref{fig:OSEconfigs}). This explains the missing transits, but would further increase the (already large) moon radius.

To sum up, we judge the river plot (the individual transit events) as rather disfavoring the moon hypothesis, without rejecting it completely. It is also worth noting that the OSE framework as introduced by \citet{Heller2014} considers only phase-folded light curves. However, the examination and validation of individual, contributing transit-like signals can be a useful addition to validate or refute a moon hypothesis.

\begin{figure}
\includegraphics[width=\linewidth]{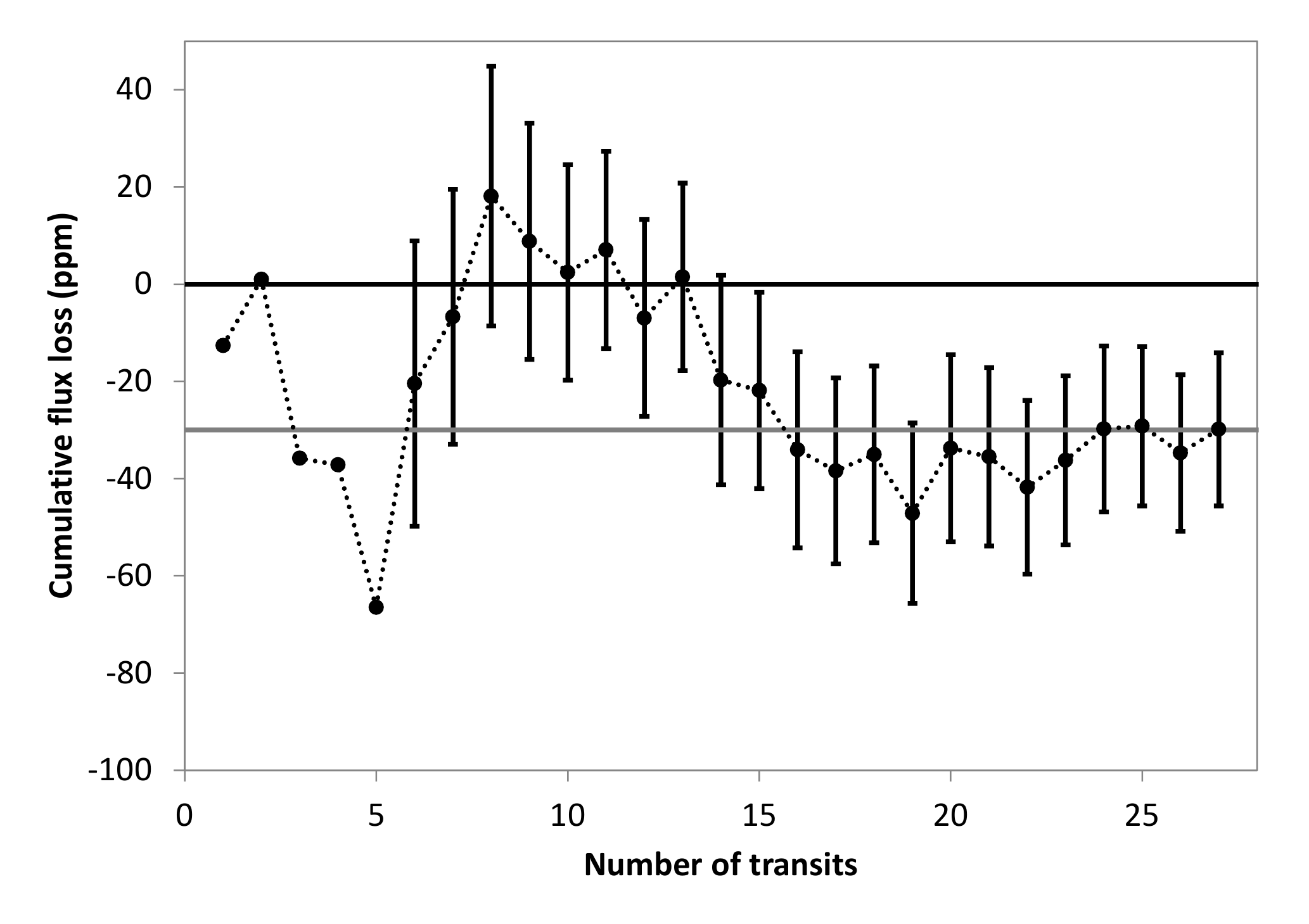}

\includegraphics[width=\linewidth]{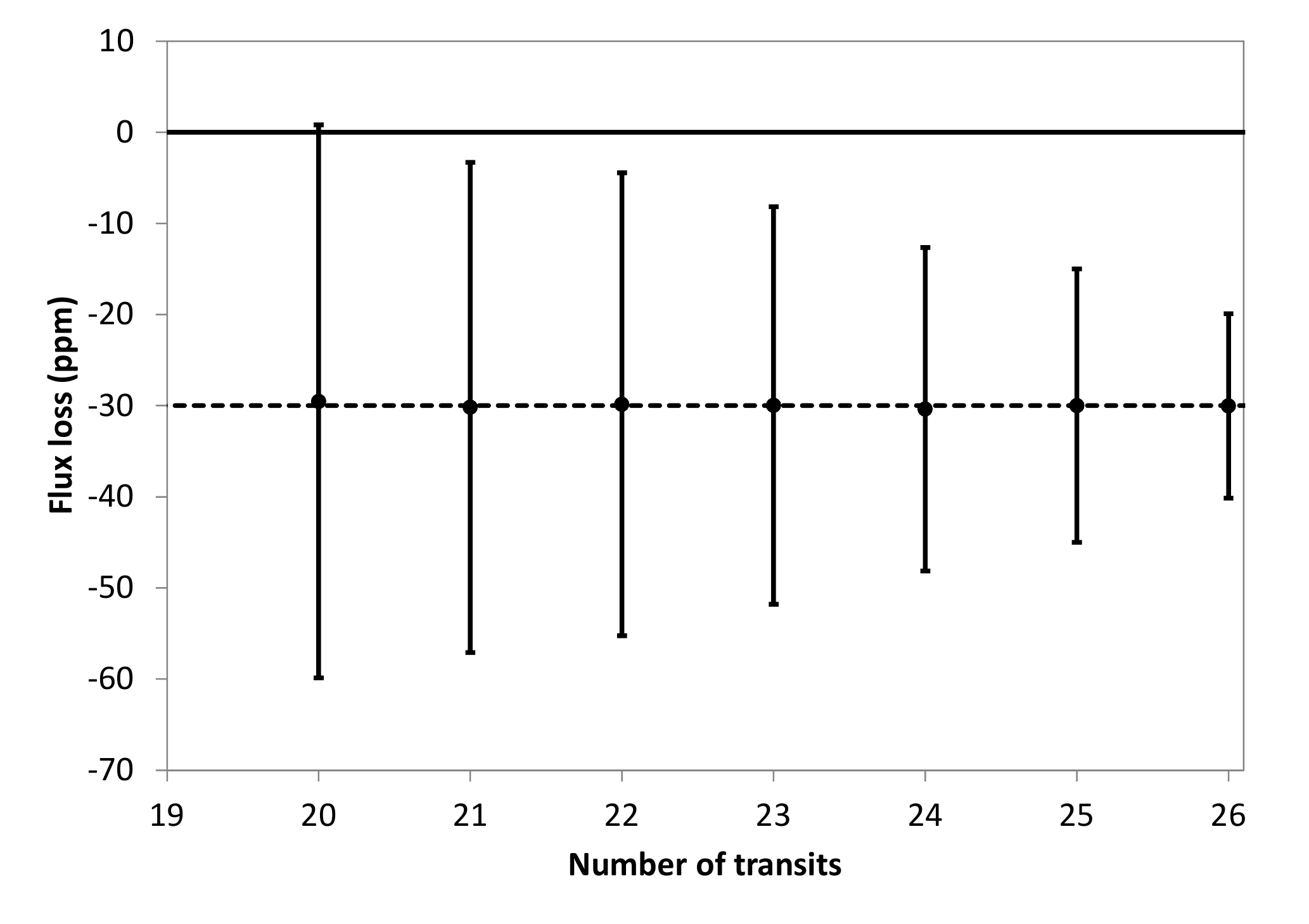}

\includegraphics[width=\linewidth]{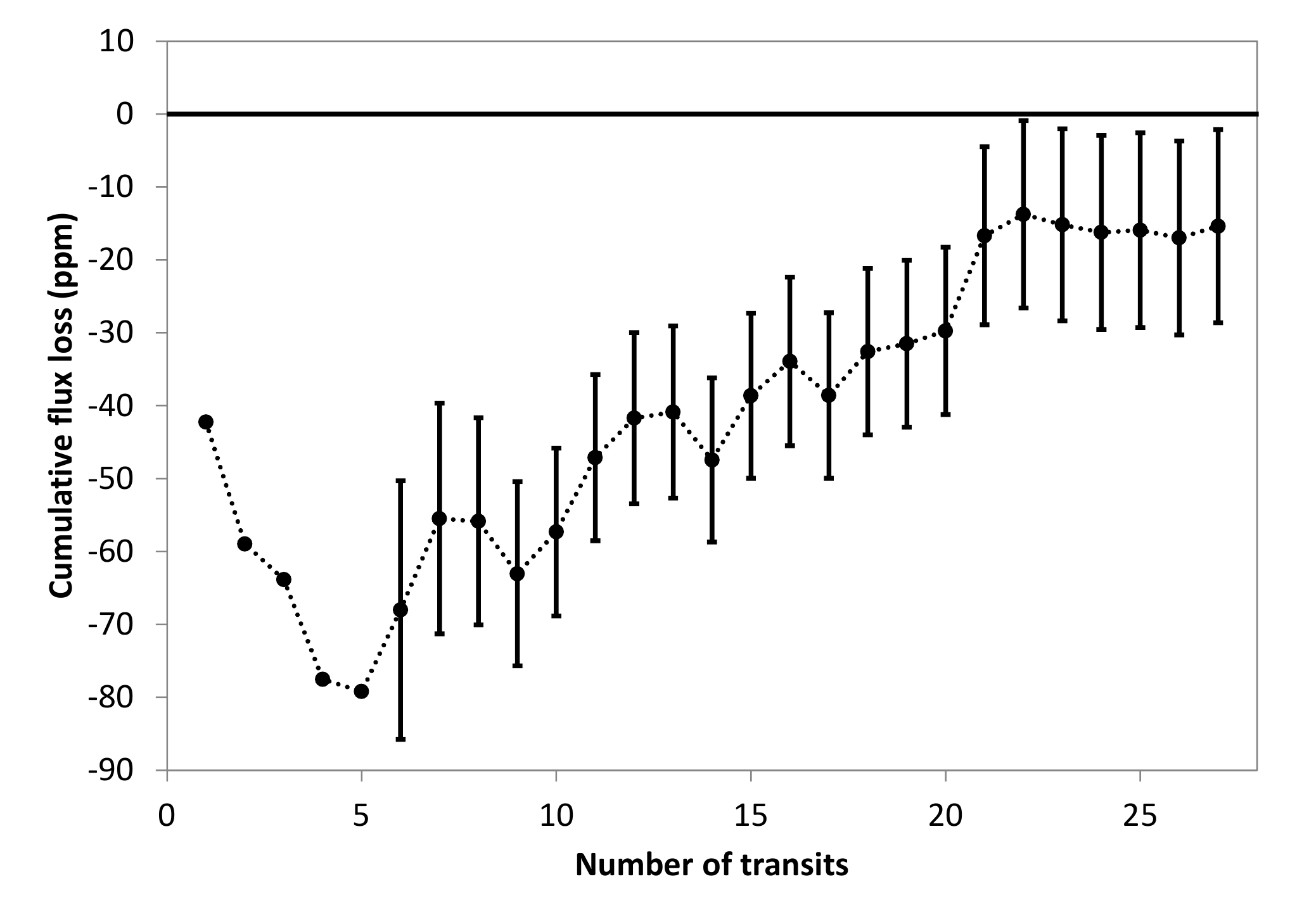}
\caption{\label{fig:buildup}Top: Real build-up for Kepler-264b flux loss over time. After a random episode, the OSE builds up from transit \#13 on and stabilizes at $\sim$30ppm after transit \#16. Uncertainties are 2$\sigma$ (95\%). Middle: Monte-Carlo test asking for the average number of transits required (result: 21) for a detection at the 95\% significance level. Bottom: Null test for the phase folded time -11 days before planetary transit, where the third largest dip (after ingress and egress) in the dataset occurs, caused by instrumental and/or stellar noise. Flux loss is only apparent at the beginning of the dataset. After only 5 transits, no OSE can have build up; proof of a false positive dip. Also, as is expected for a random-walk of noise, the trend is likely to return to zero after long enough time.}
\end{figure}

\begin{figure*}
\includegraphics[width=0.5\linewidth]{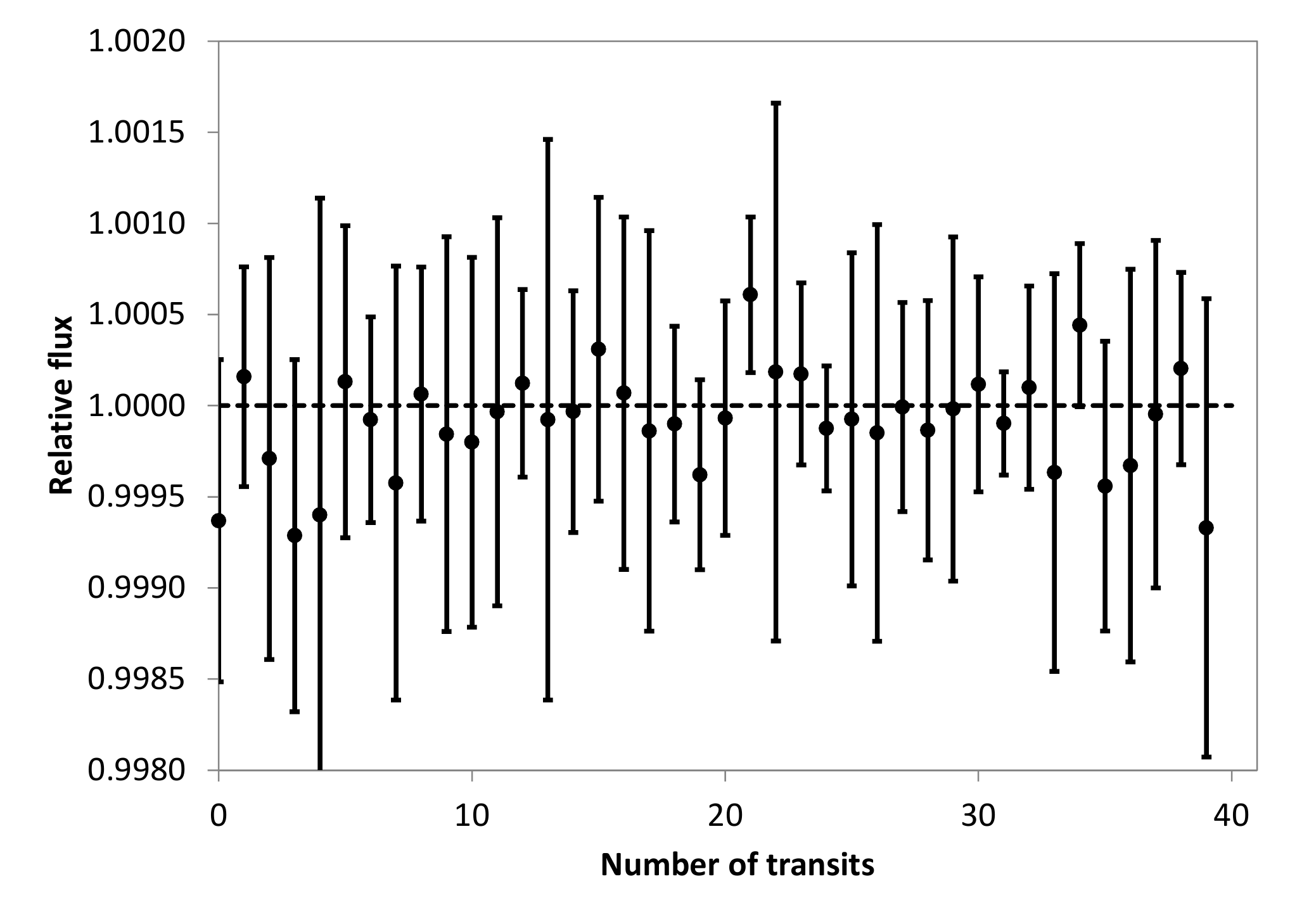}
\includegraphics[width=0.5\linewidth]{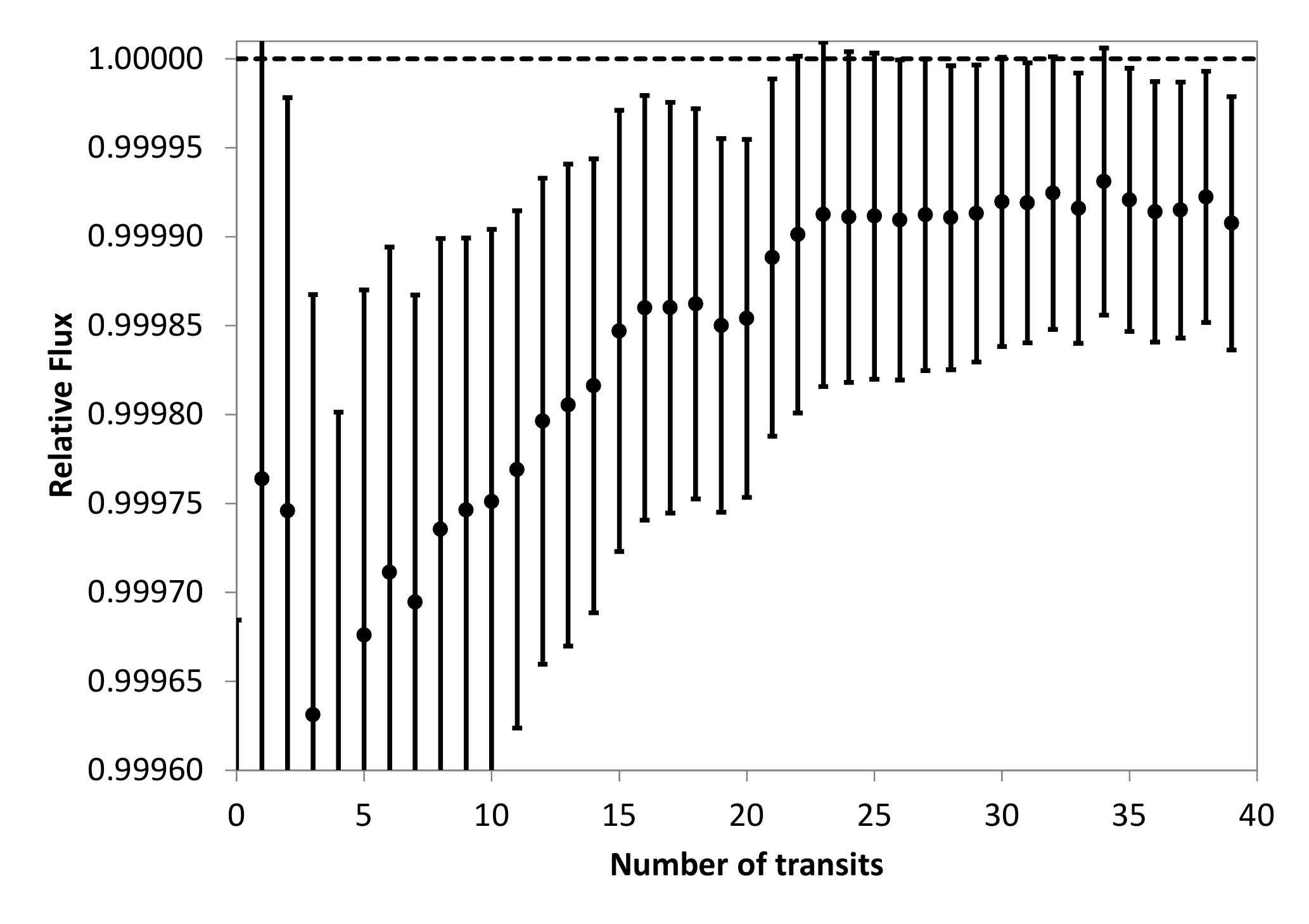}
\caption{\label{fig:buildup241c}Left: Individual flux measurements for Kepler-241c for each transit (taking the average flux for a time of 0.05d before ingress and after egress of each transit). Most periods are, by itself, consistent with no flux loss. However, it is apparent that the average is below nominal flux, indicative of another transiting body. Right: Cummulative flux loss over time. After a random episode, the OSE stabilizes after transit \#23.}
\end{figure*}

\begin{figure*}
\includegraphics[width=0.5\linewidth]{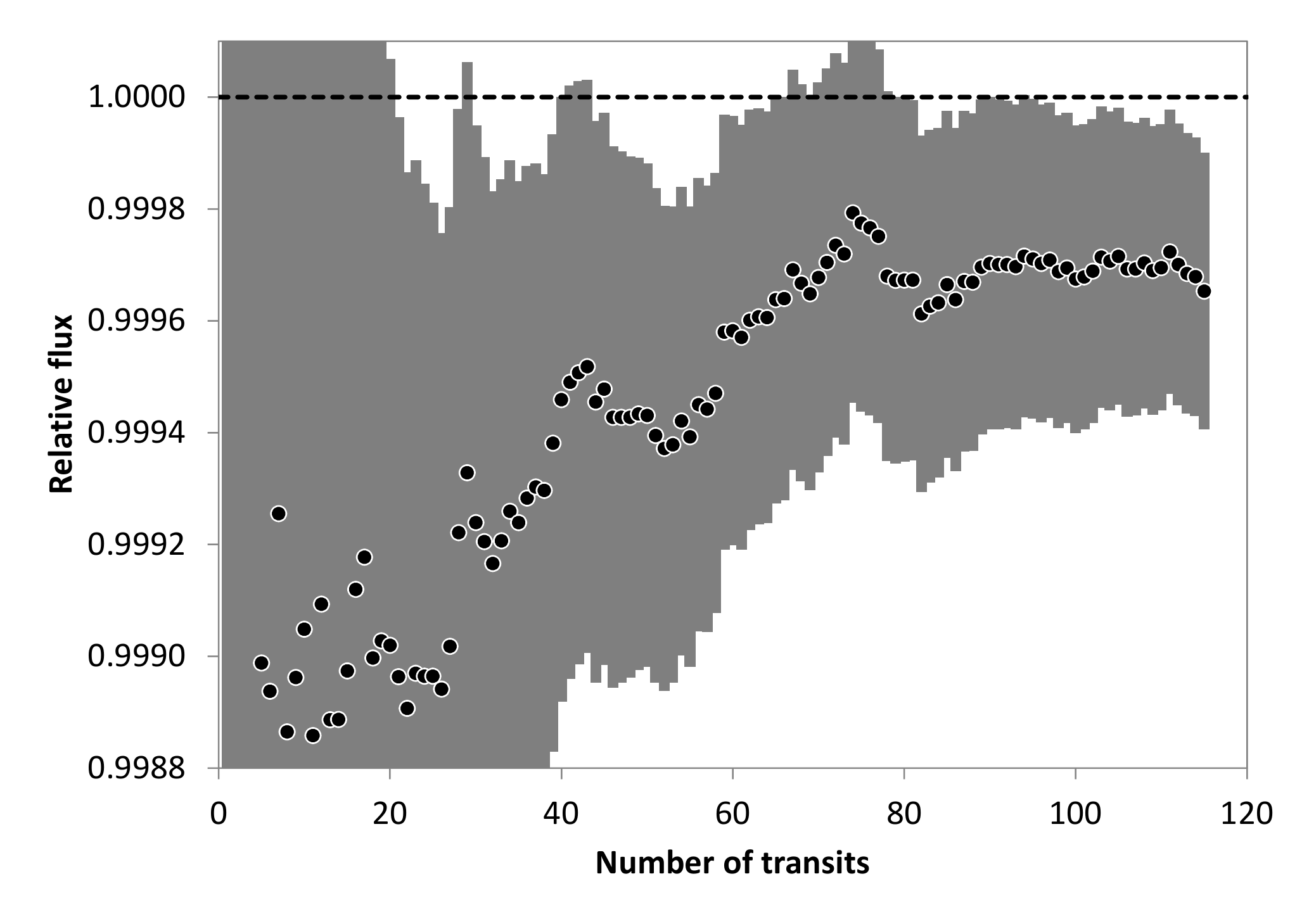}
\includegraphics[width=0.5\linewidth]{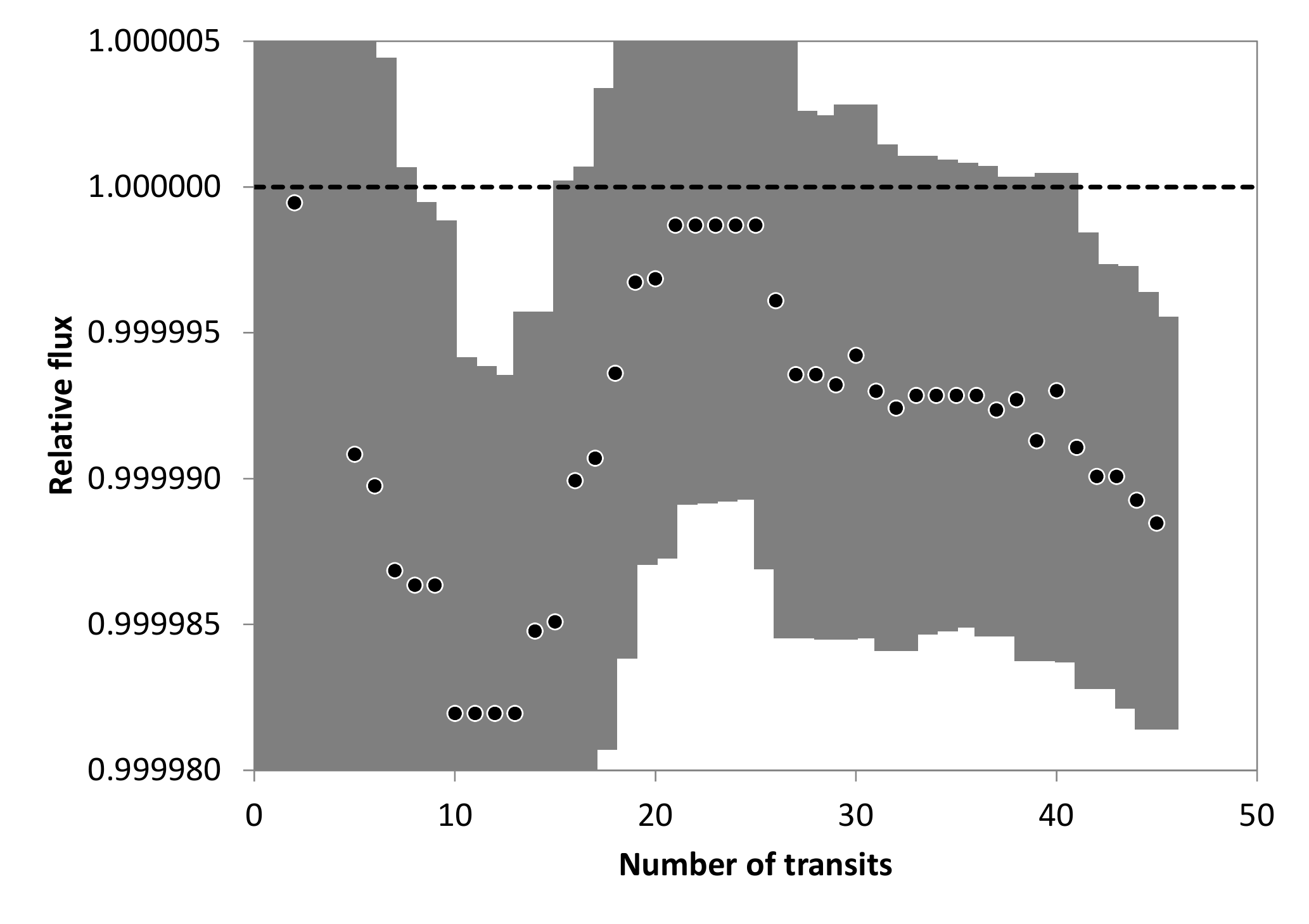}
\caption{\label{fig:buildup241b}OSE buildup for Kepler-241b (left) and KOI-367.01 (right) with 2$\sigma$ uncertainties as grey shades.}
\end{figure*}

\subsubsection{Buildup-plot}

Another test should be made to answer the question about how precisely the OSE builds up over time. Due to the very nature of the additive sampling of the effect, which includes non-transits, the dip \textit{cannot} be present after just a few transits. Such would indicate noise as the cause. On the other hand, the cumulative flux loss should build up within the first few 10s of transits. The exact number depends on the noise floor and frequency of moon transit, but should in general be between 10 and 50. After the buildup, no further (additional) flux loss should appear, but the cumulative depth of the dip should stabilize at its final level. Additional data should then only decrease the uncertainties, but not deepen the dip.

To test this, we find it useful to plot the number of transits versus the average flux loss, measured in an adequate bin before plus after planetary transit. For our example of Kepler-264b, we can see that the cumulative flux loss follows a random walk until transit \#13 (Figure~\ref{fig:buildup}). Afterwards, a dip begins in the cumulative data, becoming significantly negative from transit \#17 on (the chart shows the 2$\sigma$ (95\%) uncertainties). After transit \#19, the cumulative flux loss stabilizes at $\sim$30ppm. 

For Kepler-241c, we find that the OSE stabilizes after transit \#23 (Figure~\ref{fig:buildup241c}). The progression during the first few transits is skewed southwards due to a few outliers, but then stabilizes at the final level.

Of course, the start and the end of the observations could have happened at other points in time. To test the general build-up of the OSE, we have performed a Monte-Carlo test for Kepler-264b with $N=26 \times 10^6$ realizations to answer the question of how significant an OSE would be, on average, after only having observed randomly chosen $n$ of $N$=26 transits.  Figure~\ref{fig:buildup} shows the results for the 95\% significance level. It can be seen that, on average, 21 transits are required for a marginal detection. A more refined test for this question would be a cumulative Bayes factor map \citep{Gandolfi2015}. From a purely statistical view, one could also ask for stationarity in the data, i.e. test whether the time series follows a random-walk (null result) or converges with a trend. To test this, we employed the Augmented Dickey-Fuller test \citep{Dickey1984}. The test rejects the random walk hypothesis for Kepler-241c at the 99.9\%-level, and attributes an 80\% probability for a trend in Kepler-264b. Our null test for a random walk was the same dip of Kepler-264b at the phase-folded of -11d, for which the test gives only a 10\% probability to the existence of a real trend, as is expected. Such unit root tests are known to benefit from longer time-series, and the results for n=26 and n=40 should only be regarded as indicators.

Finally, we have repeated the tests for Kepler-241b and KOI-367.01 (Figure~\ref{fig:buildup241b}). For Kepler-241b, it can be seen that after $\sim$30 transits, a 2$\sigma$ flux loss appears. After $\sim$60 transits, it stabilizes at the final value. The behaviour for KOI-367.01 is more erratic; the significant dip after only $\sim$10 transits \textit{cannot} be real due to insufficient sampling. This must be caused by noise. The next dip occurs after $\sim$30 transits, and remains at the same level until the end of the observations. 

To sum up, we interpret these results in favor of the OSE interpretation (over noise) for our candidates.

\subsection{The scatter peak}
\label{sub:sp1}
The SP method employs a sliding boxcar median of the folded lightcurve to subtract the average transit shape. This removes any flux loss from OSE and planetary transit, and makes the SP an independent method.

\begin{figure}
\includegraphics[width=\linewidth]{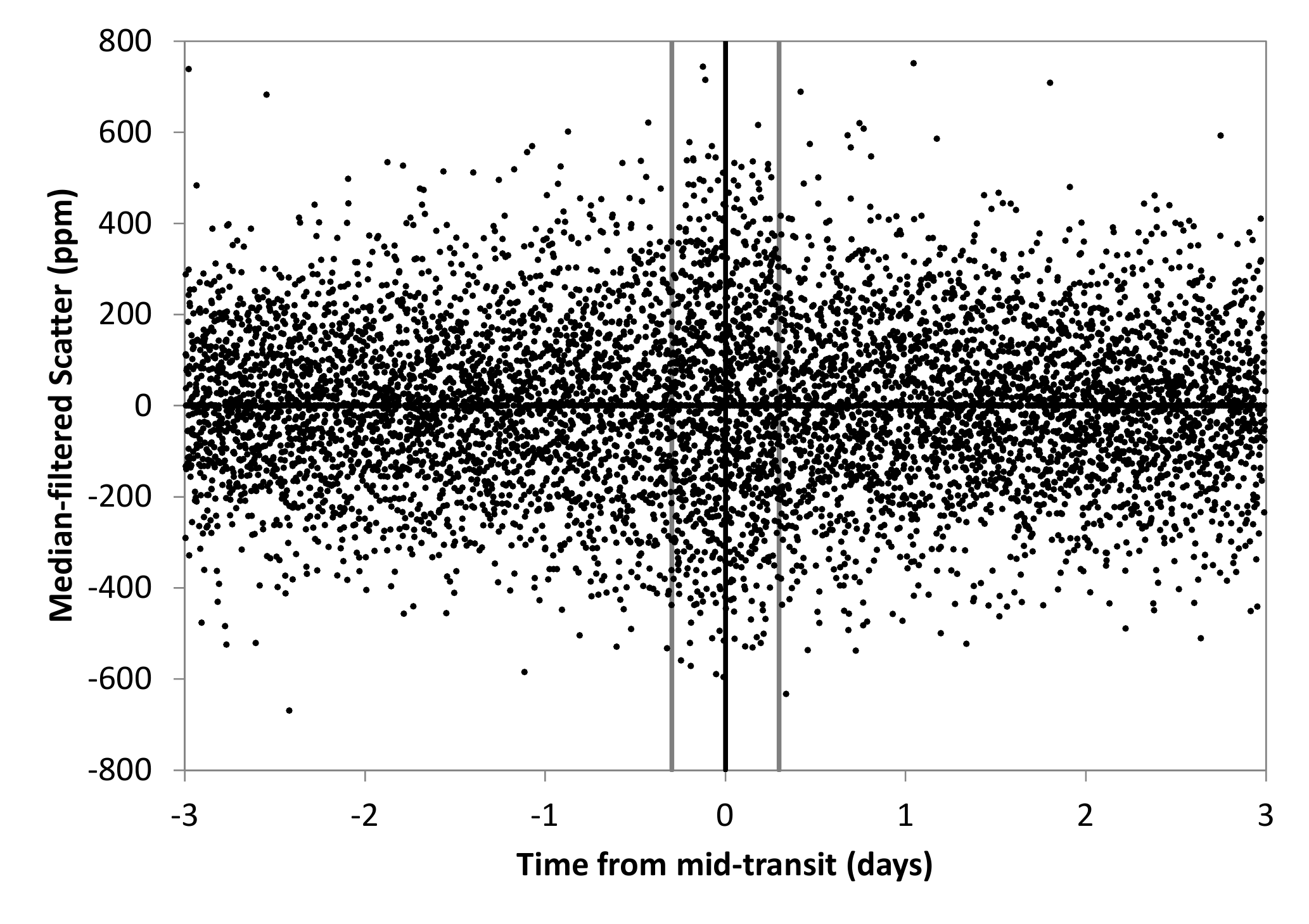}
\caption{\label{fig:scatter264b}Median-filtered scatter for Kepler-264b with false-positive scatter peaks at $>8\sigma$ significance due to improper detrending.}
\end{figure}

Unfortunately, we find that the trends in the \textit{Kepler} data make the analysis of the SP highly problematic. The raw SAP flux data contains instrumental trends on the order of $\sim$1\%, rendering any scatter on ppm scale invisible. Clearly, detrending is required. The approach described in section~\ref{sub:detrending}, where the actual transit (plus a bit of data before and after the actual transit to account for TTVs and TDVs, and to protect the moon dip) is masked, creates an artificial scatter peak in itself. This comes from the simple fact that some area is not detrended as well, as it is excluded from the fit, and has thus higher noise. To check this, one can fold to a plain area of no transit (where no scatter peak can be), and detrending with excluding some data then promptly creates a scatter peak (Figure~\ref{fig:scatter264b}).

For quiet stars, the Bayesian Maximum A Posteriori (MAP) approach (as explained in section~\ref{sub:detrending}) might be useful, as it performs the detrending using other reference stars. For our example of Kepler-264b, we are however left with noise from stellar spots on the order of $\sim$80ppm (section~\ref{sub:ose5}), which overshadows a 30ppm moon signal. 
 
We have considered two methods that can be applied in such cases.

\subsubsection{Subtraction of scatter template}
As the detrending creates a scatter peak anywhere in case of blinding the fit, this false positive peak can be created \textit{n} times at plain phase fold times, and its average height subtracted, like a template, from the time in question. In practice, unfortunately, the template shows a scatter increase of $\sim$5-50\%, while the real scatter peak within \textit{Kepler} noise is expected on the order of $\sim$1\%. We find that after subtraction of the template, all results are spurious, and the method to be useless.

\subsubsection{Temporal cleanup}
\begin{figure}
\includegraphics[width=\linewidth]{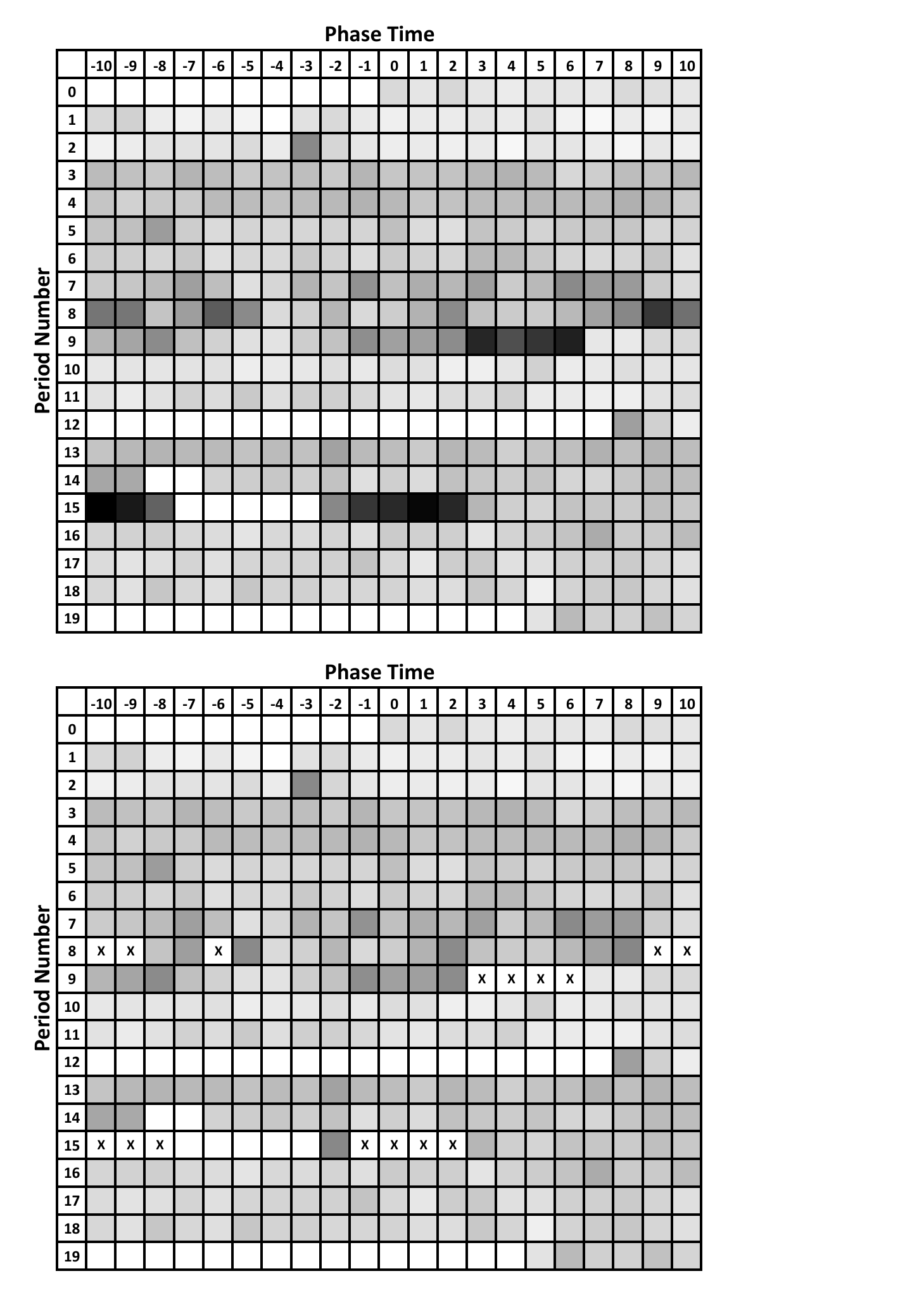}
\caption{\label{fig:scattermap}Scattermap for Kepler-264b short cadence data. On the horizontal axis, a time from -10d to +10d in the phase fold is shown, with a bin width of 1d. On the vertical axis, the 19 periods of SC data are displayed. Upper panel shows raw data, lower panel after cleanup of the 16 largest outliers (in black color before, with crosses afterwards).}
\end{figure}

If detrending is so problematic, one can try to avoid it and clean up the mess afterwards. This method uses the PDC-SAP data, without the instrumental trends but still containing occasional jumps from stellar activity. The idea is to create the scatter peak plot, and an additional map answering the question of its temporal origin (Figure~\ref{fig:scattermap}). This map can then be used to set a useful threshold, e.g. keeping the more quiet 80\% of the data. Then, one can cleanup the scatter peak plot with this decision.

For Kepler-264b, this method seems to work, as the scatter is relatively smooth afterwards. However, the noise level is still too high (at $\sim$50ppm) for the detection of the scatter peak, which is consistent with numerical simulations done by \citet{Simon2012}. Instead of 19 periods short-cadence data, about 50 to 100 are required. We conclude that no SP has been detected, for a lack of data, so that the test SP1 (significance) cannot be answered. The same is true for all other candidates. From our experience, the application of the scatter peak is not possible for stars that require heavy detrending on short ($<$10d) timescales. For longer timescales, very low amplitudes and very short transit durations, certain detrending methods might work, such as fitting polynomials or fully modeling the rotation with star-spots. The key is to exclude as little data as possible (during transits) from the fits. Such modeling, however, is beyond the scope of this paper.

\begin{figure}
\includegraphics[width=\linewidth]{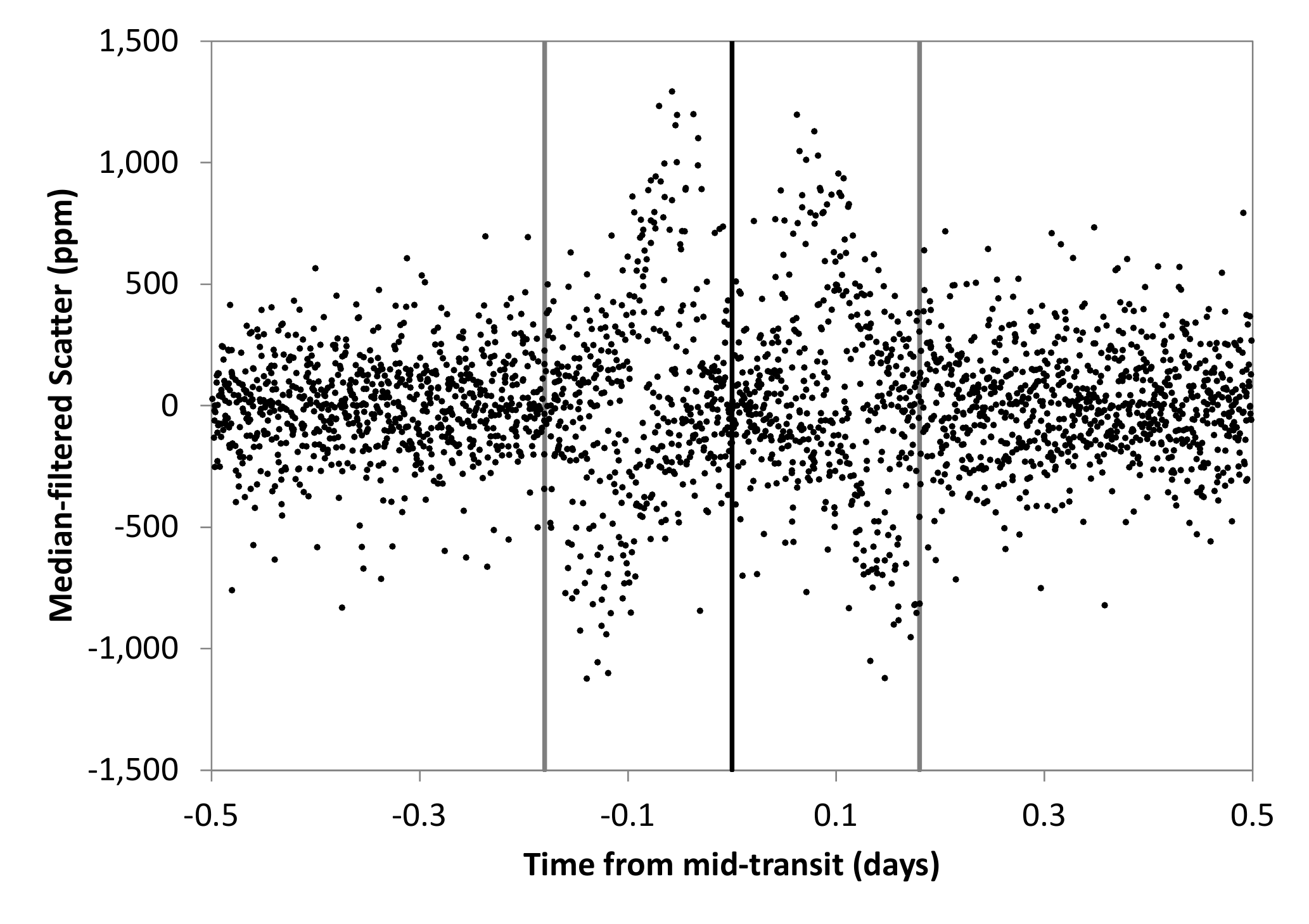}
\caption{\label{fig:scatter11d}Median-filtered scatter for Kepler-11d with false-positive scatter peaks at $>8\sigma$ significance due to an incorrect period.}
\end{figure}

\subsection{Results for criteria SP2: Period shift and SP3: TTVs+TDVs}
\label{sub:sp2}
There is one major other configuration that also causes a scatter peak: Imperfect transit folding. This can occur due to TTVs and TDVs, and also when using incorrect periods. One such case is Kepler-11d, where the published period is 22.6845$\pm$0.0009d \citep{Lissauer2013}. When using this period, the median-filtered phase fold shows clear signs of a folding error in the form of ``wings'' during ingress and egress (Figure~\ref{fig:scatter11d}). At first glance, this can be mistaken for a scatter peak, but the main difference is that it only appears on the time of ingress and egress, and not around mid-transit.  When correcting the period to P=22.687d, the scatter peak is completely gone. This makes for a good test: Shifting the period should not decrease the scatter peak, i.e. the scatter peak should not be removable by ``correcting'' the period. If it does, the period is wrong, and the scatter peak irrelevant. 

Using Kepler-264b as an example, we have also tried shifting the period, and done this for $\pm$0.1d, which is $\sim$190x the uncertainty of the period. Our step rate was 0.000001d. Around the best fit period, we found scatter peaks fluctuating by $\pm$10\%; an insignificant difference. Further away ($>|0.015|$d), the scatter peak increases significantly due to sidelobes of the folding error (see Figure~\ref{fig:scatter11d}). We conclude that the period is, within the errors, correct.

\subsection{Results for criteria SP4: Plausibility and C2: Temporal spread}
\label{sub:sp4}
\label{sub:c2}
Any detection of a scatter peak must be verified with numerical simulations, in order to assure that the peak height is in agreement with the expected value. This can be done as described in section~\ref{sub:injection}, by injecting transits into real data (without transits) and retrieving them for comparison. 

Following criteria C1, the data can be split the data in equally sized parts and performed the SP tests described above. For a lack of scatter peak detection, we have not pursued this task.

\begin{table*}
\small
\caption{Recommended test assignement\label{tab:testrecommendation}}
\begin{tabular}{lll}
\tableline
Test                       & Statistics  & Plots \\ 
\tableline
OSE1: Significance (l,r,b) & t-tests     & Phase-folds (Figure~\ref{fig:flux}) \\
OSE2: Slope (l,r)          & t-tests     & Phase-folds (Figure~\ref{fig:flux}) \\
OSE3: Stacking             & Monte-Carlo & River-plot (Figure~\ref{fig:river}), build-up plot (Figure~\ref{fig:buildup},~\ref{fig:buildup241c}) \\
OSE4: Uniqueness (a,b)     & t-tests     & Phase-fold in bins (Figure~\ref{fig:dips}) \\
OSE5: Star-spots           & Periodogram, Fourier & Time-amplitude graph (Figure~\ref{fig:rotation})\\
OSE6: Rings                & Density, Roche-radius & -- \\
OSE7: Plausibility         & --          & OSE-fit (Figure~\ref{fig:flux-ose-fits}, transit plot (Figure~\ref{fig:singletransit}) \\
C1: Temporal spread (OSE)  & Monte-Carlo, Dickey-Fuller & River-plot (Figure~\ref{fig:river}), build-up plot (Figure~\ref{fig:buildup}) \\
SP1: Significance          & t-test & SP-plot (Figure~\ref{fig:scatter264b}, smoothed) \\
SP2: Period shift          & --          & SP of phase-fold (Figure~\ref{fig:scatter11d}) \\
SP3: TTVs+TDVs             & --          & SP of phase-fold (Figure~\ref{fig:scatter11d}) \\
SP4: Plausibility          & Monte-Carlo (numerical) & -- \\
C2: Temporal spread (SP)   & Monte-Carlo & Scattermap (Figure~\ref{fig:scattermap})\\
\tableline
\end{tabular}
\end{table*}

\section{Discussion}
As summarized in Table~\ref{tab:testresults}, no OSE-like signal passes all tests. Nominally, Kepler-264b has the best result. However, its OSE fit gives the most extreme values for the moon radius ($R_{\leftmoon}=1.6\pm0.2R_{\oplus}$ orbiting a $3.33\pm0.74R_{\oplus}$ planet), raising serious doubt about its plausibility. While likely physically possible, it is unclear whether such a configuration can be real. On the plus side, different detrending methods give consistent results.

While it seems clear that Kepler-264b shows an OSE-like signal in the phase-folded light curve, one might argue that it is not caused by a moon. This poses the question, whether the feature is only due to red noise. As discussed in section~\ref{sec:req-ose}, we should expect $\sim$3 false-positives among the 958 suitable planets, from the dip depth alone. We had also argued that the consequent tests and criteria should decrease this number. For Kepler-264b, the test results are mixed, and we can thus not exclude a false-positive from time-correlated noise. After spending much time with this candidate, and looking at the data in many ways, we like to believe that perhaps part of the signal is due to moon(s), in combination with red noise, but this should be considered speculative.

The system of Kepler-241b and \textit{c} offers more plausible, but still very large putative moons ($1.1\pm0.1R_{\oplus}$ and $1.4\pm0.1R_{\oplus}$) orbiting Super-Earths. Both planets show limited significance of the individual dips (OSE1), and issues with the slopes of the dips (OSE2). On the other hand, both are easy to detrend and give similar results for all detrending methods.

Kepler-102e and 202c show marginal significance of the individual dips, and both are vulnerable to detrending options, with parabola detrending giving better OSE fits. Kepler-102e raises doubts of star-spot induced flux variations. 

For KOI-367.01 and Kepler-202c, the largest issue is certainly the fact that the parabola detrending gives a different result (no OSE), when compared to the median detrending. In this case, we have to assume a detrending artefact. The same is true for Kepler-102e, which also suffers from doubts of star-spot induced flux variations. On the other hand, it is interesting to note that planet \textit{e} is the heaviest in the Kepler-102 system; however it does not have the largest Hill radius: Using the data from \citet{Marcy2014}, the Hill radius for 102e is 433,000km, but that of 102f is larger at 483,000km. Therefore, one might argue that 102f would be more suited to host a moon. Also, in a best-fit moon orbit around 102e, at 300,000km, only retrograde moons would be stable. What is more, moon formation through accretion will only produce very small moons -- if any -- around the known planets around Kepler-102. This is because the planets are sub-Neptune in mass and very close to their star. This also makes moon formation via capture very challenging (e.g. \citet{Williams2013}). Consequently, a wide-orbit, large-radius moon around 102e is unlikely to exist.

The general mix of test passes and failures makes it very difficult to consider any one signal the most promising. From statistics, we have to assume that three of the six are false-positives, and tend towards rejecting Kepler-202c, Kepler-102e and KOI-367.01. 

For Kepler-241b and \textit{c}, and Kepler-264b, we recommend follow-up analysis using photodynamical modeling. For future work, we have summarized our test recommendations in Table~\ref{tab:testrecommendation}.

Regarding the super-stack, we should note that stars and moons are of different sizes, so that our results can only be taken as a very rough estimate. It is also unclear whether the non-detection for periods $<35d$ comes from a real lack of moons, or from different sensitivities, noise or errors. We tend to interpret the result in the way that planets with $P<35d$ have a lower probability for moons, and/or possess moons with smaller radii, so that the effect becomes hidden in the noise. Further studies will be required in the future, and the present result should be taken with caution. It can however be interpreted as an indication for the existence of exomoons among planets with periods $35d<P<80d$.

Finally, it should be noted that exomoons can reveal a planet's mass from Hill stability, using only photometry. Previous studies proposed transit timing variation \citep{Kipping2010} and the OSE \citep{Heller2014} to measure the mass of a planet, using its moon. To our knowledge, the Hill stability argument is new, and can be useful in the future as an additional test to constraint the masses of planets and moons.

\section{Conclusion}
``What is not disputed, is not particularly interesting''\footnote{``Was nicht umstritten ist, ist auch nicht sonderlich interessant.'' -- Johann Wolfgang von Goethe (1749 - 1832), German poet and philosopher.} -- as explained, we do not claim an exomoon detection in this paper, but only present methods and tests to consider them candidates. The detection of exomoons is at the limit of current photometry, and its methods and tests need to be checked, challenged and verified. The methods at work (OSE and SP) rely on phase-folded information, so that the individual TTVs, TDVs and flux loss of the transits is lost. On the positive side, time-dependent red noise is reduced by stacking. A competing search strategy, as employed by \citet{Kipping2014}, avoids phase-folding but instead uses a photodynamical approach and brute-force calculates and fits all possible configurations. Such a search is beyond the scope of this paper, but could ultimately be used to (in)validate promising results from OSE and SP. As the photodynamical method is computationally expensive, OSE+SP can deliver candidates for further investigation. Every method has its advantages and disadvantages; its assets and drawbacks. On a side note, the OSE can readily detect and characterize multi-moon systems including the radii and orbits (given high-enough data quality), as successfully demonstrated by \citet{Heller2014}, which is not possible with the current photodynamical method.

For the super-stack of all suitable planets of $35d<P<80d$, we find an average flux loss of $6\pm2\times10^{-6}$ (6ppm), corresponding to an average total exomoon radius of $2120\substack{+330\\-370}$km (0.33R${_\oplus}$) for the average star radius of 1.24$R_{\odot}$ in this sample. The analysis is less sensitive to longer-period planets with four years of \textit{Kepler} data, but there seems to be a hint of a lack of putative moon signal in planets with $P<35d$.

At first glance, the high moon radii of the individual putative moons seem implausible, as there is no such large moon in our own solar system. However, our solar system might not be the norm -- we have no Hot Jupiters, warm Neptunes, or Super-Earths in our solar system, and thus no reference for typical moons around such planets. Also, there is a strong selection bias, based on the detection limits (section~\ref{sub:sensitivity}), and in addition the simple fact that the strongest dips are most significant. The first moons to be found will likely be at the long (large/massive) end of exomoon distribution, as was the case for exoplanets.

If all candidates shown in this work turn out to be noise only (or detrending errors), then we should still consider the OSE as a powerful tool for the detection of exomoons or binary planets. Such a result would point towards necessary improvements in detrending algorithms, red noise treatment and modeling stellar noise, including spots and rotation. On the other hand, confirmation of any one candidate would make it an immensely useful method. In our solar system, a wide variety of large and small moons exists. This, together with formation theory, may lead to the conclusion that moons do exist outside of our solar system, i.e. $\eta_{Exo\leftmoon}>0$. The recent characterization of the ring system around J1407b \citep{Kenworthy2015}, potentially hosting moons, seems to support this. Consequently, a single exomoon confirmation would be a landmark in exoplanet science, and establish $\eta_{Exo\leftmoon}>0$, with $\eta_{Exo\leftmoon}$ as the fraction of exoplanets which host at least one moon. In our solar system, six of eight planets host moons (but not Mercury and Venus), giving $\eta_{Solar\leftmoon}=0.75$ (although Mars' small moons might be undetectable). As such, the potential putative moons discussed in this paper should be regarded only as \textit{candidates} (or, if the more cautious reader prefers, \textit{signals}) worth further investigation.

\acknowledgements
\section*{Acknowledgements}
We thank the anonymous referee for an extensive review, which improved the quality of this paper. This includes the idea to evaluate the occurrence rate of mutual planet-moon eclipses leading to missing moon transit dips, and the comparison to the maximum value of 6.4\% of transits. 

We thank Andrew Howard, Robert Szab\'o, Gyula Szab\'o and Attila Simon for their feedback on an earlier draft, which increased the quality of this paper. We also thank David Kipping for pointing out some of the assets and drawbacks of the OSE method.

\end{document}